\documentclass[%
preprint,
nofootinbib,
 amsmath,
 amssymb,
rmp,
]{revtex4-1}

\usepackage{graphicx}
\usepackage{dcolumn}
\usepackage{bm}
\usepackage{hyperref}
\usepackage{breakurl} 
\usepackage{mathrsfs} 
\newcommand{\olsi}[1]{\,\overline{\!{#1}}}

\begin{document}


\title{X-ray phase-contrast imaging: a broad overview of some fundamentals}

\author{David M. Paganin}
 \affiliation{School of Physics and Astronomy, Monash University, Victoria 3800, Australia}
 \email{david.paganin@monash.edu}
\author{Daniele Pelliccia}%
\affiliation{Instruments \& Data Tools Pty Ltd, Victoria 3178, Australia}%
\email{daniele@idtools.com.au}


\begin{abstract}
We outline some basics of imaging using both fully-coherent and partially-coherent X-ray beams, with an emphasis on phase-contrast imaging. We open with some of the basic notions of X-ray imaging, including the vacuum wave equations and the physical meaning of the intensity and phase of complex scalar fields. The projection approximation is introduced, together with the concepts of attenuation contrast and phase contrast.  We also outline the multi-slice approach to X-ray propagation through thick samples or optical elements, together with the Fresnel scaling theorem.  Having introduced the fundamentals, we then consider several aspects of the forward problem, of modelling the formation of phase-contrast X-ray images.  Several topics related to this forward problem are considered, including the transport-of-intensity equation, arbitrary linear imaging systems, shift-invariant linear imaging systems, the transfer-function formalism, blurring induced by finite source size, the space--frequency model for partially-coherent fields, and the Fokker--Planck equation for paraxial X-ray imaging.  Having considered these means for modelling the formation of X-ray phase-contrast images, we then consider aspects of the associated inverse problem of phase retrieval.  This concerns how one may decode phase-contrast images to gain information regarding the sample-induced attenuation and phase shift. 
  \newline\newline Fifteen video lectures, based on a preliminary version of this chapter, are available online at: \url{https://bit.ly/2GdoVg8}.
 \end{abstract}

\maketitle

\tableofcontents

\section{Introduction}

Phase contrast, namely the visualisation of transparent structures that is induced by the refraction of rays passing through them, has been known for millennia in the visible-light domain.  The heat shimmer that can be seen over hot desert sands, together with the exquisite network of sunlit caustics dancing about the floor of a clear pool of water on a sunny day, are two examples that immediately spring to mind. Such phase-contrast phenomena rely on the fact that sunlight has a reasonable degree of spatial coherence (i.e.~a sufficiently small angular diameter), once it has propagated to the surface of the earth. Twinkling starlight is another example of phase contrast in the visible-light domain.  

All of the above-listed examples of phase contrast may be classified under the moniker of ``propagation-based phase contrast'' since free-space propagation is an essential ingredient in such phenomena.  There are a plethora of other means via which phase contrast may be achieved.  Such more recently discovered modalities of phase contrast include---but are certainly not limited to---interference, inline holography, schlieren imaging and Zernike phase contrast (Born \& Wolf, 1999).

Phase contrast, propagation-based or otherwise, is not restricted to the visible-light domain.  The phenomenon can also be observed for a variety of radiation and matter wave fields, such as electrons (Cowley, 1995), neutrons (Klein \& Opat, 1976) and X-rays (Bonse \& Hart, 1965; White \& Cerrina, 1992; Snigirev, Snigireva, Kohn, Kuznetsov, \& Schelokov, 1995).

Our focus, here, is on X-ray phase-contrast imaging.  We seek to broadly introduce some aspects of this field, from a tutorial perspective. The primary intended audience is those commencing research in the field.  However, we hope this largely self-contained chapter to be both more broadly accessible and useful to a wider audience.  The coverage however is in no way comprehensive and we make extensive reference to existing literature for topics not covered in detail here. 

Section 2 deals with X-ray imaging basics, sketching a passage from the Maxwell equations of classical electrodynamics, through to the paraxial wave equation describing coherent scalar X-ray fields.  We also introduce the projection approximation, Fresnel diffraction, absorption contrast and phase contrast.  We examine, from a practical perspective, the validity conditions of the projection approximation for paraxial X-ray imaging, including the conditions under which this approximation is likely to break down.  Some attention is given to the multi-slice formulation of X-ray scattering, for samples that are sufficiently thick for the projection approximation to no longer be applicable.  We also consider the Fresnel scaling theorem, which gives a simple mapping between (i) the coherent X-ray intensity image (Fresnel diffraction pattern) recorded when a thin sample is illuminated using normally-incident plane waves, and (ii) the Fresnel diffraction pattern recorded when the same sample is illuminated using a spherical wave front in either a magnifying or de-magnifying geometry.  

Section 3 deals with elements of X-ray phase-contrast imaging, specifically with the ``forward problem'' of modelling images obtained using a variety of coherent X-ray imaging scenarios. We begin this section with an outline of the transport-of-intensity equation, which is tied to one of the common phase-contrast methods, namely propagation-based X-ray phase contrast in the regime of small object-to-detector propagation distance.  Rather than subsequently considering in detail a multiplicity of other (equally-powerful) methods for X-ray phase-contrast imaging, we instead generalise a wide class of such phase-contrast imaging systems, by considering many of them to be particular examples of shift-invariant coherent linear imaging systems. We outline the associated transfer function concept, and the realisation of X-ray phase-contrast imaging in such a general setting. We also explain how the effects of partial coherence may be introduced, at least when spatial coherence results from a finite source size, via source-size blur.  As a more sophisticated means by which the effects of partial coherence may be incorporated into the modelling of phase-contrast X-ray imaging systems, we outline the space--frequency description of partial coherence.  We close this section by considering how the transport-of-intensity equation may be extended into a Fokker--Planck equation that is able to account for the effects of both (i) coherent energy flow downstream of an illuminated sample, in the form of phase contrast as well as attenuation contrast, and (ii) diffusive energy flow downstream of the sample, in the form of position-dependent fans of small-angle X-ray scattering distributions (SAXS) that may emerge from each point on the exit surface of the sample.  Both isotropic (i.e.~rotationally symmetric) and non-isotropic (i.e.~elliptical and therefore rotationally asymmetric) position-dependent SAXS fans are considered in the Fokker--Planck extension to the transport-of-intensity equation.    

Having studied the forward problem of X-ray phase-contrast imaging, we are able to consider the corresponding ``inverse problem'' in Section 4.  Broadly speaking the inverse problem seeks to address the question of what one can infer regarding a sample (or, more generally, a wave field that is incident upon a specified imaging system) by decoding measured intensity maps, such as one or more X-ray phase-contrast images that are output by a specified system.  We indicate some key concepts in the theory of inverse problems, with particular emphasis on the idea of an inverse problem being well posed in the sense of Hadamard.  Having established this broader context for the notion of inverse problems in general, and inverse problems of imaging in particular, we then focus attention upon the inverse imaging problem of phase retrieval.  Phase retrieval, as the term implies, deals with the particular imaging inverse problem of recovering wave-field phase given one or more intensity measurements, such as those that might be output by a phase-contrast imaging system.  Some examples of phase retrieval are briefly considered.  These examples are based on the phase-contrast transfer-function model, and the transport-of-intensity equation model.  For the former case we also very briefly indicate how the effects of partial coherence may be incorporated, using (i) the idea of a coherence envelope, and (ii) smearing spatially unresolved signal over the transverse extent of a system point spread function. For the latter case, namely the transport-of-intensity model, the effects of partial coherence may be incorporated using the Fokker--Planck model described in the previous section.  Throughout these discussions, we emphasise that no one method of X-ray phase-contrast imaging is superior to all others in all circumstances, arguing rather that each have their relative strengths and limitations.  Also, we briefly mention the manner in which various means of phase retrieval may be viewed in information-optics terms, as well as in terms of the closely related concept of virtual optics. Lastly, we consider the role of spatially-unresolved object micro-structure, in the context of X-ray phase-contrast imaging, together with the connection of this concept to the previously-mentioned position-dependent SAXS fans. 

Further detail, on many of the topics presented here, is available in the textbook by Paganin (2006).  We again emphasise that we do not claim in any way to be giving a representative overview of the field. Rather, this chapter is intended as an introductory overview of some key aspects of X-ray phase-contrast imaging, which contains enough entry points to the published literature to empower a journey of further exploration and discovery.  

\section{X-ray imaging basics}

In the present section we cover some basics of coherent X-ray imaging, including the differential equations describing X-ray waves as well as their interactions with matter, the projection approximation, Fresnel diffraction and phase contrast.  We first consider the vector wave equations that govern the propagation of electromagnetic fields (i.e.~electric-field vectors  and magnetic-field vectors) as they evolve through space and time.  We then consider an often-made simplification, in which the vector-field description is replaced with a scalar-field description.  In this latter description, which ignores the effects of X-ray polarisation, the electromagnetic disturbance is modelled by a complex number at each point in space, at each instant of time.  We shall see that the magnitude and phase, respectively, of such a complex-wavefield descriptor is directly related to the intensity and wave-front profile of the X-rays.  We then introduce another important simplification, that of the fully coherent field.  We also consider the case of paraxial fields, namely beam-like X-ray fields whose associated rays (wavefront normals) may be taken as being very close to parallel to a fixed optical axis.  The diffraction of paraxial coherent fields, as modelled via the formalism of Fresnel diffraction, is also considered.  The related concepts of absorption-contrast imaging and phase-contrast imaging are then introduced in fairly general terms, as means of imaging that are respectively sensitive to the magnitude and phase, of the complex field used to describe X-ray light in the scalar approximation mentioned above.  The manner in which X-rays interact with matter, such as the material in samples being imaged or in X-ray optical elements, is also considered.  Two important approximations for such light--matter interactions are studied, namely the projection approximation and its generalisation to the multi-slice approximation.  Taken together, the above-listed suite of related concepts constitutes the ``X-ray imaging basics'' referred to in the title to this section.

\subsection{Vector vacuum wave equations} 

The Maxwell equations, which govern the evolution of classical electromagnetic fields in space and time, lead to the following d'Alembert equations for the electric field ${\bf E}(x,y,z,t)$ and magnetic field ${\bf B}(x,y,z,t)$ in free space:
\begin{eqnarray}\label{eq:WaveEquationsElectric}
\left(\frac{1}{c^2}\frac{\partial^2}{\partial
t^2}-\nabla^2\right){\bf E}(x,y,z,t)={\bf 0},
\\ \label{eq:WaveEquationsMagnetic}\left(\frac{1}{c^2}\frac{\partial^2}{\partial
t^2}-\nabla^2\right){\bf B}(x,y,z,t)={\bf 0}.
\end{eqnarray}
\noindent Here, $(x,y,z)$ are Cartesian spatial coordinates, $t$ is time, $c$ is a speed given by the  Maxwell relation
\begin{equation}\label{eq:SpeedOfLight}
    c=\frac{1}{\sqrt{\mu_0\varepsilon_0}},
\end{equation}
\noindent $\varepsilon_0$ is the electrical permittivity of free space and $\mu_0$ is the magnetic permeability of free space. The Laplacian in three spatial dimensions is
\begin{equation}
\nabla^2=\frac{\partial^2}{\partial x^2}+\frac{\partial^2}{\partial y^2}+\frac{\partial^2}{\partial z^2}. 
\end{equation}
Syst\`{e}me-Internationale (SI) units are used throughout. 
  
The d'Alembert equations imply two facts which were quite revolutionary when first discovered: (i) electromagnetic disturbances propagate as waves in vacuum; (ii) the speed $c$ of these electromagnetic waves, given by the Maxwell relation (Eq.~(\ref{eq:SpeedOfLight})), coincides so closely with the speed of light in vacuum, as to very strongly suggest that light is an electromagnetic wave.  In the late nineteenth-century context in which it was derived, this then-radical observation unified what were previously thought to be three separate bodies of physics knowledge: electricity, magnetism and (visible-light) optics.  

This was indeed a colossal moment in the history of physics, that is worth savouring just a little further. Before the advent of the Maxwell equations and the associated discovery that visible light is an electromagnetic disturbance, there were five separate mathematical theories describing aspects of the physical world: electricity, magnetism, (visible-light) optics, thermodynamics and mechanics.  With both the Maxwell equations and the discovery that light is an electromagnetic wave, the first three of these theories united into one overarching theory of electromagnetism and electromagnetic waves.  Such a unification remains a guiding light in much contemporary high-energy physics, such as the Lagrangian-field-theory formulation of the electro-weak model and quantum chromodynamics, as well as forming an inspiration for modern quests to unify quantum theory with Einstein's general theory of relativity (Freund, 1986; Maggiore, 2005; Mandl \& Shaw, 2010; Thomson, 2013). 

Returning to the main thread of our argument, we now know that the class of electromagnetic waves is not exhausted by those that are visible to the human eye. Of particular focus to us are hard X-ray electromagnetic waves. From now on, unless specified otherwise, we will deal with hard X-ray waves. While the underlying formalism is obviously the same as for visible light, at least in the fundamental sense that both visible light and X-rays may be described at a classical level via the Maxwell equations, a number of approximations are peculiar to the short-wavelength regime we are focusing on here.   

\subsection{Scalar vacuum wave equation and complex wave-function} 

Equations~(\ref{eq:WaveEquationsElectric}) and (\ref{eq:WaveEquationsMagnetic}) are a pair of vector equations, or, equivalently, a set of six scalar equations: three for the Cartesian components of the electric field, and three for the Cartesian components of the magnetic field.  Each of these six scalar vacuum field equations has the form:  
\begin{eqnarray}\label{eq:WaveEquationComplexDisturbance}
\left(\frac{1}{c^2}\frac{\partial^2}{\partial
t^2}-\nabla^2\right)\Psi(x,y,z,t)=0.
\end{eqnarray}

It is convenient to treat $\Psi(x,y,z,t)$ as a complex function, termed the ``wave-function'', which describes the X-ray field.  Only the real part of this wave-function is physically meaningful, but we will not need to make use of this fact in the formalism described here. By transitioning from a vector-wave description to a scalar-wave description of the X-ray field, polarisation is implicitly neglected (or a single linear polarisation is implicitly assumed). This assumption is often reasonable in many paraxial imaging and diffraction contexts.  Notwithstanding this, there are many cases (e.g.~magnetic scattering of circularly-polarised X-rays, and dynamical diffraction from near-perfect crystals) where the effects of X-ray polarisation must be taken into account.  Such polarisation-dependent phenomena, while important and interesting, will not be treated further here.

\subsection{Physical meaning of intensity and phase}

At each point $(x,y,z)$ in space, for each instant of time $t$, $\Psi(x,y,z,t)$ is a complex number. As such, it has magnitude and phase, so we may write:
\begin{equation}\label{eq:MadelungDecomposition}
\Psi(x,y,z,t)=\sqrt{I(x,y,z,t)}\exp[i\phi(x,y,z,t)].
\end{equation}
In the above expression, the magnitude of $\Psi(x,y,z,t)$ has been written as $\sqrt{I(x,y,z,t)}$, so that
\begin{equation}
I(x,y,z,t)=|\Psi(x,y,z,t)|^2,
\end{equation}
\noindent where $I(x,y,z,t)$ is the intensity of the field.  The phase of $\Psi(x,y,z,t)$ has been denoted by $\phi(x,y,z,t)$. For the instant of time $t$, surfaces of constant phase may be identified with wave-fronts of the X-ray field.  These fronts, in free space, move at very close to the speed $c$ of a plane wave of light in vacuum.  Note that the speed of a structured light beam in vacuum is in general slightly different from the speed of a plane wave in vacuum (Giovannini et al., 2015).  The wave-fronts move in a direction that is typically away from the source generating the waves.  See Fig.~1(a) for a pictorial representation of these concepts.  

\begin{figure}
\includegraphics[scale=0.38]{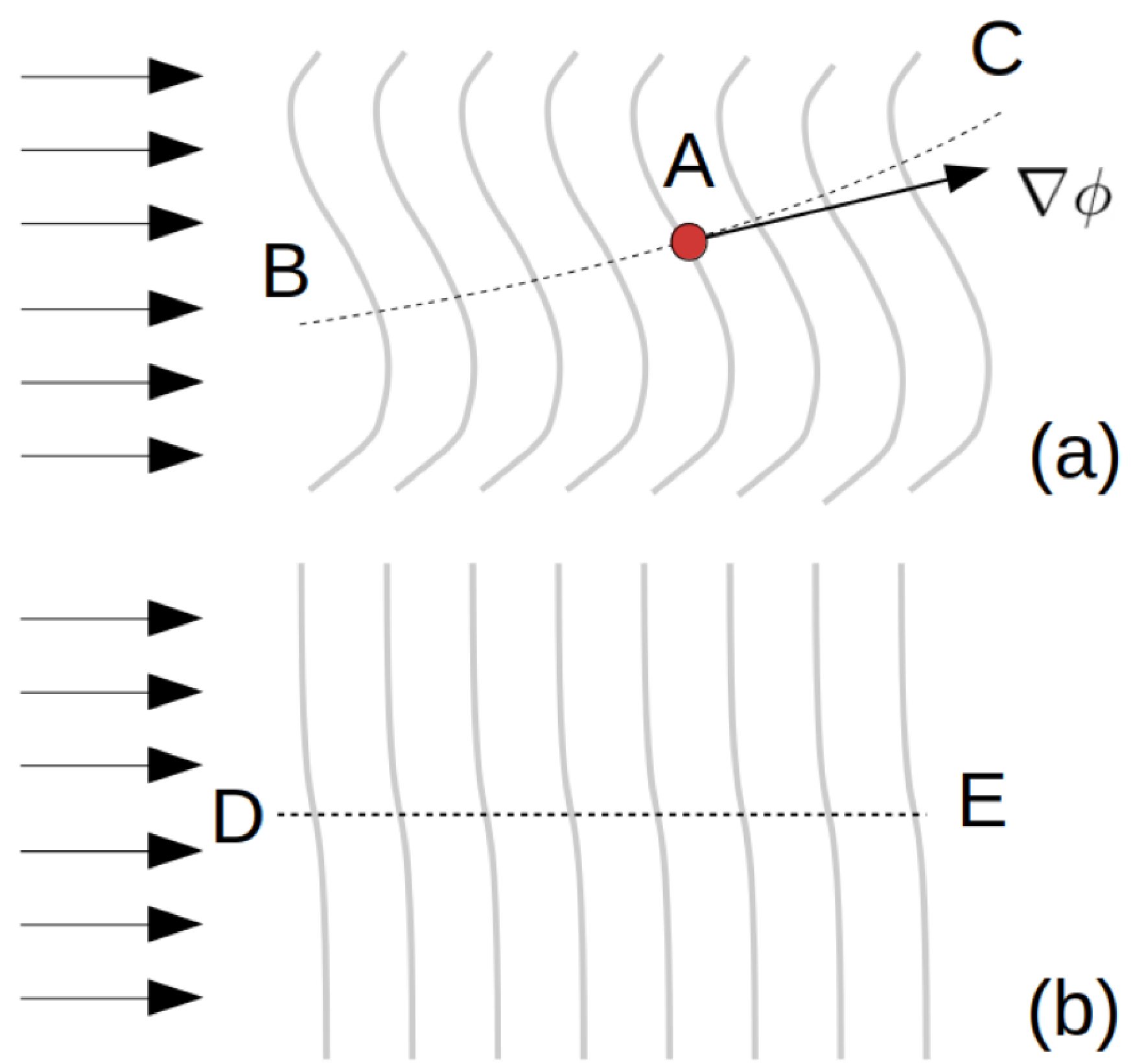}
\caption{X-ray wave-fronts.  (a) The wave-fronts, namely the surfaces of constant phase, are indicated by the series of curved surfaces.  At any point $A$ the wave-fronts move away from the source as time increases.  The direction of energy flow at point $A$ is $\nabla\phi(x,y,z,t=t_0)$. The current density (Poynting vector) is everywhere tangent to the associated streamlines, such as the streamline $BAC$ passing through the point $A$.  (b) A depiction of a paraxial field. Its wave-fronts differ only slightly from planar surfaces perpendicular to the optical axis (i.e.~the $z$ axis, which runs from left to right). The associated streamlines such as $DE$ are well approximated by straight lines parallel to the optical axis.  Image adapted from Paganin (2006).}
\label{fig:XRayWaveFronts}
\end{figure}

The moving X-ray wavefronts transport energy through space and time. This flow of optical energy, which may be thought of as a linear momentum density or local current density, may be described by the Poynting vector at the instant of time $t=t_0$: 
\begin{eqnarray}
{\bf S}(x,y,z,t=t_0)\propto I(x,y,z,t=t_0) \nabla \phi(x,y,z,t=t_0). \quad \label{eq:PoyntingVector}
\end{eqnarray}
This Poynting vector, or energy-flow vector, is  perpendicular to each wave-front, and has a magnitude that is proportional to the intensity of the flowing X-ray wave.  

The X-ray current density, as embodied in Eq.~(\ref{eq:PoyntingVector}), has a very strong analogy with the current density used in classical fluid mechanics.  It is worth dwelling on this connection so as to build up some intuitive understanding of the physical meaning of Eq.~(\ref{eq:PoyntingVector}).  In a classical fluid, such as flowing water, the current density $\bf{j}$ at a given point will be proportional to the product of (i) the mass density (mass per unit volume) $\rho$ at that point, and (ii) the velocity of the fluid $\bf{v}$ at that point.  Moreover, if the fluid flow is irrotational (no local ``twisting'') then this velocity can be written as the gradient of a scalar ``velocity potential'' $\xi$.  Hence, for a classical fluid that transports mass as it flows through space and time, the current density can be written as $\bf{j}=\rho\nabla\xi$.  

Moving back to the domain of X-ray energy flow, the intensity $I$ of the X-rays is analogous to the mass density $\rho$ of a flowing fluid.  The velocity vector $\bf{v}=\nabla\xi$ of the classical fluid is analogous to the local X-ray direction $\nabla\phi$.  Also, as a shift in terminology, we speak of the Poynting vector $\bf{S}$ for the evolving electromagnetic field, rather than the current density $\bf{j}$ for the evolving classical fluid.  With these identifications, we see that Eq.~(\ref{eq:PoyntingVector}) for the X-ray Poynting vector is exactly analogous to the corresponding expression $\bf{j}=\rho\nabla\xi$ for the current density of an irrotational classical fluid.  

Another connection, between flowing fluids and propagating X-ray beams, arises via the concept of streamlines associated with mass flow (or energy flow).  See Fig.~1 once again, together with the explanations in the associated caption.  Conceptual connections, between mass flow in a classical fluid and energy flow in a propagating X-ray beam, could be pursued further, e.g.~by pointing out that both the classical-fluid current density and the X-ray Poynting vector have an energy-conservation property that is modelled by continuity equations that are mathematically identical in form.  However such further conceptual connections, between flowing mass in classical fluids and flowing energy in X-ray fields, will not be examined in further detail here.  For additional information on this topic, see e.g.~Berry (2009) as well as Paganin \& Morgan (2019), together with references contained therein.         

\subsection{Fully coherent fields}

Assume the field to be strictly monochromatic, and therefore perfectly coherent, so that its time development at any point in space is given by oscillations with a fixed angular frequency $\omega$:
\begin{equation}\label{eq:MonochromaticField}
\Psi(x,y,z,t)=\psi_{\omega}(x,y,z)\exp(-i\omega t).
\end{equation}
\noindent Here,
\begin{equation}
\omega = 2\pi f = ck,  
\end{equation}
\noindent $f$ denotes temporal frequency, and $k$ is the wave-number corresponding to the vacuum wavelength $\lambda$:
\begin{equation}
k=2\pi/\lambda.  
\end{equation}

Substitution of Eq.~(\ref{eq:MonochromaticField}) into Eq.~(\ref{eq:WaveEquationComplexDisturbance}) gives the Helmholtz equation in vacuum:
\begin{equation}\label{eq:Helmholtz_Equation}
    \left(\nabla^2+k^2\right)\psi_{\omega}(x,y,z)=0.
\end{equation}

This vacuum wave equation for coherent scalar electromagnetic waves may be generalised to account for the presence of material media.  Such media are here assumed to be static, non-magnetic, and sufficiently slowly spatially varying, so that they may be described by a position-dependent refractive index $n_{\omega}(x,y,z)$.  This refractive index alters the vacuum wavelength as follows:
\begin{equation}
\lambda\longrightarrow\frac{\lambda}{n_{\omega}(x,y,z)},
\end{equation}
\noindent hence 
\begin{equation}
k\longrightarrow k \, n_{\omega}(x,y,z). 
\end{equation}
The vacuum Helmholtz equation (Eq.~(\ref{eq:Helmholtz_Equation})) therefore becomes the Helmholtz equation in the presence of non-magnetic static scattering media:
\begin{eqnarray}\label{eq:ScalarHelmholtzEquationInRefractiveMedium}
    \left[\nabla^2+k^2n_{\omega}^2(x,y,z) \right]
    \psi_{\omega}(x,y,z)=0.
\end{eqnarray}
See e.g.~Paganin (2006) for a full derivation of the above equation, which elaborates on the key assumptions that the scattering medium be (i) linear, (ii) isotropic, (iii) static, (iv) non-magnetic, (v) have zero charge density and (vi) zero current density, and (vii) be spatially slowly varying in its material properties.

As an interesting aside, note that Eq.~(\ref{eq:ScalarHelmholtzEquationInRefractiveMedium}) is mathematically identical in form to the time-independent Schr\"{o}dinger equation for non-relativistic electrons in the presence of a scalar scattering potential (this latter equation assumes that the effects of electron spin can be ignored, and that the material with which the electron interacts is non-magnetic (Bransden \& Joachain, 1989)).  Hence, the research fields of coherent X-ray optics and transmission electron microscopy have much in common.  An analogous comment may be made with regard to the connections between coherent X-ray optics and coherent neutron optics, since the time-independent form of the Klein--Gordon equation (for neutrons, ignoring their spin) is again mathematically identical in form to Eq.~(\ref{eq:ScalarHelmholtzEquationInRefractiveMedium}) (Mandl \& Shaw, 2010).  We round off this brief indication of the strong foundational connections that exist between the fields of X-ray optics, electron optics and neutron optics, with a quote from J.M.~Cowley's famous book on diffraction physics.  Here, the author speaks of ``the possibility for a unified treatment of the different branches of diffraction physics, employing electrons, X-rays or neutrons'' (Cowley, 1995).   

As a second aside, we recall the statement invoked in deriving Eq.~(\ref{eq:ScalarHelmholtzEquationInRefractiveMedium}), namely the three assumptions that the scattering media are ``static, non-magnetic, and sufficiently slowly spatially varying, so that they may be described by a position-dependent refractive index''.  The breakdown of any or all of these three key assumptions leads to extremely interesting generalisations.  For example, 
\begin{itemize}
    \item the breakdown, of the assumption of a static sample, enters us into the realm of time-dependent samples, including those that experience radiation damage during the act of X-ray imaging; 
    \item the breakdown, of the assumption of a non-magnetic sample, is key to the study of magnetic materials using, for example, circularly polarised X-rays; 
    \item the breakdown, of the assumption of a slowly spatially varying sample, will become progressively more important as X-ray imaging is pushed more and more often to regions of high resolution, e.g.~on nanometre and smaller length scales.
\end{itemize}

\subsection{Coherent paraxial fields} 

A paraxial field is a special case of propagating field, in which all wave-fronts may be obtained by slightly deforming planar surfaces perpendicular to the optical axis (i.e.~the $z$ axis).  With reference to Fig.~1(b), all Poynting vectors are close to being parallel to the optical axis (hence the term ``paraxial''). The associated streamlines such as $DE$ are well approximated by straight lines parallel to the optical axis.

Assume our monochromatic complex scalar X-ray wave-field to be paraxial, in the sense just described. Under this approximation it is natural to express the complex disturbance ${\psi}_{\omega}(x,y,z)$ as a product of a $z$-directed plane wave $\exp(ikz)$, and a perturbing envelope  $\tilde{\psi}_{\omega}(x,y,z)$.  We then have:
\begin{equation}\label{eq:EnvelopeForPlaneWave}
{\psi}_{\omega}(x,y,z)\equiv\tilde{\psi}_{\omega}(x,y,z)\exp(ikz).
\end{equation}
In the above expression, we have an underlying plane wave $\exp(ikz)$ that is gently distorted, with this distortion being modelled via multiplication of the said plane wave by an ``envelope'' $\tilde{\psi}_{\omega}(x,y,z)$.  While it is true that there is no loss of generality in the above construction, since any monochromatic scalar wave $Q$ (paraxial or non-paraxial) could always be expressed as the product of $Q\exp(-ikz)$ and $\exp(ikz)$, the factorisation in Eq.~(\ref{eq:EnvelopeForPlaneWave}) is really only physically meaningful (at least in our present context) when describing waves that are paraxial with respect to the positive $z$ axis (optical axis). 

Conveniently, 
\begin{equation}
|\tilde{\psi}_{\omega}(x,y,z)|^2=|\psi_{\omega}(x,y,z)|^2=I_{\omega}(x,y,z), 
\end{equation}
\noindent so that the intensity of the envelope $\tilde{\psi}_{\omega}(x,y,z)$ is the same as the intensity of ${\psi}_{\omega}(x,y,z)$.

Now, if Eq.~(\ref{eq:EnvelopeForPlaneWave}) is substituted into Eq.~(\ref{eq:ScalarHelmholtzEquationInRefractiveMedium}), and the term containing the second $z$ derivative of the envelope is discarded as being small compared to the other terms on account of the paraxial assumption, we obtain the inhomogeneous paraxial equation: 
\begin{equation}\label{eq:ParaxialEquationInhomogeneous}
 \left(2ik\frac{\partial}{\partial z}+\nabla_{\perp}^2+k^2[n_{\omega}^2(x,y,z)-1]\right)\tilde{\psi}_{\omega}(x,y,z)=0.
\end{equation}
\noindent Here,
\begin{equation}
    \nabla_{\perp}^2\equiv\frac{\partial^2}{\partial x^2}+\frac{\partial^2}{\partial
    y^2}
\end{equation}
\noindent is the transverse Laplacian, namely the Laplacian in the $(x,y)$ plane perpendicular to the optical axis $z$.  Thus we may write: 
\begin{equation}
    \nabla^2=\nabla_{\perp}^2+ \frac{\partial^2}{\partial z^2}.
\end{equation}

We again draw a parallel with quantum mechanics, noting that Eq.~(\ref{eq:ParaxialEquationInhomogeneous}) is mathematically identical in form to the time-{\em dependent} Schr\"{o}dinger equation in 2+1 dimensions (i.e.~two space dimensions $(x,y)$ and one time dimension $t$), in the presence of a time-dependent scalar potential $V(x,y,t)$, if one replaces $z$ with $t$, and considers $n_{\omega}^2(x,y,z\rightarrow t)-1$ to be proportional to $-V(x,y,t)$ (Bransden \& Joachain, 1989).   

\subsection{Projection approximation and absorption contrast}\label{subs:projection_approximation}

Consider Fig.~2.  Here, $z$-directed monochromatic complex scalar X-ray waves illuminate a static non-magnetic object, from the left.  By assumption, the object is totally contained within the slab of space between $z=0$ and $z=z_0\ge 0$.  The object is described by its refractive index distribution $n_{\omega}(x,y,z)$, which will only differ from unity (i.e.~the refractive index of vacuum) within the volume occupied by the object.

We wish to determine the complex disturbance (wave-function) over the plane $z=z_0$, which is termed the ``exit surface'' of the object, as a function of both (i) the complex disturbance over the ``entrance surface'' $z=0$ and (ii) the refractive index distribution of the object. 

\begin{figure}
\includegraphics[scale=0.3]{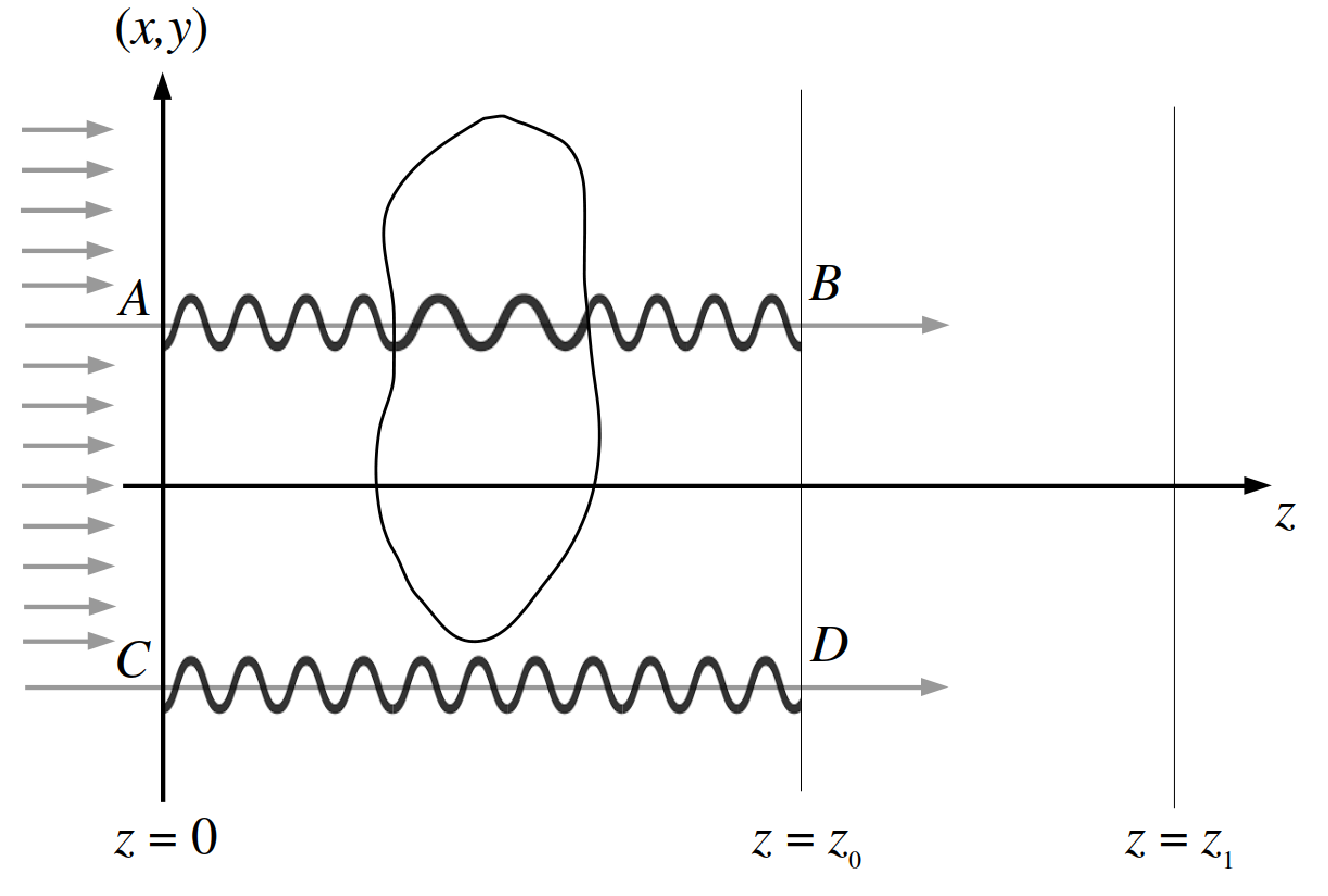}
\caption{From entrance-surface to exit-surface wave-field, under the projection approximation. Adapted from Paganin (2006).}
\label{fig:ProjectionApproximation}
\end{figure}

We assume the object to be sufficiently slowly varying in space, that all streamlines of the X-ray flow may be well approximated by straight lines parallel to $z$.  Under this so-called ``projection approximation'', the validity conditions for which are further discussed in Sec.~2.8 below, the spread of the X-ray waves in the transverse plane can be neglected.  This high-energy approximation enables us to discard the transverse Laplacian in Eq.~(\ref{eq:ParaxialEquationInhomogeneous}).  Thus, for the purposes of deriving the projection approximation, we may write:
\begin{equation}
\frac{\partial}{\partial
z}\tilde{\psi}_{\omega}(x,y,z)\approx\frac{k}{2i}[1-n_{\omega}^2(x,y,z)]\tilde{\psi}_{\omega}(x,y,z).
\label{eq:ParaxialProjectionApprox}
\end{equation}

Since we are neglecting the transverse spread of the field as it propagates from $z=0$ to $z=z_{0}$, we are effectively assuming the influence of the sample, upon the X-rays that traverse it, to be well approximated by the projection of that sample onto the plane $z=z_{0}$. As a consequence, Eq.~(\ref{eq:ParaxialProjectionApprox}) is, for each fixed point $(x,y)$, a simple linear first-order ordinary differential equation, which can be immediately integrated with respect to $z$ (this is the \textit{projection} taking place) to give:
\begin{eqnarray}\label{eq:zz7}
    \tilde{\psi}_{\omega}(x,y,z=z_0) \approx \exp
    \left\{\frac{k}{2i}\int_{z=0}^{z=z_0}[1-n_{\omega}^2(x,y,z)]dz\right\}\tilde{\psi}_{\omega}(x,y,z=0).
\end{eqnarray}

At this point, it is convenient to introduce a complex form for the refractive index.  The real part of the complexified refractive index corresponds to the refractive index in the conventional sense.  The imaginary part of the complexified refractive index is a measure of the absorptive properties of a sample.  With the above indications in mind, we write the complex refractive index as
\begin{equation}\label{eq:ComplexRefractiveIndex}
    n_{\omega}=1-\delta_{\omega}+i\beta_{\omega},
\end{equation}

\noindent where 
\begin{equation}
|\delta_{\omega}|, |\beta_{\omega}| \ll 1
\end{equation}
since the complex refractive index for hard X-rays is typically extremely close to unity.  Note that the negative sign is included in Eq.~(\ref{eq:ComplexRefractiveIndex}) since the real part of the X-ray refractive index is typically (slightly) less than unity.  Hence:
\begin{eqnarray}\label{eq:OneMinusRefrIndexSquaredApproximate}
  1-n_{\omega}^2(x,y,z)\approx
  2[\delta_{\omega}(x,y,z)-i\beta_{\omega}(x,y,z)],
\end{eqnarray}
where we have discarded terms containing $\delta_{\omega}^2$, $\beta_{\omega}^2$ and $\delta_{\omega}\beta_{\omega}$ since these will be much smaller than the terms that have been retained on the right side of Eq.~(\ref{eq:OneMinusRefrIndexSquaredApproximate}).

If the above expression is substituted into Eq.~(\ref{eq:zz7}), we obtain the ``projection approximation'':
\begin{eqnarray}\label{eq:zz8}
\tilde{\psi}_{\omega}(x,y,z=z_0)\approx  \tilde{\psi}_{\omega}(x,y,z=0) \exp
\left\{-ik\int_{z=0}^{z=z_0}[\delta_{\omega}(x,y,z)-i\beta_{\omega}(x,y,z)]dz\right\}.
\end{eqnarray}
This shows that the exit wave-field $\tilde{\psi}_{\omega}(x,y,z=z_0)$ may be obtained from the entrance wave-field $\tilde{\psi}_{\omega}(x,y,z=0)$ via multiplication by the following complex-valued position-dependent ``transmission function'' ${\mathcal T}_{\omega}(x,y)$:  
\begin{equation}\label{eq:complex-tranmission-function}
{\mathcal T}_{\omega}(x,y) = \exp
\left\{-ik\int_{z=0}^{z=z_0}[\delta_{\omega}(x,y,z)-i\beta_{\omega}(x,y,z)]dz\right\}.
\end{equation}

The position-dependent phase shift
\begin{equation}
\arg\left[{\mathcal T}_{\omega}(x,y)\right]\equiv \Delta\phi_{\omega}(x,y),
\end{equation}
due to the object, is:
\begin{equation}\label{eq:PhaseShiftProjectionApproximation}
    \Delta\phi_{\omega}(x,y)=-k\int\delta_{\omega}(x,y,z)dz.
\end{equation}

\noindent Note that, to avoid clutter, we no longer explicitly indicate the limits of integration.  The above expression quantifies the deformation of the X-ray wave-fronts due to passage through the object.  Physically, for each fixed transverse coordinate $(x,y)$, phase shifts (and the associated wave-front deformations) are continuously accumulated along energy-flow streamlines (loosely, ``rays'') such as $AB$ in Fig.~2.  In making all of these statements, it is useful to look back to Fig.~1 and recall the direct connection between the phase of a complex wave-field, and its associated wave-fronts.  The phase shifts---associated with passage of an X-ray wave through an object---quantify the wave-front deformations and associated refractive properties of the object.  Also, since we are working with a wave picture rather than the less-general ray picture for X-ray light, refraction is both modelled by, and conceptualised as being associated with, wave-front deformation rather than ray deflection.

Refraction, due to the object, is an attribute  that may be augmented by the attenuation due to the object.  This latter quantity may be obtained by taking the squared modulus of Eq.~(\ref{eq:zz8}), to give the Beer--Lambert law:
\begin{eqnarray}\label{eq:zz8a}
    I_{\omega}(x,y,z=z_0) =\exp\left[-\int\mu_{\omega}(x,y,z)dz\right]I_{\omega}(x,y,z=0). \end{eqnarray}
Above, we have used the following expression relating the imaginary part $\beta_{\omega}$ of the refractive index, to the associated linear attenuation coefficient $\mu_{\omega}$: 
\begin{equation}
\label{eq:AbsorptionCoefficient}
    \mu_{\omega}=2k\beta_{\omega}.
   \end{equation}
Note that Eq.~(\ref{eq:zz8a}) may also be written in the logarithmic form:
\begin{eqnarray}\label{eq:zz8a_log_form}
\log_e\left[\frac{I_{\omega}(x,y,z=z_0)}{I_{\omega}(x,y,z=0)}\right]=-\int\mu_{\omega}(x,y,z)dz.
\end{eqnarray}

Equation~(\ref{eq:zz8a}) forms the basis for ``absorption contrast imaging''.  In particular, if a two-dimensional position sensitive detector is placed in the plane $z=z_0$ in Fig.~2, and the illuminating radiation has an intensity $I_{\omega}(x,y,z=0)$ that is approximately constant with respect to $x$ and $y$, then all contrast in the resulting ``contact'' image will be due to local absorption of rays such as $AB$ in Fig.~2.  While the logarithm of this image is sensitive to the projected linear attenuation coefficient $\int\mu_{\omega}(x,y,z)dz$, the contact image contains no contrast that is due to the phase shifts quantified by Eq.~(\ref{eq:PhaseShiftProjectionApproximation}).  This lack of phase contrast, in conventional contact X-ray imaging, is unfortunate.  This is because many structures of interest (such as soft biological tissues illuminated by hard X-rays) are close to being non-absorbing, meaning that they are poorly visualised, or not visualised at all, in absorption-contrast X-ray imaging.     

\subsection{Fresnel diffraction and propagation-based phase contrast}

Consider Fig.~3, which shows a source $A$ radiating into free space.  Optical elements and samples, which may lie between $A$ and the plane $z=0$, are not shown. The ``diffraction problem'' seeks to determine the wave-field over the plane $z>0$, given the disturbance over the plane $z=0$.  The space $z\ge 0$ is assumed to be vacuum, and all waves in this space are assumed to be both paraxial with respect to the optical axis $z$, and monochromatic.

\begin{figure}
\includegraphics[scale=0.45]{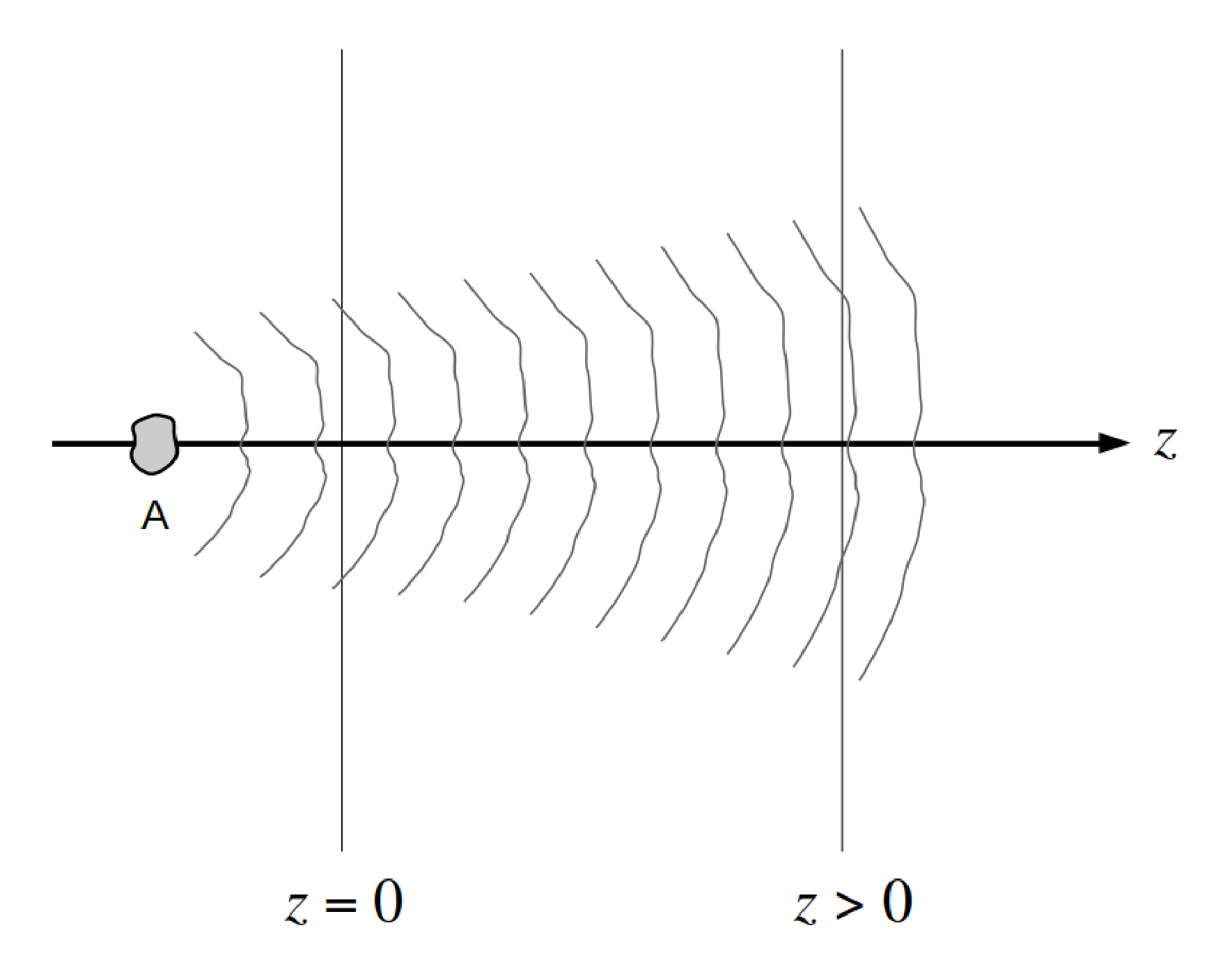}
\caption{Free-space propagation of X-ray waves.}
\end{figure}

In the space $z\ge0$ the waves will obey the ``$n_{\omega}(x,y,z)=1$'' special case of Eq.~(\ref{eq:ParaxialEquationInhomogeneous}), namely:
\begin{equation}\label{eq:ParaxialEquation}
 \left(2ik\frac{\partial}{\partial
 z}+\nabla_{\perp}^2\right)\tilde{\psi}_{\omega}(x,y,z)=0.
\end{equation}
The above equation is often referred to as the free-space paraxial equation, the paraxial wave equation or the parabolic equation for paraxial waves.  It can also be thought of as a diffusion-type equation having a purely imaginary diffusion coefficient.

The solution to the diffraction problem, based on the above free-space paraxial equation, may be written as:
\begin{equation}\label{eq:FresnelDiffraction}
    \psi_{\omega}(x,y,z=\Delta)={\mathcal{D}}_{\Delta}\psi_{\omega}(x,y,z=0),\hspace{1em} \Delta\ge0.
\end{equation}
\noindent Here, ${\mathcal{D}}_{\Delta}$ is a (Fresnel) diffraction operator, which acts on the unpropagated forward-travelling field $\psi_{\omega}(x,y,z=0)$, propagating it a distance $\Delta$, to give $\psi_{\omega}(x,y,z=\Delta)$.  An expression for ${\mathcal{D}}_{\Delta}$, which may be readily derived from the free-space paraxial equation, will be given later. Note, also, that explicit expressions for the Fresnel diffraction operator are given in a number of optics textbooks, such as Lipson \& Lipson (1981), Hecht (1987), Cowley (1995), Born \& Wolf (1999) and Goodman (2005).

From the squared magnitude of Eq.~(\ref{eq:FresnelDiffraction}), it is clear that the intensity of the propagated field depends on both the intensity and phase of the unpropagated field.  This point is both trivial---because the right side, of the squared modulus of Eq.~(\ref{eq:FresnelDiffraction}), obviously depends on the phase---and of profound importance, since it implies that the Fresnel diffraction pattern, namely the propagated intensity over the plane $z > 0$ in Fig.~3, provides the phase contrast that was missing from the contact image.  

This mechanism, for obtaining intensity contrast (in the plane $z>0$) that is sensitive to phase variations (in the plane $z=0$), is known as propagation-based phase contrast. As has already been pointed out, this phenomenon has been known (albeit under different names) for millennia, in the context of visible-light optics.  Furthermore, this effect has been known for many decades in both visible-light microscopy (e.g.~Zernike, 1942; Bremmer, 1952) and electron microscopy (e.g.~Cowley, 1959).  The X-ray-imaging community became particularly interested in this phenomenon in the 1990s (see e.g.~White \& Cerrina, 1992), with the main pioneering studies being performed in the mid 1990s (Snigirev, Snigireva, Kohn, Kuznetsov, \& Schelokov, 1995; Snigirev, Snigireva, Kohn, \& Kuznetsov, 1996a; Cloetens, Barrett, Baruchel, Guigay, \& Schlenker, 1996; Wilkins, Gureyev, Gao, Pogany, \& Stevenson, 1996; Nugent, Gureyev, Cookson, Paganin, \& Barnea, 1996).         

\begin{figure}
\includegraphics[scale=0.95]{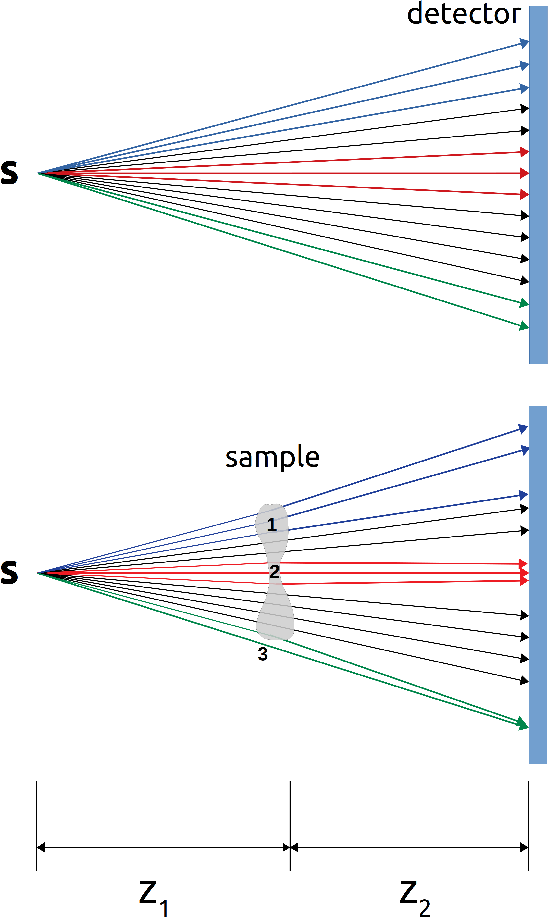}
\caption{Geometric-optics diagram to aid understanding of propagation-based X-ray phase-contrast imaging.  Here, $s$ is a small X-ray source, which illuminates an object shown in grey.  The imaging plane (at the detector position) is shown to the right of the image.  The source-to-object and object-to-detector distances are denoted by $z_1$ and $z_2$ respectively. By comparing the upper panel (object absent) with the lower panel (object present), it can be seen that the object deviates the incoming rays in a manner that is governed by the object shape and density. Adapted from Gureyev et al.~(2009).}
\label{fig:RRR1}
\end{figure}

For the remainder of this sub-section, we seek to further develop our intuition, regarding the qualitative nature of propagation-based X-ray phase contrast. With this in mind, consider Fig.~\ref{fig:RRR1}, in which a small X-ray source $s$ illuminates an object shown in grey.  The source-to-object distance is denoted by $z_1$ and the object-to-detector distance is denoted by $z_2$.  The distance $z_2$ is assumed to be large enough that propagation-based phase contrast is manifest over the detector plane, but not so large that multiple Fresnel diffraction fringes are present.  More precisely, we are here assuming the Fresnel number (Saleh \& Teich, 2007)
\begin{equation}\label{eq:FresnelNumberFreeSpacePropagation}
N_F = \frac{M a^2}{\lambda z_2}     
\end{equation}
to be much greater than unity. Here, $a$ corresponds to the smallest transverse characteristic feature size in the object that is not smeared out by the finite size of the source, $z_2$ is the object-to-detector distance and 
\begin{equation}\label{eq:GeometricMagnificationM}
M=\frac{z_1+z_2}{z_1}     
\end{equation}
is the geometric magnification.  Assuming the object in Fig.~\ref{fig:RRR1} to be sufficiently thin that the projection approximation holds, one may identify three different features within the object, which are  here labelled \textbf{1}, \textbf{2} and \textbf{3} and identified by their action on the incoming rays, which are colour-coded in blue, red and green respectively.

\begin{itemize}

\item Features such as \textbf{1} correspond to either the thin object in projection behaving locally like a convex lens, or a point within the volume of the object which has a local peak of density.  Because the real part of the complex refractive index is less than unity for X-rays, convex X-ray lenses are defocusing optical elements (cf.~the case for visible light, where convex lenses are focusing optics since the real part of the refractive index is greater than unity).  Since \textbf{1} may be considered as a defocusing feature in the object, the local ray density (identified by the blue rays) on the detector will be lessened via the refractive effects of \textbf{1} compared to the situation without the sample; hence these points in the detector plane will have reduced brightness, on account of the propagation distance $z_2$ that lies between \textbf{1} and the detector plane.

\item Features such as \textbf{2}, which may be either a concave feature in the projected thickness of the sample or a feature within the sample that has a local trough of density, will act as a converging lens for X-rays.  Hence the intensity at the detector (locally-converging red rays) will be increased by the effects of refraction by feature \textbf{2}, provided that $z_2$ is large enough for the intensity-increasing effects of the ``focusing element'' \textbf{2} to be manifest at the detector. Again note the crucial role played by the object-to-detector distance $z_2$, through which the wave propagates before reaching the detector.

\item One also has propagation-based phase contrast due to features such as \textbf{3}, which correspond to points on the edge of the object.  Here, ``edge'' refers to the edge of the object when projected along the optical axis $z$.  On account of Fresnel diffraction in the slab of vacuum between the object and the detector, the propagation-based phase-contrast signature of an edge such as \textbf{3} will be a dark/white band (green rays) where intensity is removed from the edge of the geometrical projection of the sample and deflected towards the outside rim of that edge.  Such ``edge contrast'' is a characteristic feature of propagation-based X-ray phase contrast.

\end{itemize}

Before proceeding, we strongly recommend to readers who have not previously seen propagation-based X-ray phase-contrast images, that they briefly study some of the images in one or more of the classic early papers (Snigirev, Snigireva, Kohn, Kuznetsov, \& Schelokov, 1995; Snigirev, Snigireva, Kohn, \& Kuznetsov, 1996a; Cloetens, Barrett, Baruchel, Guigay, \& Schlenker, 1996; Wilkins, Gureyev, Gao, Pogany, \& Stevenson, 1996; Nugent, Gureyev, Cookson, Paganin, \& Barnea, 1996).  This will further develop the reader's intuition for the qualitative nature of such contrast, beyond what has been sketched here.

\subsection{Validity of the projection approximation}\label{Sec:ValidityOfProjApprox}

As discussed in Sec.~2.6, when employing the projection approximation we are assuming the X-ray diffraction effects within the sample to be negligible. In the limit in which the transverse spread of paraxial X-rays within the sample is negligible, we can treat the sample as being infinitely thin along the direction specified by the optical axis.  Of course, the object cannot be infinitely thin in reality, but it can be assumed so for the purposes of computing the influence that this sample has upon the scattered X-rays.  This amounts to ``squashing'' or ``projecting'' its 3D distribution of complex refractive index, to approximate the sample by its projection onto the plane immediately after that sample.  Under this high-energy approximation, the actual sample will create approximately the same distribution of scattered X-ray intensity downstream of its scattering volume, as the scattered X-ray intensity due to a different sample that is obtaining by ``squashing'' the actual sample so that it becomes a zero-thickness phase--amplitude screen that is perpendicular to the optical axis.
\begin{figure}
\includegraphics[scale=2.7]{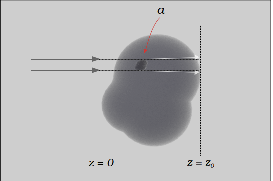}
\caption{The projection approximation is valid insofar we can neglect the diffraction effects, within the volume occupied by the sample, that are caused by the smallest resolvable feature in the sample. In the figure, the transverse diffraction spread caused by a feature of average size $a$ (white arrows) can be neglected if the maximum transverse deviation $\lambda z_{0}/a$ cannot be resolved.}
\label{fig:cell}
\end{figure}

It is natural to enquire into the limits of validity of the projection approximation, as well as some  consequences of the breakdown of such an assumption.  The first thing to ask, along these lines, is: ``What does it mean to state that the diffraction effects---within the volume that is occupied by the sample---must be negligible, in order for the projection approximation to be a reasonably accurate approximation?'' Or, more precisely: ``With respect to what, must these diffraction effects be negligible?'' The yardstick for this comparison is the smallest features in the sample that can be confidently detected by our imaging system. This is a key point: broadly speaking, for any given X-ray illumination, the validity of the projection approximation depends on the spatial resolution of the imaging system. A coarse enough imaging resolution would prevent the detection of fine diffraction effects, hence the projection approximation would hold even in a case where, for the same sample under the same illumination conditions, the projection approximation would be invalid when the scattered intensity is measured with a position-sensitive detector having a spatial resolution that is less coarse.
 
Let us make this argument more quantitative. Suppose we want to image a sample which is fully contained within the slab of space between $z=0$ and $z=z_{0}$, as shown in Fig.~\ref{fig:cell}, with the said slab of space being perpendicular to the optical axis $z$. More specifically, we are aiming at resolving a feature of average size $a$ that is embedded within the larger sample (think for instance of resolving an organelle inside a cell, or a cell inside a larger volume of biological tissue). In other words, we require an imaging system with resolution better than $a$. As explained previously, the projection approximation will be valid if we can assume that X-rays propagate along straight lines within the volume that is occupied by the sample, i.e.~diffraction from features of size $a$ (or larger) is negligible (within the volume occupied by the sample). Radiation of wavelength $\lambda$ scattered by such features will have a typical (maximum) diffraction angle of the order of
\begin{equation}
\Delta \theta = \frac{\lambda}{a}.
\label{eq:DiffractionAngle}
\end{equation}
Therefore, as sketched in Fig.~\ref{fig:cell}, the maximum spread of the radiation at the exit face of the sample (assuming, say, an organelle to be close to the entrance face) will be 
$\Delta \theta \, z_{0}$.
The projection approximation is valid if we can neglect the diffraction spread when compared to the resolution, namely if 
\begin{equation}
\frac{\lambda}{a} \, z_{0} \ll a.
\end{equation}
The previous inequality can be redefined in terms of the Fresnel number
\begin{equation}
N_F = \frac{a^2}{\lambda z_{0}} \gg1.
\label{eq:FresnelNumber}
\end{equation}

\begin{figure}
\includegraphics[scale=0.6]{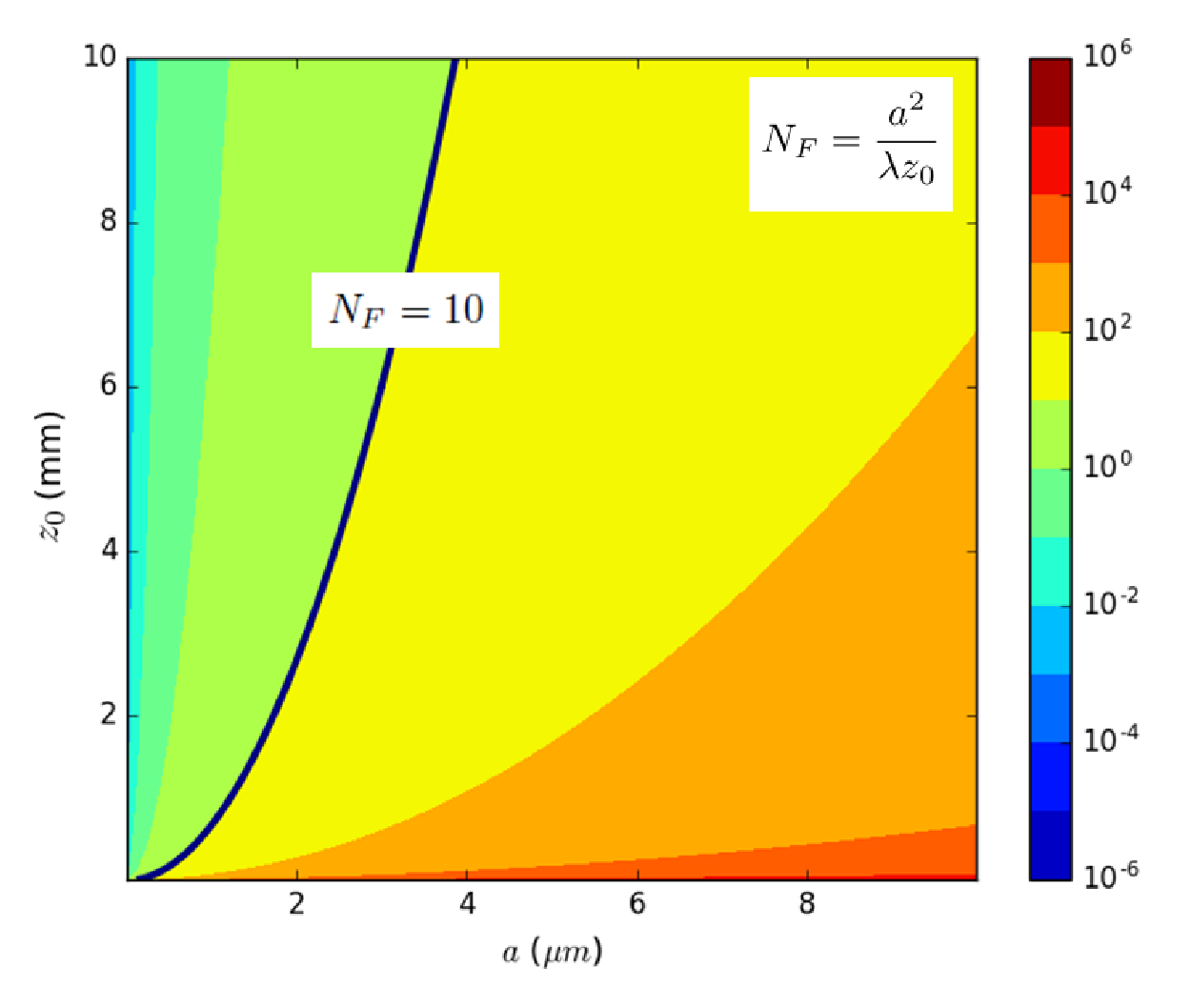}
\caption{Contour map of the Fresnel number $N_F$, as a function of resolution $a$ and sample thickness $z_{0}$, calculated at the X-ray wavelength $\lambda = 1.5$ \AA. The black line marks the contour $N_F = 10$. As a rule of thumb, a Fresnel number below this value is a situation where the projection approximation might not hold.}
\label{fig:PAContour}
\end{figure}

Figure  \ref{fig:PAContour} shows a contour map of the Fresnel number as a function of both $a$ and $z_0$, calculated setting the wavelength to $\lambda = 1.5$ \AA. Somewhat arbitrarily, the contour $N_F = 10$, marked with a thicker line, is chosen as a boundary for the validity of the projection approximation. The region on the right of this line (warm colours such as yellow, orange and red) is where the projection approximation is generally valid. This region corresponds to the range of values attained, for instance, by modern micro-CT (computed tomography) systems, where resolutions of a few micrometres and sample thicknesses of a few millimetres are the state of the art.

The region on the left of the $N_F = 10$ contour (cold colours) is where the projection approximation is at risk. In this region, namely the domain of ultra high resolution X-ray microscopy systems (Jacobsen, 2019) typical of X-ray synchrotron (Duke, 2000) beam-lines, the sample thickness becomes very large compared to the resolution. Admittedly, in this region lays one of the major strength of X-rays when compared with other probes for microscopy: X-rays can visualise minute details within larger samples---for instance single cells within a larger matrix of embedding tissue---in a less invasive fashion.

As one moves to progressively higher resolution, e.g.~in tomography, the projection approximation will become increasingly ill behaved.  An example of this situation is plotted in Fig.~\ref{fig:PAContour_hires}.  One will eventually need to embrace fully dynamical models such as the multi-slice approximation (Cowley \& Moodie, 1957, 1959), in the context of the inverse problem (Sabatier, 2000) of computed tomography (Natterer, 1986) or diffraction tomography (Wolf, 1969; M\"uller, Sch\"urmann, \& Guck, 2016), to a larger degree than is the case at present.  This transition is already in progress, as even a cursory survey of the recent literature will show.  On a related note, and again as one moves to ever-higher resolution, the scalar approximation for the X-ray wave-field may begin to break down e.g.~when large scattering angles or magnetic phenomena are considered (Detlefs, 2019).  In all of this, much guidance is to be gleaned from the existing literature on electron tomography, which has---very broadly speaking---been forced to grapple with such problems at an earlier stage than has been the case with the X-ray tomography community.  Much can also be learned from the very well established field of X-ray scattering and absorption by magnetic materials---see e.g.~the text by Lovesey \& Collins (1996)---together with any aspects of X-ray physics in which the effects of magnetism and/or polarisation are important.

\subsection{Describing the propagation through thick samples: multi-slice approach}

\begin{figure}
\includegraphics[scale=0.6]{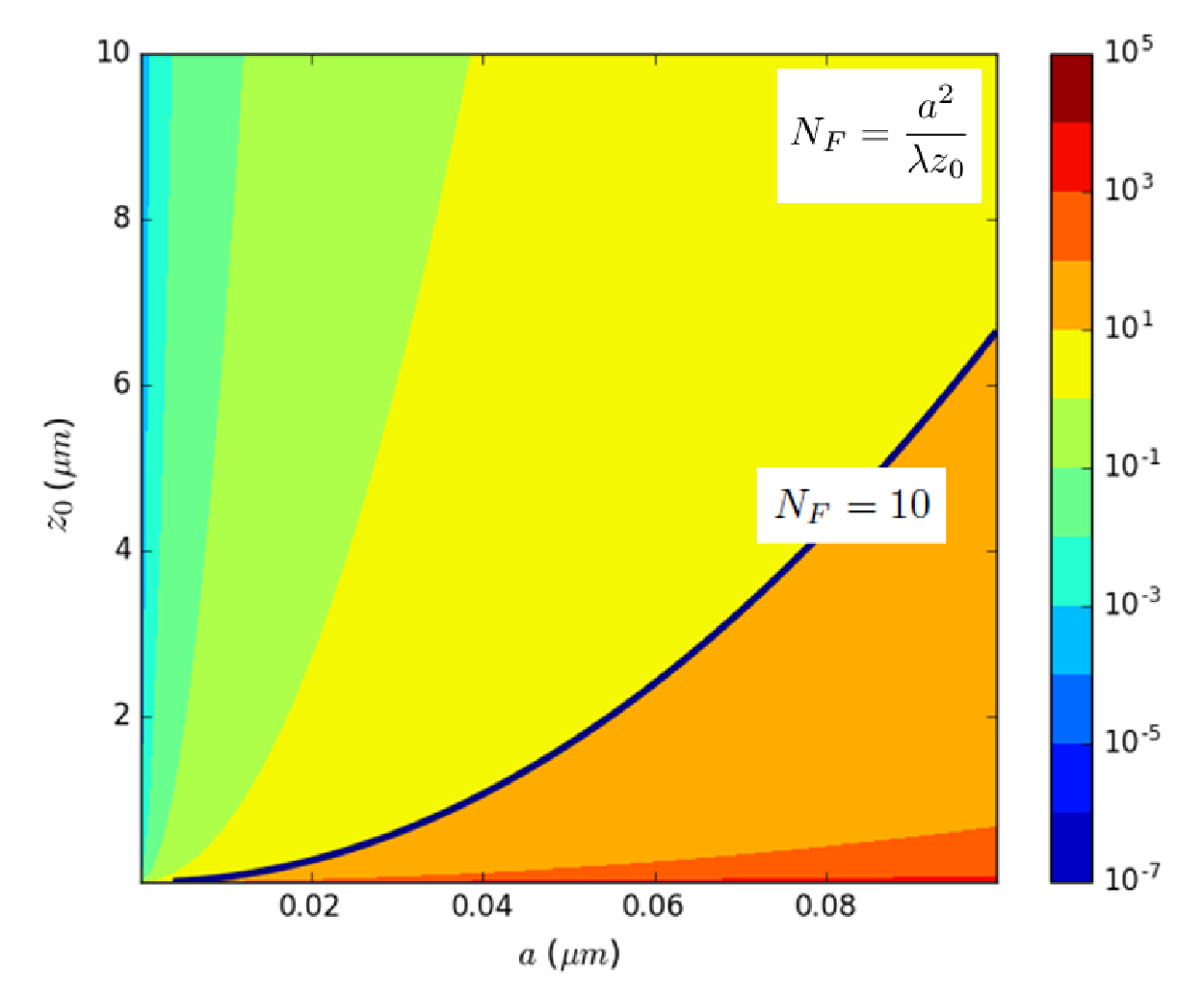}
\caption{Contour map of the Fresnel number $N_F$, as a function of resolution $a$ and sample thickness $z_0$, calculated at the X-ray wavelength $\lambda = 1.5$ \AA, for the high resolution case applicable to modelling X-ray diffractive optics. As before, the black line marks the contour $N_F = 10$.}
\label{fig:PAContour_hires}
\end{figure}

Simulating and modelling high resolution transmission X-ray optics, or reflective optics, is an example of a situation where the projection approximation generally does not hold. Transmission optics such as compound refractive X-ray lenses (Kirkpatrick \& Baez, 1948; Tomie, 1994; Snigirev, Kohn, Snigireva, \& Lengeler, 1996b; Tomie, 2010) or Bragg--Fresnel lenses (Erko, Aristov, \& Vidal, 1996) can be, to some extent, considered thin in the medium-resolution range. Here, a thin optical element is considered to be synonymous with an optical element for which the projection approximation holds. High resolution applications, however, demand extremely fine X-ray optical structures.  For instance, the widths of the outermost zones of Fresnel or Bragg--Fresnel lenses can be in the nanometre range.  Crystalline optical elements also fall into this category of optical structure, for which the projection approximation is inadequate.  This breakdown of the projection approximation can be appreciated in Fig.~\ref{fig:PAContour_hires}, which is a close-up view of the contour map in Fig.~\ref{fig:PAContour}, applicable in the region relevant to high resolution X-ray optics. \\

Modelling X-ray propagation through such finely-structured optical elements (and/or samples) requires dropping the projection approximation in favour of a more accurate approach that has a broader domain of validity. Furthermore, conventional reflective X-ray optics (Ehrenberg, 1947; Kirkpatrick \& Baez, 1948) must be considered ``thick'' in all cases, as obviously the beam angular deviation in reflection is always significant. In all such cases---reflective optical elements, thick optical elements, finely structured optical elements etc.---the multi-slice approximation is a very useful, and very general, approach.  In the following paragraphs we briefly explain the basics of this approximation.

Introduced by Cowley \& Moodie (1957 and 1959) in the context of transmission electron microscopy, and subsequently studied over many decades within that context (Kirkland, 2010), the multi-slice method has been brought to a very high state of development in the electron-optics community.  It is only relatively recently, however, that the same idea has been used to both analytically model and numerically simulate high resolution X-ray optics and imaging scenarios (Paganin, 2006; Martz et al., 2007; D\"oring et al., 2013; Li, Wojcik, \& Jacobsen, 2017; Munro, 2019; Du, Nashed, Kandel, G\"{u}rsoy, \& Jacobsen, 2020).   

In the multi-slice approach, the thick sample is decomposed into a number of slices along the optical axis direction. The thickness of each slice is chosen to guarantee that such a slice can be considered optically thin. This corresponds to $N_F \gg 1$ for each individual slice. Therefore, for each slice (in a multiply-sliced, i.e.~``multi-sliced'', scattering structure) one can assume the projection approximation to be valid.   

\begin{figure}
\includegraphics[scale=1.2]{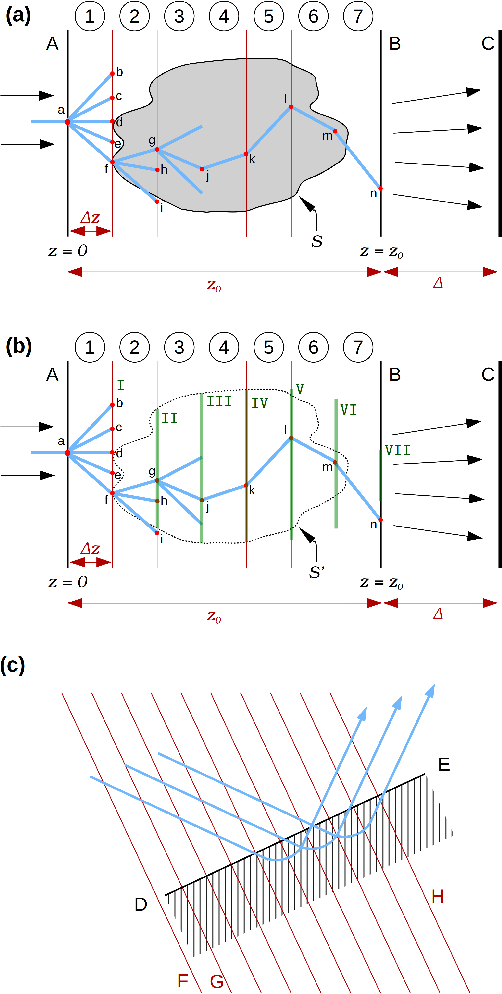}
\caption{Multi-slice formulation of X-ray scattering. (a) A scattering object $S$, shaded in grey, lies between planes $A$ and $B$.  (b) The scatterer $S$ is approximated via a series of infinitely thin phase--amplitude screens $S'$.  (c) Multi-slice geometry for reflective optical elements.}
\label{fig:Multi-slice}
\end{figure}

The essential idea behind the multi-slice approach is illustrated in Fig.~\ref{fig:Multi-slice}(a).  Here, an incident X-ray wave---which is here taken to be a $z$-directed monochromatic scalar plane wave, for simplicity---is incident upon a static sample $S$ whose scattering volume is shown shaded in grey.  Outside this grey region, the complex refractive index $n(x,y,z)$ corresponds to vacuum (i.e.~$n(x,y,z)=1$).  Inside the grey region, $n(x,y,z)$ can deviate from unity.  Note that we are here ceasing to notate the explicit dependence of quantities such as $n$ on $\omega$, even though this dependence is still present.  The scattering volume is considered to be localised to the slab of space between the planes labelled $A$ and $B$.  This slab of space is then considered to be chopped into a number $N$ of contiguous parallel slabs, hereafter termed ``slices''.  If the slab between planes $A$ and $B$ has a thickness of $z_0$, then the thickness of each slice will be
\begin{equation}
\Delta z=\frac{z_0}{N}.    
\end{equation}
In the diagram, seven such slices are indicated, labelled ``1'' through to ``7'' respectively.  For the purposes of illustration, the first slice (slice 1) is considered to contain only vacuum.  Next, we imagine that the scatterer $S$ is replaced by a different scatterer, $S'$, as represented by the set of parallel equally-spaced infinitely-thin phase--amplitude screens II, III, $\cdots$, VII indicated in green in Fig.~\ref{fig:Multi-slice}(b).  Each of the infinitely-thin screens is obtained by projecting the complex refractive index, in any given slice, so as to be ``squashed'' against the exit surface of the slice.  Thus, for example, within slice 2, the slice of the sample shaded in grey in Fig.~\ref{fig:Multi-slice}(a) is approximated, in Fig.~\ref{fig:Multi-slice}(b), by (i) vacuum within the interior of slice 2, and (ii) an infinitely thin phase--amplitude screen labelled II at the exit surface of slice 2.  Such sequential slicing is the key approximation underpinning the multi-slice model, which approximates the complex wave-field over the exit surface $B$ in Fig.~\ref{fig:Multi-slice}(a) by the complex wave-field over the same plane in Fig.~\ref{fig:Multi-slice}(b).  The sequence of wave-field propagation steps, in either the transmission geometry sketched in Fig.~\ref{fig:Multi-slice}(b) or the reflection geometry of Fig.~\ref{fig:Multi-slice}(c), may then be described as follows:
\begin{enumerate}
\item The complex wave-field, over the entrance surface $A$, is specified.  
\item This wave-field is then propagated, using e.g.~the Fresnel diffraction formalism, from the entrance surface of slice 1 (containing point $a$) to the exit surface of slice 1 (containing points $b,c,d,e,f$). 
\item The resulting complex wave-field is then multiplied by the phase--amplitude screen labelled I in Fig.~\ref{fig:Multi-slice}(b). 
\item This wave-field is then propagated, from the entrance surface of slice 2 (containing points $b,c,d,e,f$) to the exit surface of slice 2 (containing points $g,h,i$).
\item The resulting complex wave-field is then multiplied by the phase--amplitude screen labelled II in Fig.~\ref{fig:Multi-slice}(b).
\item The above slice-by-slice process is iterated until the exit surface $B$ of the sample is reached. 
\item The complex wave-field, over plane $B$, may then be propagated (e.g.~using the Fresnel diffraction formalism) to the downstream surface $C$.  The squared magnitude of the resulting field gives the intensity distribution registered by a planar position-sensitive detector with surface $C$.
\end{enumerate}

We now supplement the above verbal description, of the multi-slice process, with some corresponding equations.  Following Eq.~(\ref{eq:complex-tranmission-function}), and again dropping the subscript $\omega$ for clarity, the transmission function of the slice $j$ can be written as:
\begin{equation}
\mathcal{T}_{j}(x,y) = \exp \left\{ i k \, [\tilde{n_j}(x,y)-1] \, \Delta z \right\}.
\label{eq:TranFunctSlice}
\end{equation}
In Eq.~(\ref{eq:TranFunctSlice}), 
\begin{eqnarray}
\tilde{n_j}(x,y) = \frac{1}{\Delta z}\int_{z_{j-1}}^{z_j} \!\!\! n(x,y,z) dz \approx n(x,y, z=z_j) \label{eq:OneSliceOfMultiSlice}
\end{eqnarray}
is the complex refractive index of slice $j$, located at the longitudinal position $z=z_j$.  With similar notation, 
\begin{equation}
\psi_{j}(x,y) \equiv \psi(x,y, z=z_j). 
\end{equation}
The slice thickness is 
\begin{equation}
\Delta z = z_{j}-z_{j-1}.
\end{equation}
Note that, by passing from the middle part to the right-most part of Eq.~(\ref{eq:OneSliceOfMultiSlice}), we have assumed the refractive index of each slice to be independent of $z$, within the volume occupied by the said slice.  This will be a good approximation if the slices are thin enough, compared to the length scale over which $n(x,y,z)$ varies appreciably, in the longitudinal direction $z$. 

Under these assumptions, the propagation of the wave field to the next slice can be performed using Fresnel propagation in vacuum, via Eq.~(\ref{eq:FresnelDiffraction}):
\begin{equation}
\psi_{j+1}(x,y)={\mathcal{D}}_{\Delta z} \, \left[ \psi_{j}(x,y) \mathcal{T}_{j}(x,y) \right]. 
\label{eq:OneSliceOfMultislice}
\end{equation}
Note that we can also consider the phase--amplitude screen to be applied after each propagation step, rather than before each such step, and thereby modify Eq.~(\ref{eq:OneSliceOfMultislice}) to $\psi_{j+1}(x,y)=\mathcal{T}_{j}(x,y)  \, \left[ \mathcal{D}_{\Delta z} \psi_{j}(x,y)  \right]$.  This latter approach is consistent with Figs.~\ref{fig:Multi-slice}(a) and (b).  In the limit of infinitely-thin slices, both approximations converge to the same result, and for sufficiently thin slices both approximations will yield very similar results for the numerical evaluation of scattered X-ray wave fields.  Note, furthermore, that a convenient explicit form for $\mathcal{D}_{\Delta z}$ will be given later (see Eq.~(\ref{eq:DiffractionOperatorFresnelDiffraction}) below).  Such a form for the wave-field diffraction operator is convenient since, being based on Fourier transformations (Goodman, 2005), it can be rapidly and stably implemented in numerical models employing the multi-slice formalism.  In such numerical models, one would almost always make use of the so-called ``fast Fourier transform'' (FFT), which, as its name implies, is computationally rapid to execute (Press, Teukolsky, Vetterling, \& Flannery, 2007).

As we have already mentioned, the multi-slice algorithm applies its ``propagate and project'' procedure iteratively, to evolve the X-ray wave field through all slices of the thick sample.  The previously-cited papers (Martz et al., 2007; D\"oring et al., 2013; Li, Wojcik, \& Jacobsen, 2017; Munro, 2019; Du, Nashed, Kandel, G\"{u}rsoy, \& Jacobsen, 2020) give excellent examples of the application of this very powerful and general method for modelling X-ray interactions with such thick samples, in situations where the projection approximation has broken down.  In addition to the above papers, we note that, for those seeking to apply the multi-slice method in an X-ray setting, much can be learned from the previously cited electron-optics text by Kirkland (2010).  

We close this section with three remarks.  

{\em Remark \# 1.} To date, the primary utility of the multi-slice method in a coherent-X-ray-optical setting appears to be in the numerical modelling of X-ray interactions with thick samples and/or optical elements. It remains to be seen whether the method may also yield useful insights of a fundamental nature, for example in relation to its evident connection to the path-integral concepts outlined in the next remark. 

{\em Remark \# 2.}  The multi-slice method has very strong links with the path-integral concept (Dirac, 1935, 1945; Feynman, 1948), which is usually associated with quantum mechanics but which is also applicable to coherent X-ray optics.  With this in mind, return attention to Fig.~\ref{fig:Multi-slice}(b).  Consider the incident X-ray that strikes the point $a$ on the entrance surface $A$ of slice 1.  For this particular point $a$, propagation through the distance $\Delta z$ of slice 1, via application of ${\mathcal{D}}_{\Delta z}$, may be considered as convolution with a Huygens-type wavelet (Green function, real-space propagator) which, from a complex-ray perspective (Keller, 1962), corresponds to the fan of rays $ab,ac,ad,ae,af$.  Next, consider one member of this fan of complex rays, say $af$.  At the point $f$ the complex ray will undergo a phase--amplitude shift associated with the infinitely-thin phase--amplitude screen labelled I in the diagram.  This process (propagation via a fan of complex rays through the slab of vacuum between slices, followed by a phase--amplitude shift due to each infinitely-thin scattering screen) can now be iterated.  Thus we have another fan of complex rays emanating from the point $f$, here labelled $fg,fh,fi$, and so on, through all of the slices, until one reaches the exit surface $B$.  One possible complex-ray path is $afgjklmn$, but clearly there are infinitely many possible paths, with the total disturbance over the plane $B$ being a coherent superposition of the complex amplitudes (these would be called ``probability amplitudes'' in a quantum-mechanics context) associated with each of the infinitely-many paths through the volume of space between the planes $A$ and $B$.  Thus interpreted, the multi-slice process is an example (albeit one that is restricted to a forward-scattering geometry) of a path-integral formulation, in the sense of the ``path integrals'' associated with Dirac (1935, 1945) and Feynman (1948).  For more recent treatments regarding path integrals in a quantum-mechanics context, see Feynman \& Hibbs (1965), Maggiore (2005) and Mandl \& Shaw (2010).  For further information on the connection between the multi-slice formalism and path integrals, in the context of electron scattering, see e.g.~Jap \& Glaeser (1978) and Van Dyck (1975, 1985). 

{\em Remark \# 3.} In the ``mono-slice'' limiting case where multi-slice is performed using only a single slice to encompass the entire sample, multi-slice reduces to the projection approximation.  In this sense, the multi-slice procedure can be viewed as a generalisation of the projection approximation, since the former model reduces to the latter in the one-slice limit. 

\subsection{Fresnel scaling theorem}

Return consideration to the spherical-wave-illumination geometry sketched in the lower panel of Fig.~\ref{fig:RRR1}.  As a quick reminder, in this figure we illustrated a thin object being illuminated by a point source $s$, with $z_1$ being the distance from the source to the sample, and $z_2$ being the distance from the sample to a planar detector.  We saw that this point-source illumination implies a geometric magnification $M$ given by Eq.~(\ref{eq:GeometricMagnificationM}).
  
Remarkably, the thin-object and paraxial approximations together imply the measured intensity over the planar detector (for the case of illumination by a spherical wave) to be equal to the intensity that would be measured if the same thin object were instead to be illuminated by $z$-directed plane waves and the detector moved to a distance of $z_2/M$ downstream of the object.  This equality holds up to a transverse magnification and a multiplicative scaling whose precise form will be made clearer in the next paragraph.    

Let $z=0$ denote the nominally-planar exit surface of the sample.  Let  $I_{\textrm{spherical-wave}}(x,y,z=z_2)$ denote the intensity over the detector plane $z=z_2$, as illustrated for the case of spherical-wave illumination shown in the lower panel of Fig.~\ref{fig:RRR1}.  Let $I_{\textrm{plane-wave}}(x,y,z=z_2/M)$ denote the intensity that would have been measured, for the same object, had it instead been illuminated with $z$-directed plane waves and the detector placed at a distance $z=z_2/M$ downstream of the thin sample.  Clearly, $I_{\textrm{plane-wave}}(x,y,z=z_2/M)$ will be a smaller image than $I_{\textrm{spherical-wave}}(x,y,z=z_2)$, so let us now analytically ``stretch'' the plane-wave-illumination image to be the same size as the point-source-illumination image, by considering $I_{\textrm{plane-wave}}(x/M,y/M,z=z_2/M)$.  The Fresnel scaling theorem asserts $I_{\textrm{spherical-wave}}(x,y,z=z_2)$ and $I_{\textrm{plane-wave}}(x/M,y/M,z=z_2/M)$ to be the same image, up to a multiplicative factor of $M^2$:
\begin{eqnarray}
M^2 I_{\textrm{spherical-wave}}(x,y,z=z_2) = I_{\textrm{plane-wave}}\left(\frac{x}{M},\frac{y}{M},z=\frac{z_2}{M}\right).
\label{eq:FresnelScalingTheorem}
\end{eqnarray}
The factor of $M^2$ accounts for energy conservation, since geometric magnification by a factor of $M$ will increase the area of an image by a factor of $M^2$, thereby reducing the intensity by a factor of $1/M^2$.

While we will not be employing the Fresnel scaling theorem at any point in the present chapter, it is worth briefly remarking upon why this result is useful. For numerical computation of magnified or de-magnified images, the phase excursions associated with expanding or contracting spherical waves can often be so large that proper numerical sampling, of such phase excursions on a pixellated grid that is parallel to the optical axis, becomes problematic.  This problem arises from the Nyquist-limit requirement (Press, Teukolsky, Vetterling, \& Flannery, 2007), which in the present context requires that the phase change by less than $\pi$ radians between adjacent pixels, in order for the complex disturbance to be adequately sampled.  This requirement for adequate sampling  may necessitate an impracticably large number of pixels in a given numerical computation.  The Fresnel scaling theorem sidesteps this issue by reducing the computation, of a Fresnel diffraction pattern in an expanding-wave or contracting-wave geometry, to an equivalent calculation in a parallel-wave geometry.  Note that for a contracting-wave geometry, namely one in which there is de-magnification rather than magnification, one can simply set $M$ to lie between zero and unity, in the form of the Fresnel scaling theorem given in Eq.~(\ref{eq:FresnelScalingTheorem}).  Another use, of the Fresnel scaling theorem, is that it allows one to take analytical or numerical results that are obtained under the assumption of plane-wave illumination of thin objects, and readily generalise them to the case of spherical-wave-illumination geometry without needing to perform any further computation or calculation.  Note, also, that the Fresnel scaling theorem is applicable to both monochromatic and polychromatic paraxial radiation.  For more information on the Fresnel scaling theorem, including a derivation of the results that have been merely quoted here, see e.g.~Paganin (2006).

\section{The forward problem: modelling X-ray phase-contrast images}

So-called ``forward problems'', in physics, seek to determine effects from causes. In any situation where a mathematical model is available, one can use it to predict the evolution of a physical system, within the limits implied by the domain of validity of that model. Examples of such forward problems include determining the spectrum of different sound pitches that would be created if a guitar string of a given length and tension etc.~were to be plucked at a particular position, or solving the Schr\"{o}dinger equation of non-relativistic quantum mechanics in order to determine the allowed energy levels of a hydrogen atom.

This section is dedicated to the particular forward problem of modelling X-ray phase-contrast images.  As a first example of such images, we consider a model based on the transport-of-intensity equation, in a regime where the object-to-detector propagation distance is sufficiently small.  We then consider generalised phase-contrast X-ray imaging systems, these being an infinite variety of imaging systems that yield phase contrast in the sense that they are sensitive to the refractive (phase) effects of X-ray-transparent samples.  We pay particular attention to the class of linear imaging systems, including but not limited to the shift-invariant linear imaging systems.  A key theme that emerges, here, is that ``perfect'' imaging systems cannot yield phase contrast, since by definition they reproduce the intensity distribution that is input into them.  Thus, optical aberrations, in the sense defined by Born \& Wolf (1999), of an imaging system are required in order for it to produce phase contrast.

We then show how arbitrary aberrated linear shift-invariant imaging systems, including but not limited to those that are employed for X-ray phase-contrast imaging, may be described in a Fourier-space manner using the idea of a transfer function.  This allows us to write down a simple representation of such phase-contrast systems, that may be readily implemented numerically.  In addition to its numerical utility in modelling a very broad class of X-ray phase-contrast imaging systems, the transfer-function concept is often useful for the clear analytical insights it provides.  This section also includes an introduction to effects of spatial coherence that cause image blurring due to non-zero source size.   A more sophisticated formalism for incorporating the effects of partial coherence is then considered, namely the space--frequency description of partial coherence (Wolf, 1982; Mandel \& Wolf, 1995; Wolf, 2007).  Here, one models a partially coherent disturbance (at each angular frequency or energy) using a statistical ensemble of strictly monochromatic fields.  This formalism, the historical development of which was inspired by analogous constructions in the field of statistical mechanics (Sears \& Salinger, 1975; Huang, 1987), is numerically efficient to implement as well as being a powerful conceptual lens via which one can study partial coherence in the context of the formation of X-ray phase-contrast images.  In this section, we also offer some comments regarding the connection between (i) various models for partially-coherent X-ray fields, and (ii) the notion of unresolved speckle.  We close this section by extending the transport-of-intensity formalism for X-ray phase-contrast imaging, to include the effects arising from unresolved speckle that is associated with position-dependent small-angle X-ray scattering fans.  Such fans of diffuse scatter emerge from the exit surface of the sample, when it contains fine-level structural detail that is not directly resolved by a position-sensitive detector, on account of that detector having effective pixels that are significantly larger than the unresolved X-ray speckles.  This extension of the transport-of-intensity equation, namely the Fokker--Planck equation for paraxial X-ray imaging (Morgan \& Paganin, 2019; Paganin \& Morgan, 2019), is able to model the effects of both rotationally-symmetric and rotationally-asymmetric fans of position-dependent small-angle X-ray scatter.

\subsection{Transport-of-intensity equation}    

Substitute Eq.~(\ref{eq:MadelungDecomposition}) into Eq.~(\ref{eq:ParaxialEquation}), expand, cancel a common factor, and then take the imaginary part.  This gives a continuity equation expressing local conservation of optical energy (Teague, 1983; cf.~Madelung, 1927), called the ``transport-of-intensity equation'' (TIE):
\begin{eqnarray}
-\nabla_{\perp}\cdot\left[I(x,y,z)\nabla_{\perp}\phi(x,y,z)\right]
=k\frac{\partial I(x,y,z)}{\partial z}. \label{eq:TIE}
\end{eqnarray}

By recalling Eq.~(\ref{eq:PoyntingVector}), the TIE may be interpreted as stating that the divergence of the transverse Poynting vector (transverse energy-flow vector) ${\bf S}_{\perp} \propto I \nabla_{\perp}\phi$ governs the longitudinal rate of change of intensity.  Such an interpretation is very intuitive, from the physical perspective that we now describe.  If the divergence of the Poynting vector is positive, because the paraxial wave-field is locally behaving as an expanding (diverging!) wave, optical energy will be moving away from the local optical axis and so the longitudinal derivative of intensity will be negative (local defocusing; see point ${\bf{1}}$ in the lower panel of Fig.~\ref{fig:RRR1} for an example).  Conversely, if the divergence of the Poynting vector is negative, because the wave-field is locally contracting, optical energy will be moving towards the local optical axis and so the longitudinal derivative of intensity will be positive (local focusing; see point ${\bf{2}}$ in the lower panel of Fig.~\ref{fig:RRR1} for an example). 

Indeed, if we speak of the negative divergence ``$-\nabla_{\perp}\cdot$'' as the ``convergence'', then the TIE merely makes the intuitive statement that the convergence of the transverse Poynting vector is proportional to the rate of change of intensity along the propagation direction.  Under this very simple view, a converging wave (positive convergence or negative divergence) has a positive rate of change of intensity with respect to $z$, because optical energy is being concentrated (focused) as $z$ increases (see, once again, point ${\bf{2}}$ in the  lower panel of Fig.~\ref{fig:RRR1}).  Similarly, a diverging wave (negative convergence or positive divergence) has a negative rate of change of intensity with respect to $z$, because optical energy is being rarefied (defocused) as $z$ increases (see point ${\bf{1}}$ in the lower panel of Fig.~\ref{fig:RRR1}).

The above comments also pertain to the form of the TIE obtained when using the first-order finite-difference approximation
\begin{equation}
\frac{\partial I(x,y,z)}{\partial z} \approx \frac{I(x,y,z+\delta
z)-I(x,y,z)}{\delta z}.
\label{eq:FiniteDifference}
\end{equation}
\noindent When Eq.~(\ref{eq:FiniteDifference}) is substituted into Eq.~(\ref{eq:TIE}), with the resulting expression then being re-arranged to isolate the propagated intensity, we obtain the following approximate description for propagation-based phase contrast, in the regime of sufficiently small propagation distances $\delta z$:
\begin{eqnarray}\label{eq:TIE_for_defocus}
I(x,y,z+\delta z) \approx I(x,y,z)  -\frac{\delta
z}{k}\nabla_{\perp}\cdot\left[I(x,y,z)\nabla_{\perp}\phi(x,y,z)\right]. 
\end{eqnarray}

Propagation-based methods are only one of many means by which X-ray phase contrast can be achieved.  Other methods include:
\begin{itemize}
\item the use of one or more crystals as diffractive optical elements, in both an interferometric geometry (Bonse \& Hart, 1965) and a non-interferometric geometry (F\"{o}rster, Goetz, \& Zaumseil, 1980);
\item diffractive imaging from far-field patterns of both crystalline samples (Hammond, 2009) and non-crystalline samples (Miao, Charalambous, Kirz, \& Sayre, 1999);
\item methods employing transmissive periodic gratings (Momose et al., 2003; Weitkamp et al.,~2005; Pfeiffer et al.,~2008);
\item methods employing transmissive random gratings (B\'{e}rujon, Ziegler, Cerbino, \& Peverini, 2012; Morgan, Paganin, \& Siu, 2012);
\item methods using edge illumination (Olivo, Ignatyev, Munro, \& Speller, 2011; Munro et al., 2013; Pelliccia \& Paganin, 2013a; Diemoz {\em et al.}, 2017);
\item ptychographic methods (Pfeiffer, 2018);
\item methods employing Zernike phase contrast (Neuh\"{a}usler et al., 2003);
\item methods based on Fourier holography (Eisebitt et al., 2004).
\end{itemize}
Due to space limitations, these will not be reviewed here, but we note that (i) some of these methods will be briefly touched upon at several later points in this chapter; (ii) many of these methods can be considered to be special cases of the set of all possible linear shift-invariant phase-contrast imaging systems, which will be treated below. Taken together, the previously listed suite of methods forms a powerful toolbox for the X-ray phase-contrast imaging of samples, with each method having its particular strengths and limitations.  We emphasise that no method is superior to all others in all scenarios and circumstances.

\subsection{Arbitrary imaging systems}

We have already seen that the act of free-space propagation, from plane to downstream plane, can achieve phase contrast in the sense that the propagated image (over the downstream plane, such as that given by the detector in the lower panel of Fig.~\ref{fig:RRR1}) has a transverse intensity distribution that depends on the transverse X-ray phase shifts in an upstream plane (such as the plane at the exit-surface of the object in Fig.~\ref{fig:RRR1}).  What happens if we generalise this propagation-based X-ray phase-contrast-imaging scenario, to a more general X-ray phase-contrast-imaging setup, by interposing an optical imaging system in between the object and the detector? 

\begin{figure}
\includegraphics[scale=0.35]{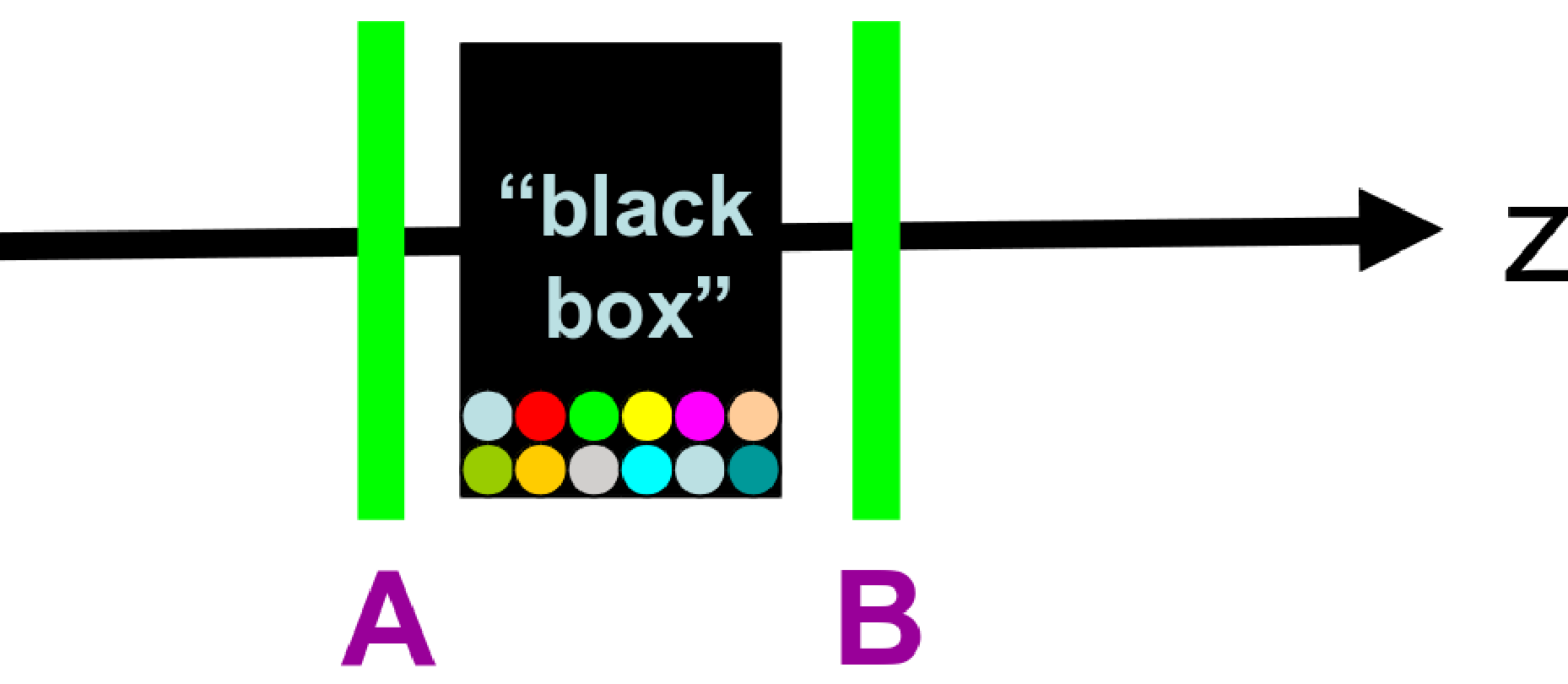}
\caption{Generalised phase-contrast imaging system.  X-rays from a source to the far left (not shown) pass through a sample (not shown), the exit surface of which is denoted $A$.  The wave-field over this plane $A$, which is assumed to be a paraxial beam travelling in the $z$ direction, is then input into an arbitrary imaging system denoted by the black box. The corresponding output complex wave-field exists over the plane $B$.  The state of the black box is schematically denoted by the coloured dials, representing the control parameters $\tau_1,\tau_2$ etc.}  
\label{fig:AbbimagingSystem}
\end{figure}

With this aim in mind, let us consider an arbitrary coherent X-ray imaging system that takes a two-dimensional monochromatic paraxial complex X-ray wave-field $\psi_{\rm IN}(x,y)$ as input.  This input corresponds to the $(x,y)$ plane labelled $A$ in Fig.~\ref{fig:AbbimagingSystem}, which is perpendicular to the optical axis $z$.  Assume also that the state of the imaging system can be characterised by a set of real control parameters $\tau_1, \tau_2, \cdots$, with $\psi_{\rm OUT}(x,y,\tau_1,\tau_2,\cdots)$ being the corresponding complex output wave-field.  Examples of particular control parameters might include the defocus of an imaging system, the spherical aberration of an X-ray lens, the astigmatism of an X-ray lens, the curvature of an X-ray mirror in each of two transverse directions, the angular orientation of a crystal beam-splitter, the thickness of a beam-splitter, the radius of an aperture, the diameter of a rotationally symmetric X-ray source etc.    

We may consider the action of our imaging system, in operator terms.  Before proceeding along these lines, we need to say a few words regarding the operator concept in an imaging setting.  For our purposes, an operator ``acts'' on a given function to give a new function, as we have already seen in the context of the diffraction operator $\mathcal{D}_{\Delta}$.  Thus, if the operator $\mathcal{A}$ acts on the function $f$ to give a different function $g$, this would be written as
\begin{equation}
\mathcal{A}f=g.    
\end{equation}
We follow the usual convention that each
operator acts on the element to the right of it, with the rightmost operator acting first.  For example, if $\mathcal{A},\mathcal{B}$ are two operators, then $\mathcal{BA}f$ is the same as $\mathcal{B}(\mathcal{A}f)$, so that $f$ is first acted upon by $\mathcal{A}$ to give $\mathcal{A}f$, with the result being subsequently acted upon by $\mathcal{B}$ to give $\mathcal{BA}f$.  Returning to the main thread of the argument, we may consider an imaging system to be described by a generalised diffraction operator ${\mathcal{D}}(\tau_1,\tau_2,\cdots)$ that acts on the input field to give the output field.  This may be written in the following way:
\begin{eqnarray}
\psi_{\rm OUT}(x,y,\tau_1,\tau_2,\cdots)
={\mathcal{D}}(\tau_1,\tau_2,\cdots)\psi_{\rm IN}(x,y).
\label{eq:ArbitraryCoherentImagingSystem}
\end{eqnarray}

At this stage our imaging system has a very high degree of generality.  Its arbitrariness is limited only by the assumptions associated with a forward-propagating monochromatic scalar input (over plane $A$ in Fig.~\ref{fig:AbbimagingSystem}) being mapped to a forward-propagating monochromatic scalar output (over plane $B$ in Fig.~\ref{fig:AbbimagingSystem}) that has the same energy.
\subsection{Arbitrary linear imaging systems}

Before proceeding any further, we note that, here and henceforth, the reader is assumed to be familiar with the following basics of Fourier analysis:
\begin{itemize}
    \item forward and inverse Fourier transformation;
    \item the Dirac delta and its associated sifting property;
    \item the concept of convolution and the associated convolution theorem of Fourier analysis;
    \item the Fourier derivative theorem. 
\end{itemize}  
For a coverage of these basics in our specific context of optical physics, see e.g.~the books by Lipson \& Lipson (1981) and Hecht (1987), together with the much more detailed treatments in Bracewell (1986) and Goodman (2005). For a compressed overview that employs a notation consistent with this chapter, see Appendix A in Paganin (2006).  

After this short introduction, we are now ready to delve into the main topic of this section. Let us make the further assumption (beyond those made in the preceding sub-section) that an imaging system is linear, i.e.~that the output field is a linear function of the input field.  Stated differently, we are here assuming the superposition principle to hold: if the input field is given by the sum of two particular input fields $\alpha\psi_{\rm IN}^{(1)}(x,y)+\beta\psi_{\rm IN}^{(2)}(x,y)$, where $\alpha$ and $\beta$ are arbitrary complex weighting coefficients, then the output field will (by assumption) always be equal to the sum of corresponding outputs, i.e.
\begin{eqnarray}
 \mathcal{D}[\alpha\psi_{\rm IN}^{(1)}(x,y)+\beta\psi_{\rm IN}^{(2)}(x,y)]  =\alpha\mathcal{D}\psi_{\rm IN}^{(1)}(x,y)+\beta\mathcal{D}\psi_{\rm IN}^{(2)}(x,y)+\gamma.
 \label{eq:LinearityProperty}
\end{eqnarray}
The complex constant $\gamma$, while consistent with the assumption of linearity, will be set to zero since it is natural to assume that a zero input field corresponds to a zero output field.  Assume further that any magnification, rotation and shear is taken into account via an appropriate choice of coordinates for the plane occupied by the output wave-field.  

Bearing all of the above points in mind, the action of the imaging system can then be described by the following linear integral transform, which may be viewed as a continuous form of matrix multiplication:  
\begin{eqnarray}\label{eq:ArbitraryLinearCoherentImagingSystem}  
\psi_{\rm
OUT}(x,y,\tau_1,\tau_2,\cdots) =\iint dx' dy'
G(x,y,x',y',\tau_1,\tau_2,\cdots)\psi_{\rm IN}(x',y').
\end{eqnarray}
Note, in the present context, that an integral transform may be here understood as ``an integral that transforms one function into another''.  A {\em linear} integral transform is an integral that (i) transforms one function into another, and (ii) has the property of linearity.  The linearity property, by definition, requires the linear integral transform of a sum of two functions, to be equal to the sum of the corresponding transforms (cf.~Eq.~(\ref{eq:LinearityProperty})). For example, given a function $f(x)$ of one variable $x$, an arbitrary linear integral transform could be written as 
\begin{equation}
g(x)=\int f(x') K(x,x') dx' + L(x),     
\end{equation}
where $K(x,x')$ and $L(x)$ are arbitrary functions.  For the purposes of the present overview, we use linear integral transforms such as that in Eq.~(\ref{eq:ArbitraryLinearCoherentImagingSystem}) to represent the action of linear imaging systems.  Here, the linear integral transform serves to change (transform!) the field input into the imaging system, into the field that is output from the imaging system.  Finally, we note that: (i) The function $K(x,x')$ is often called the ``kernel'' of the linear integral transform; (ii) if one can assume that a zero input gives a zero output, then $L(x)=0$.

The kernel of the linear integral transform in Eq.~(\ref{eq:ArbitraryLinearCoherentImagingSystem}) has been denoted by $G$, since it is a Green function (Bracewell, 1986; Goodman, 2005; Paganin, 2006).  It may also be interpreted as a generalised Huygens wavelet (Lipson \& Lipson, 1981; Hecht, 1987; Born \& Wolf, 1999).  Other terms, that may be used for exactly the same concept, include (i) ``real-space propagator''; (ii) ``wave-field propagator''; (iii) ``impulse response''; and (iv) ``complex point-spread function''. To further understand the key physical idea that is underpinned by this concept, choose the special case 
\begin{equation}
\psi_{\rm IN}(x',y')=\delta(x'-x_0,y'-y_0)
\end{equation}
for the input field in Eq.~(\ref{eq:ArbitraryLinearCoherentImagingSystem}), where $\delta(x,y)$ is a two-dimensional Dirac delta, corresponding to a single point 
\begin{equation}
(x,y)=(x_0,y_0)
\end{equation}
being illuminated in the input plane of the imaging system.  Via the sifting property of the Dirac delta, Eq.~(\ref{eq:ArbitraryLinearCoherentImagingSystem}) gives the associated output field as $G(x,y,x_0,y_0,\tau_1,\tau_2,\cdots)$.  Therefore $G(x,y,x_0,y_0,\tau_1,\tau_2,\cdots)$ is the output field as a function of $x$ and $y$ coordinates in the output plane, which would be obtained if a unit-strength point source were to be located at position $(x_0,y_0)$ in the input plane, and the imaging system interposed between input and output plane were to have the state characterised by the particular control parameters $\tau_1,\tau_2,\cdots$.  Hence $G(x,y,x_0,y_0,\tau_1,\tau_2,\cdots)$ is indeed a generalised Huygens-type wavelet, with the form of the wavelet depending on both the state of the imaging system and on the position $(x_0,y_0)$ of the input ``pinpoint of X-ray light''.  

We close this sub-section by reversing the above chain of logic, so as to physically motivate the writing down of Eq.~(\ref{eq:ArbitraryLinearCoherentImagingSystem}) for an arbitrary linear imaging system.  We characterise such an imaging system by the fact that, if the input is a ``pinpoint of X-ray light'' $\delta(x-x',y-y')$ at some point $(x',y')$ in the entrance plane $A$ of Fig.~\ref{fig:AbbimagingSystem}, then the corresponding output field---considered as a function of coordinates $(x,y)$ over the output plane $B$---will be given by $G(x,y,x',y',\tau_1,\tau_2,\cdots)$.  In this expression for the output field $G$, the coordinates $(x',y')$ of the input ``pinpoint of X-ray light'' are considered to be fixed, with the parameters $\tau_1,\tau_2,\cdots$ describing the state of the imaging system also being fixed. To proceed further, we can use the sifting property of the Dirac delta to decompose an arbitrary input field $\psi_{\rm IN}(x,y)$ as a superposition (described by the continuous sum, namely an integral) of X-ray pinpoints of light, each such pinpoint having the form $\delta(x-x',y-y')$, so that:
\begin{equation}\label{eq:PixelDecomposition}
\psi_{\rm IN}(x,y)=\iint dx' dy' \psi_{\rm IN}(x',y') \delta(x-x',y-y').
\end{equation}
In order to map the input field to the output field, namely to convert $\psi_{\rm IN}(x,y)$ on the left side of the above integral into $\psi_{\rm OUT}(x,y)$, we need only replace each of the pinpoint inputs $\delta(x-x',y-y')$ under the integral sign, with its corresponding output $G(x,y,x',y',\tau_1,\tau_2,\cdots)$.  This invocation of the superposition principle---which is justified on account of our key assumption that the imaging system is linear---leads directly from Eq.~(\ref{eq:PixelDecomposition}) to Eq.~(\ref{eq:ArbitraryLinearCoherentImagingSystem}).

\subsection{Arbitrary linear shift-invariant imaging systems}

We specialise still further, by assuming the linear imaging system to be ``shift invariant''.  This augments the previous assumptions, with the additional assumption that, if there is a transverse shift of the input wave-field, this merely serves to transversely shift the output wave-field. Such an assumption cannot hold for arbitrarily large transverse shifts, but is often approximately true for a sufficiently small range of transverse shifts in the vicinity of the centre of the field of view of a coherent linear imaging system.  In this approximation, the Green function depends only on coordinate differences, in the sense that the action of the system does not change under transverse translation.  Mathematically, the assumption of transverse-shift invariance implies that Eq.~(\ref{eq:ArbitraryLinearCoherentImagingSystem}) may be simplified to:    
\begin{eqnarray}\label{eq:ArbitraryShiftInvariantLinearCoherentImagingSystem1}
\qquad \psi_{\rm
OUT}(x,y,\tau_1,\tau_2,\cdots) =\iint dx' dy'
G(x-x',y-y',\tau_1,\tau_2,\cdots)\psi_{\rm IN}(x',y').
\end{eqnarray}
This is a two-dimensional convolution integral, and hence may be more compactly written as:
\begin{eqnarray}\label{eq:ArbitraryShiftInvariantLinearCoherentImagingSystem2}
\psi_{\rm
OUT}(x,y,\tau_1,\tau_2,\cdots)  = \psi_{\rm IN}(x,y) \otimes G(x,y,\tau_1,\tau_2,\cdots),
\end{eqnarray}
\noindent where $\otimes$ denotes two-dimensional convolution.  

A very rich variety of imaging systems in coherent X-ray optics may be described using the formalism based on Eq.~(\ref{eq:ArbitraryShiftInvariantLinearCoherentImagingSystem1}), including propagation-based X-ray phase contrast, analyser-crystal-based phase contrast, imaging/microscopy using compound refractive lenses, imaging/microscopy using Fresnel zone plates, inline holography, off-axis holography, Zernike phase-contrast imaging and imaging/microscopy using Kirkpatrick--Baez mirrors.   However, systems such as X-ray wave-guides, where transverse shift invariance is inapplicable, need the more general form given by  Eq.~(\ref{eq:ArbitraryLinearCoherentImagingSystem}).

\subsection{Transfer function formalism}

Before proceeding any further, let us establish the convention for Fourier transforms that will be used for the remainder of this chapter.  As previously mentioned, we use the Fourier-transform convention and notation from Appendix A of Paganin (2006). In one spatial dimension, the Fourier transform $\mathcal{F}$ of a function $g(x)$ is denoted by
\begin{equation}
\mathcal{F}[g(x)]\equiv\breve{g}(k_x),    
\end{equation}
where $k_x$ is the Fourier coordinate corresponding to $x$, and
\begin{equation}\label{eq:FourierTransformForward}
\breve{g}(k_x)=\frac{1}{\sqrt{2\pi}}\int_{-\infty}^{\infty} g(x) \exp(-ik_xx)dx.
\end{equation}
Furthermore,  
\begin{equation}\label{eq:FourierTransformInverse}
g(x)=\frac{1}{\sqrt{2\pi}}\int_{-\infty}^{\infty} \breve{g}(k_x)\exp(ik_xx)dk_x    
\end{equation}
denotes the corresponding inverse Fourier transform.  In two spatial dimensions, and in an obvious extension of the above notation, the forward Fourier transform becomes
\begin{equation}
\breve{g}(k_x,k_y)=\frac{1}{2\pi}\iint_{-\infty}^{\infty} g(x,y) \exp[-i(k_xx+k_yy)]dxdy.
\end{equation}
Similarly, the inverse transform becomes
\begin{equation}
g(x,y)=\frac{1}{2\pi}\iint_{-\infty}^{\infty} \breve{g}(k_x,k_y)\exp[i(k_xx+k_yy)]dk_xdk_y.    
\end{equation}

The forward and inverse Fourier transforms may be viewed in rather physical terms as embodying the related concepts of synthesis and decomposition.  In many areas of mathematical physics, complicated objects are often synthesized via a linear combination of simpler objects (``basis elements'' or ``basis functions'').  Examples include the decimal-number system in which an arbitrary real number is expressed as a linear combination of all different integer powers of 10, the idea of a beam of sunlight being composed of a sum of light beams having all of the different colours of the rainbow, the Taylor series representation of a function $g(x)$ as a linear combination of different powers of $x$, and the Fourier series representation of the same function as a linear combination of oscillatory sine and cosine terms of varying period.  In all of these examples, synthesis and decomposition go hand-in-hand. Thus, returning to the example of the beam of sunlight, we may (i) decompose the said polychromatic beam into a linear combination of monochromatic beams, each of which have a particular energy (colour), or we may instead (ii) synthesise the polychromatic beam by ``building'' it as a linear superposition of monochromatic beams.  The processes of synthesis and decomposition are inverses of one another, both conceptually and analytically.  Under this synthesis--decomposition viewpoint, the inverse Fourier transformation in Eq.~(\ref{eq:FourierTransformInverse}) synthesises a function $g(x)$ by expressing it as a linear combination of oscillatory basis functions $\exp(ik_xx)$.  The multiplicative weight factors, in this synthesis, are the coefficients $\breve{g}(k_x)/\sqrt{2\pi}$. The converse process of decomposition occurs in the forward Fourier transformation shown in  Eq.~(\ref{eq:FourierTransformForward}), since this integral takes the function $g(x)$ as input, thereby decomposing it into the weighting coefficients $\breve{g}(k_x)$.  It is also worth remarking that, in some sense, $g(x)$ and $\breve{g}(k_x)$ can be viewed as different representations of the same entity, since either can be uniquely converted to the other via the forward or inverse Fourier transformations.

We digress!  Having established both our conventions and notation for Fourier transformation, let us Fourier transform both sides of Eq.~(\ref{eq:ArbitraryShiftInvariantLinearCoherentImagingSystem2}) with respect to $x$ and $y$.  Invoke the convolution theorem of Fourier analysis, to convert convolution in $(x,y)$ space to multiplication in $(k_x,k_y)$ space. The inverse Fourier transform of the resulting expression is the following operator-type description for the action of the imaging system (cf.~Eq.~(\ref{eq:ArbitraryCoherentImagingSystem}), which has the same form but is more general):
\begin{eqnarray}
\psi_{\rm
OUT}(x,y,\tau_1,\tau_2,\cdots) = {{\mathcal{D}}}(\tau_1,\tau_2,\cdots) \psi_{\rm IN}(x,y).
\label{eq:ArbitraryShiftInvariantLinearCoherentImagingSystem3}
\end{eqnarray}
\noindent Here, the Fourier transform of our generalised Huygens-type wavelet $G(x,y,\tau_1,\tau_2,\cdots)$ is termed the ``transfer function'': 
\begin{eqnarray}
T(k_x,k_y,\tau_1,\tau_2,\cdots)\equiv 2\pi\mathcal{F}
[G(x,y,\tau_1,\tau_2,\cdots)],
\label{eq:ArbitraryShiftInvariantLinearCoherentImagingSystem4}
\end{eqnarray}
\noindent $(k_x,k_y)$ denote Fourier-space (spatial frequency)  coordinates corresponding to real-space coordinates $(x,y)$, and the generalised diffraction operator quantifying our imaging system is the following Fourier-space filtration:
\begin{eqnarray} {{\mathcal{D}}}(\tau_1,\tau_2,\cdots)=
\mathcal{F}^{-1}T(k_x,k_y,\tau_1,\tau_2,\cdots)\mathcal{F}.
\label{eq:ArbitraryShiftInvariantLinearCoherentImagingSystem5}
\end{eqnarray}

\noindent Once again, it is important to recall that all operators act from right to left: i.e.~if the operator ${{\mathcal{D}}}(\tau_1,\tau_2,\cdots)$ is applied to an input field, that input field is first acted upon by the Fourier transform $\mathcal F$, then multiplied by the transfer function ${T}$, and then inverse Fourier transformed.  

In words, Eqs.~(\ref{eq:ArbitraryShiftInvariantLinearCoherentImagingSystem3}) and (\ref{eq:ArbitraryShiftInvariantLinearCoherentImagingSystem5}) state the following. In order to map the input field $\psi_{\rm IN}(x,y)$ to the corresponding field $\psi_{\rm OUT}(x,y)$ output by a linear shift-invariant imaging system, a sequence of three steps may be used:
\begin{enumerate}

    \item Apply the Fourier transform operator $\mathcal F$ to the input field (decomposition of the input wave);

    \item Multiply the resulting object, which will be a function of the Fourier coordinates $(k_x,k_y)$, by the transfer function $T(k_x,k_y,\tau_1,\tau_2,\cdots)$ corresponding to the linear shift-invariant imaging system being in a state described by the control parameters $(\tau_1,\tau_2,\cdots)$ (filtration of the Fourier components of the input wave);

    \item Apply the inverse Fourier transform operator (synthesis of the output wave).

\end{enumerate}
This verbal description may be considered as pseudo code for a computational simulation of any linear shift-invariant imaging system.  The corresponding computer codes are typically rendered extremely efficient by the use of the fast Fourier transform (FFT) to implement both the forward and inverse Fourier transform operators (Press, Teukolsky, Vetterling, \& Flannery, 2007).  From a more physical perspective, we re-iterate the following: Step 1 is a {\em decomposition} of the input field into its constituent plane-wave components (Fourier components), Step 2 is a {\em filtration} of these plane-wave components in which each such plane-wave component is weighted by a different multiplicative factor that is given by the transfer function $T$, and Step 3 is a {\em synthesis} in which all of the resulting re-weighted plane waves are added up to give the output field.   

An important special case, of a linear shift-invariant imaging system, is the previously considered case of paraxial free-space propagation through vacuum by a distance $\Delta$.  In this case we can write ${{\mathcal{D}}}(\tau_1,\tau_2,\cdots)\longrightarrow {\mathcal{D}}_{\Delta}$, with
\begin{equation}\label{eq:DiffractionOperatorFresnelDiffraction}
   {\mathcal{D}}_{\Delta}=\exp(ik\Delta)\mathcal{F}^{-1}
   \exp\left[\frac{-i\Delta(k_x^2+k_y^2)}{2k}\right]\mathcal{F}.
\end{equation}
This is a two-Fourier-transform version of the Fresnel diffraction integral, using an operator notation that is consistent with Eq.~(\ref{eq:FresnelDiffraction}).  For more detail on Eq.~(\ref{eq:DiffractionOperatorFresnelDiffraction}), together with a derivation based on the more general angular-spectrum formulation for non-paraxial wave-field propagation, we refer the reader to Paganin (2006).  

A second important special case corresponds to analyser-based X-ray phase contrast, where the X-ray field transmitted through a sample is reflected from the surface of a near-perfect crystal before having its intensity registered by a position-sensitive detector.  In this case, upon suitable rotation of the $(x,y)$ coordinates, 
\begin{eqnarray} {{\mathcal{D}}}(\tau_1,\tau_2,\cdots)\longrightarrow \mathcal{A} = \mathcal{F}^{-1} A(k_x) \mathcal{F},
\label{eq:ArbitraryShiftInvariantLinearCoherentImagingSystem6}
\end{eqnarray}
\noindent where the analyser-crystal transfer function $A(k_x)$ is a polarisation-dependent function of $k_x$ whose exact form is not needed here.

For a third and final special case, we may generalise Eq.~(\ref{eq:DiffractionOperatorFresnelDiffraction}) so as to characterise a wide class of shift-invariant coherent linear imaging systems.  This may be done in a very natural way by noting that the exponential in Eq.~(\ref{eq:DiffractionOperatorFresnelDiffraction}) is given by the complex unit $i$ multiplied by a quadratic function of spatial frequency.  Evidently, this can be generalised, by replacing the exponent with $i$ multiplied by an arbitrary Taylor series expansion in all possible powers $(k_x)^m(k_y)^n$ of the spatial frequency, where $m=0,1,2,\cdots$ and $n=0,1,2,\cdots$.  Letting the coefficients in such a Taylor-series expansion be denoted via the set of complex aberration coefficients $\{\alpha_{m,n}\}$, we can then write: 
\begin{eqnarray} {{\mathcal{D}}}(\tau_1,\tau_2,\cdots)\longrightarrow{{\mathcal{D}}}(\{\alpha_{m,n}\})=  \mathcal{F}^{-1} \exp \left[i \sum_{m=0}^{\infty} \sum_{n=0}^{\infty}  \alpha_{m,n} (k_x)^m (k_y)^n \right]\mathcal{F}.
\label{eq:ArbitraryShiftInvariantLinearCoherentImagingSystem7}
\end{eqnarray}
We speak of the control parameters  $\{\alpha_{m,n}\}$ as ``aberration coefficients'' on account of their very close connection to the theory of aberrated imaging systems (Born \& Wolf, 1999).  These aberrations include ``coherent aberrations'' such as defocus, spherical aberration, coma and astigmatism, all of which correspond to real non-zero values of certain $\alpha_{m,n}$ coefficients.  We should also point out the ``incoherent aberrations'' associated with the imaginary parts of certain $\alpha_{m,n}$ coefficients, which relate to Fourier-space damping such as that which is induced by source-size blur and/or the finite acceptance angle of a specified imaging system.  Non-zero values, for any one or more of the aberration coefficients $\{\alpha_{m,n}\}$, typically yields phase contrast.  Conversely, for ``perfect'' imaging systems in which all of the aberration coefficients vanish, no phase contrast is operative.  We will explore this last-mentioned point in more detail, in the next section.  For further information on the aberration-coefficient representation for imaging systems, in the specific context of phase-contrast imaging, see e.g.~Paganin (2006), Paganin \& Gureyev (2008), Beltran, Kitchen, Petersen, \& Paganin (2015) and Paganin, Petersen, \& Beltran (2018), together with references therein.

\subsection{Phase contrast}

The squared magnitude of the input--output equation
\begin{equation}
\psi_{\rm
OUT}(x,y,\tau_1,\tau_2,\cdots)= {\mathcal D}(\tau_1, \tau_2, \cdots) \psi_{\rm IN}(x,y)
\end{equation}
gives the intensity, output by a shift-invariant linear imaging system, as:
\begin{eqnarray}\label{eq:IntensityTransfer}
I_{\rm
OUT}(x,y,\tau_1,\tau_2,\cdots)= |{\mathcal D}(\tau_1, \tau_2, \cdots) \psi_{\rm IN}(x,y)|^2. \quad
\end{eqnarray}

Evidently---and with the important exception of the ``perfect imaging system'' case where $\mathcal{D}$ is equal to unity---the output intensity typically depends upon both the intensity and phase of the input, since the right side of Eq.~(\ref{eq:IntensityTransfer}) will typically couple the phase of the input field to the intensity of the output field.  Any state $(\tau_1,\tau_2,\cdots)$ of the imaging system, which generates an output intensity that is influenced by the input phase, is said to exhibit ``phase contrast''. Again, most states of an ``imperfect'' imaging system described by the operator $\mathcal{D}\ne 1$ will yield both intensity contrast and phase contrast. 

Let us re-iterate this particularly important point.  If we define a ``perfect'' imaging system as one which perfectly reproduces the input field, up to a magnification and transverse translation that are both irrelevant in the present context since such effects may always be taken into account via a suitable change of spatial-coordinate system, then such a system will have 
\begin{equation}
{\mathcal D}(\tau_1, \tau_2, \cdots) = 1.
\end{equation}
Equation~(\ref{eq:IntensityTransfer}) thereby reduces to 
\begin{equation}
I_{\rm
OUT}(x,y,\tau_1,\tau_2,\cdots)=I_{\rm IN}(x,y).  
\end{equation}
Hence, an imaging system which is perfect at the field level---in the sense that the generalised diffraction operator mapping input field to output field is given by $\mathcal{D}=1$---yields no phase contrast.  Perfect imaging systems are maximally imperfect for the purposes of phase contrast. Conversely,  imperfect (aberrated) imaging systems typically do exhibit phase contrast.  Evidently, ``perfect imaging system'' and ``aberrated imaging system'' are misleading misnomers in the context of phase-contrast imaging.  Nevertheless, we shall continue to use these terms, since their meaning is somewhat entrenched in optical-physics usage.   

We close this section by elaborating, a little further, on the link between optical aberrations and phase contrast.  Given the observation that the defocus-generated ``aberration'' (e.g.~via free-space propagation) has an associated paraxial wave equation and transport-of-intensity equation for describing propagation-based phase contrast, it naturally follows that there will be generalised wave equations and generalised intensity-transport equations associated with arbitrary optical aberrations and aberration-based phase contrast, in which one propagates through an abstract ``aberration space'' rather than a physical free-space vacuum (Allen, Oxley, \& Paganin, 2001; Paganin \& Gureyev, 2008; Beltran, Kitchen, Petersen, \& Paganin, 2015; Paganin, Petersen, \& Beltran, 2018).  These papers, together with references therein, give further information regarding the nature of the phase contrast that may be associated with tuning any one control parameter, or set of control parameters, that characterise arbitrary aberrated  linear shift-invariant imaging systems.

\subsection{Introducing partial coherence: source blurring}

\begin{figure}
\includegraphics[scale=2.0]{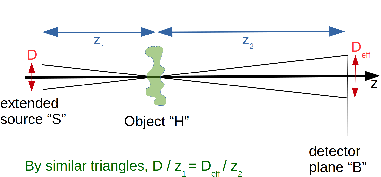}
\caption{An extended incoherent source $S$, with diameter $D$, leads to source-size blurring of spatial width $D_{\rm eff}=D \, z_2 / z_1$ over the detector plane $B$.}
\label{fig:Blurring}
\end{figure}

We open this section by briefly outlining the idea of image blurring due to non-zero source size. The fact, that the X-ray source is not a point source, will lead to some blurring of images formed using such a source: see Fig.~\ref{fig:Blurring}.  For a so-called ``extended incoherent source'', say a planar source of diameter $D$, one can by definition consider each point on the source to be an independent radiator (Zernike, 1938) of X-rays.  Provided that the source is not too large, and that the assumption of independent radiators is reasonable, each of the radiators will form a separate image of the object, with images due to separate points on the source being transversely displaced from one another.  The net effect, of adding all of the slightly-displaced images formed by each point on the extended incoherent source, is to blur the resulting image obtained over the plane $B$.  The transverse length scale, over which this blurring takes place, may be obtained via the similar-triangles construction in Fig.~\ref{fig:Blurring}.  Here, we see that the transverse spatial extent $D_{\rm eff}$ of the source-size-induced blurring, of the image of the object $H$ that is obtained over the detector plane $B$, is given by
\begin{equation}
D_{\rm eff}=\frac{D \, z_2}{z_1}.       
\end{equation}
Note that the source-size-induced blurring becomes progressively worse as $z_2$ increases, for fixed $D$ and $z_1$.  For a simple example of this effect, one can hold one's hand against a flat horizontally-oriented matt surface on a clear day while the sun is directly overhead, and observe how the initially-sharp shadow becomes progressively more blurred as the hand is lifted progressively higher above the flat surface.

\begin{figure}
\includegraphics[scale=1.6]{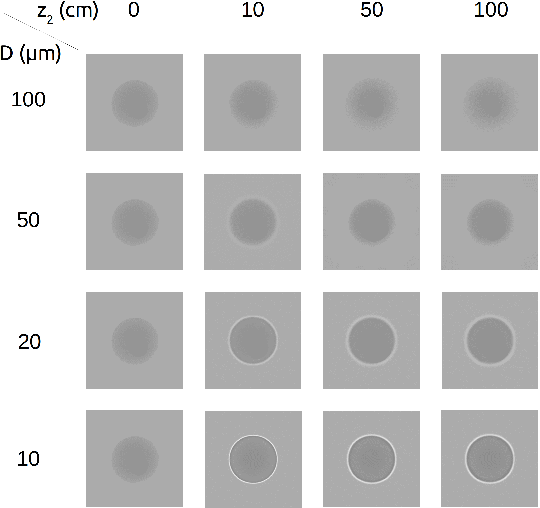}
\caption{Simulated propagation-based X-ray phase-contrast images of a solid carbon sphere of diameter 0.5 mm, corresponding to X-rays with wavelength 0.5 \AA. The source-to-object distance $z_1$ is fixed at 10 cm.  The object-to-detector distance $z_2$ is increased as one moves from left to right.  The source diameter $D$ decreases as one moves from top to bottom.  Adapted from Gureyev et al.~(2009).}
\label{fig:RRR2}
\end{figure}

We now return to the main thread of our discussion regarding propagation-based phase contrast, by considering the effect of source-size blur in such a setting.  To develop intuition regarding how the image-sharpening effect of propagation-based phase contrast competes with the image-smearing effect of source-size blurring, consider the simulations shown in Fig.~\ref{fig:RRR2}.  These simulations correspond to an X-ray wavelength of 0.5 \AA, and a fixed source-to-object distance $z_1$ of 10 cm.  The two variables are (i) the diameter of the source $D$, which decreases from top to bottom in the figure, and (ii) the object-to-detector propagation distance $z_2$, which increases from left to right.  The simulated sample is a solid carbon sphere with diameter 0.5 mm.  When the source has the relatively large diameter of $D=100\,\mu$m, corresponding to the top row of Fig.~\ref{fig:RRR2}, increasing the object-to-detector distance $z_2$ has the expected effect of progressively blurring the image of the sphere.  A similar trend is seen in the second row, corresponding to halving the source diameter.  All of the images in the top two rows of Fig.~\ref{fig:RRR2} may be taken as demonstrating absorption contrast alone, with a dark ``shadow'' of the carbon sphere corresponding to the absorption of X-rays that pass through the sphere.  However, in the bottom two rows of the figure, the source size is sufficiently small to have reduced the source-size-induced blurring to such a degree that propagation-based phase contrast is manifest.  Edge contrast, in the sense described earlier, is clearly evident in both of the bottom rows, for object-to-detector propagation distances of 10 cm or greater (columns 2, 3 and 4 of the bottom two rows).  If the object-to-detector propagation distance is zero, however, one has a contact image that displays no propagation-based X-ray phase contrast (column 1).  In all of the above, one has an evident trade-off: (i) $z_2$ must be sufficiently large to obtain sufficiently strong propagation-induced phase contrast; (ii) $z_2$ must be sufficiently small for the source-size blurring to be sufficiently mild that it does not wash out the sharpening effect of propagation-based phase contrast.
\subsection{Partial coherence}

The reader is assumed to be familiar with elementary concepts of partial coherence, including the notions of spatial and temporal coherence, at the level of introductory texts such as those of Lipson \& Lipson (1981) or Hecht (1987). Here we build upon such foundations via a more advanced (albeit very intuitive, if one thinks about it for long enough!) perspective based on the coherence-optics equivalent of the density-matrix formalism of quantum mechanics (Bransden \& Joachain, 1989).  This formalism is known as the space--frequency description of partial coherence (Wolf, 1982; Mandel \& Wolf, 1995; Wolf, 2007).

Our intent here is rather modest.  We will briefly introduce the space--frequency description of partial coherence, in which a given partially-coherent field may be described via a two-point correlation function known as the ``cross-spectral density''.  Two points in space are needed, here, since the concept of partial coherence may be very closely tied to the ability of two separate optical disturbances to form interference fringes when combined, as explained in Zernike's groundbreaking paper (Zernike, 1938). It is worth bearing in mind, while swimming through the abstractions of two-point correlation functions such as the cross-spectral density, that such constructs may always be ultimately tied back to the idea that the degree of coherence is essentially about the capacity of wave-fields to form interference fringes: the higher the degree of visibility of such interference fringes, the higher the degree of coherence (Zernike, 1938; Born \& Wolf, 1999; Wolf, 2007). The double-pinhole experiment may be regarded as the archetypal interference experiment.  In the idealised case of two point-like pinholes, the contrast of the interference fringes is directly related to the two-point correlation function.   Historical remarks aside, the cross-spectral density may be obtained at any given angular frequency via a suitable averaging procedure over a statistical ensemble of strictly monochromatic fields, all of which have the same angular frequency.  We use this space--frequency description of partially-coherent X-ray fields to describe the action of any phase-contrast imaging system, provided that the system is linear and the effects of polarisation may be neglected.  As one further aside, we note that the process, of partial-coherence calculations that involve averaging over a statistical ensemble, has a rather strong conceptual connection with the use of Gibbs-type statistical ensembles in the formalism of statistical mechanics and thermodynamics (Sears \& Salinger, 1975; Huang, 1987).     

An advantage of the particular formalism adopted here---apart from its broad applicability to a rich variety of both existing and future phase-contrast X-ray imaging scenarios---is the readiness with which the resulting mathematical expressions may be implemented in computer code. Indeed, much of the mathematics that will follow is, in essence, computer pseudo-code rather than explicit calculations.

Since the context is X-ray phase-contrast imaging, or X-ray imaging more broadly, we describe a given paraxial complex scalar partially-coherent X-ray field via a stochastic process that can be characterised via a statistical ensemble of strictly monochromatic fields 
\begin{equation}
\{\psi_\omega^{(j)}(x,y),c_j\}
\label{eq:EnsembleStrictlyMonochromaticFields1}
\end{equation}
at each angular frequency $\omega$, with all fields in the ensemble having the same angular frequency $\omega$ (Wolf, 1982; Mandel \& Wolf, 1995; Wolf, 2007).  Here, the {\em j} index labels each member of the ensemble, with the associated real statistical weight $c_j$.  Each of these weights lies between zero and unity inclusive, with 
\begin{equation}
\sum_j c_j = 1.  
\label{eq:EnsembleStrictlyMonochromaticFields2}
\end{equation}
In general, these weights will depend on angular frequency, although for compactness our notation does not explicitly indicate this dependence. 

The ``cross-spectral density'' $W_{\omega}(x_1,y_1,x_2,y_2)$ is a two-point correlation function that quantifies the degree of correlation between the optical disturbance at the pair of points $(x_1,y_1)$ and $(x_2,y_2)$, at the specified angular frequency $\omega$.  It is given by the ensemble average over all of the $N$ members of the statistical ensemble, with this averaging process being indicated via angular brackets:
\begin{equation}
W_{\omega}(x_1,y_1,x_2,y_2) = \sum_{j=1}^{j=N} c_j \psi_\omega^{(j)*}(x_1,y_1) \psi_\omega^{(j)}(x_2,y_2)  \equiv \langle \psi_\omega^{*}(x_1,y_1) \psi_\omega(x_2,y_2)\rangle.
\label{eq:Coherence1}
\end{equation}
In the above expression, the superscript asterisk denotes complex conjugation.

The associated spectral density, which may be viewed as the ``diagonal of the cross-spectral density'', is:
\begin{equation}
S_{\omega}(x,y) \equiv W_{\omega}(x,y,x,y) = \langle |\psi_\omega(x,y)|^2 \rangle.
\label{eq:GettingSfromW}
\end{equation}

Note that the spectral density is simply the average intensity, being a weighted average over the intensities of the individual fields that make up the statistical ensemble $\{\psi_{\omega}^{(j)}(x,y),c_j\}$.  We note also that the measured intensity $I(x,y)$, registered by a detector which integrates over angular frequencies $\omega$, will be given by the following weighted average of the spectral density:
\begin{equation}\label{eq:Coherence7}
I(x,y)=\int S_{\omega}(x,y) \aleph(\omega) d\omega ,
\end{equation}

\noindent where $\aleph(\omega)$ is a measure of the variable efficiency of the detector, as a function of angular frequency $\omega$.  The form of the energy spectrum is implicitly taken into account in the cross-spectral density itself, considered as function of energy 
\begin{equation}
E=\hbar\omega,     
\end{equation}
where $\hbar$ is Planck's constant $h$ divided by $2\pi$.  Note also that the set of weights $\{c_j\}$, used to construct each of the $\omega$-dependent quantities $S_{\omega}(x,y)$ in the above spectral sum (Eq.~(\ref{eq:Coherence7})), will in general be different for each $\omega$.  Thus, as previously mentioned, $c_j$ could be more precisely notated as $c_{j,\omega}$.  Notwithstanding this, for the sake of notational simplicity we will continue to denote the statistical weights at a given fixed angular frequency $\omega$, as $c_j$.  Lastly, to expand on a point that was made a little earlier in the present paragraph, we note that the energy spectrum in Eq.~(\ref{eq:Coherence7}) is implicit in the normalisation of the $\psi_{\omega}^{(j)}(x,y)$ functions.  Thus, if $\omega$ takes a value $\tilde{\omega}$ which is such that there is very little X-ray power at energy $\hbar\tilde{\omega}$, all of the $\psi_{\tilde{\omega}}^{(j)}(x,y)$ functions will be multiplied by a suitably small normalisation factor.

We will take the cross-spectral density as the descriptor of partially-coherent X-ray fields whose statistics are independent of time (a property known as statistical stationarity).  We have also implicitly made the assumption of ergodicity, namely the assumed equality of ensemble averages with time averages. Moreover, if Gaussian statistics may be assumed, then all higher-order correlation functions may be determined from the two-point field correlation function $W$ with which we are working. Proper treatments of these important points will not be given here, with the reader being referred to standard texts such as those of Goodman (1985) and Mandel \& Wolf (1995).  For a more elementary treatment of these underpinning concepts, see Wolf (2007).   

\subsection{Modelling a wide class of partially-coherent X-ray phase-contrast imaging systems}

We now turn to the question of how the space--frequency description for partial coherence may be used to study the manner in which cross-spectral densities are transformed upon passage through linear imaging systems utilising partially-coherent X-ray radiation.  Given the previously-mentioned relation between cross-spectral density and spectral density, and the fact that spectral density may be averaged over angular frequency to give the total detected intensity distribution, the formalism outlined below will allow one to determine the intensity distribution output by any linear phase-contrast imaging system.  This covers a rather broad class of X-ray phase-contrast imaging systems, and is likely to also cover a rather large class of coherent imaging systems that may be developed in the future.  While we ignore the effects of polarisation, this attribute can be readily taken into account by suitably generalising the ideas presented here, if required.  Such a generalisation involves replacing the complex scalar cross-spectral density with a second-rank or third-rank tensor: see Mandel \& Wolf (1995) and Wolf (2007), together with references therein, for information regarding such tensorial cross-spectral densities. 

\subsubsection{Space--frequency model for cascaded systems}

The following steps enable one to calculate the intensity distribution, produced by a large class of linear X-ray phase-contrast imaging systems, using the space--frequency model:

\begin{enumerate}
\item Assume a particular realistic model of the X-ray source after its radiation has passed through whatever conditioning optics, such as monochromators, that may be present in a given X-ray phase-contrast imaging system.  This will imply a given statistical ensemble of strictly monochromatic fields
\begin{equation}\label{eq:EnsembleOfFields}
\{\psi_\omega^{(j)}(x,y),c_j\}, 
\end{equation}
at the entrance surface of an object that is to be imaged. 

\item Assume that the object is static, non-magnetic, elastically scattering and sufficiently gently spatially varying for the projection approximation to be valid.  The ensemble of fields $\{\psi_\omega^{(j)}(x,y),c_j\}$ at the entrance surface of the object then leads to the ensemble of fields
\begin{equation}
\quad\quad \{\psi_\omega'^{(j)}(x,y),c_j\}=\{\psi_\omega^{(j)}(x,y)\mathcal{T}^{(j)}_{\omega}(x,y),c_j\} 
\end{equation}
at the nominal exit surface of the object, where $\mathcal{T}^{(j)}_{\omega}(x,y)$ is the complex transmission function corresponding to illumination of the object by the strictly monochromatic field $\psi_\omega^{(j)}(x,y)$.  

\item Each member in the ensemble of fields $\{\psi_\omega'^{(j)}(x,y),c_j\}$ at the exit surface of the object may then be individually propagated through a specified linear imaging system, to give the ensemble of strictly monochromatic fields
\begin{eqnarray}
\{\psi_\omega''^{(j)}(x,y),c_j\}  =\{\mathcal{D}^{(j)}_{\omega}(\tau_1(\omega),\tau_2(\omega),\cdots)\psi_\omega'^{(j)}(x,y),c_j\}
\end{eqnarray}
at the exit surface of the imaging system. Here,
\begin{eqnarray}
\mathcal{D}^{(j)}_{\omega}(\tau_1(\omega),\tau_2(\omega),\cdots) 
\end{eqnarray}
denotes the operator which maps the {\em j}th input field at angular frequency $\omega$, to the {\em j}th output field at the same angular frequency, when the imaging system is in a state characterised by the control parameters $\tau_1(\omega),\tau_2(\omega),\cdots$.  Note that the control parameters of the imaging system will in general vary with angular frequency, even though the said imaging system would typically have the same physical configuration for all angular frequencies.  However, for paraxial fields, one could often assume $\mathcal{D}^{(j)}_{\omega}(\tau_1(\omega),\tau_2(\omega),\cdots)$ to be independent of $j$.

\item The exit-surface, of the previously mentioned generalised imaging system, is assumed to coincide with the surface of the detector.  The cross-spectral density $W$, the spectral density $S$, and the average intensity $I$, over the surface of the detector, can then be calculated using the ensemble $\{\psi_\omega''^{(j)}(x,y),c_j\}$, together with the previously specified formulae for $W$, $S$ and $I$.     
\end{enumerate}

\begin{figure}
\includegraphics[scale=0.28]{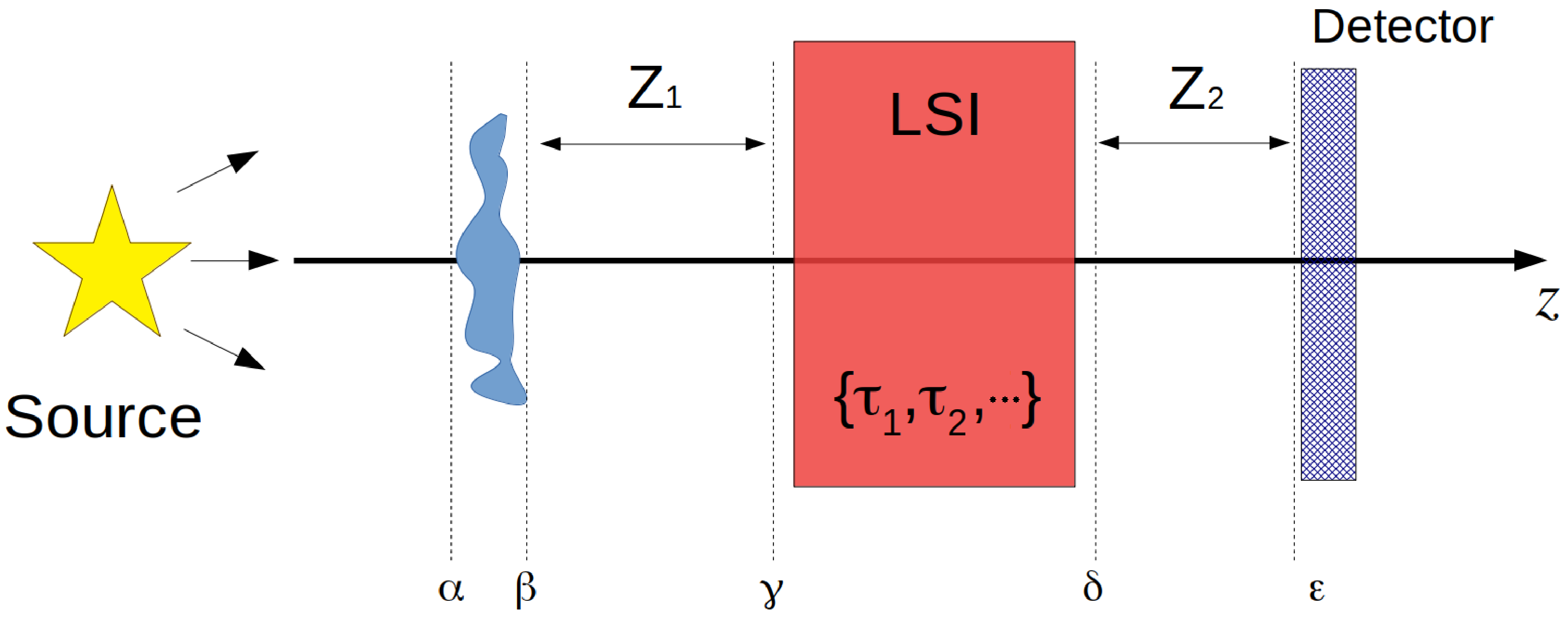}
\caption{Sample experimental setup, to illustrate the means for modelling cascaded X-ray imaging systems that is outlined in the main text.}
\label{fig:ExampleImagingSystem}
\end{figure}

As an example of the logic outlined in general terms above, but with the added complication of chaining together several linear imaging-system operators, consider the cascaded X-ray phase-contrast imaging setup that is sketched in Fig.~\ref{fig:ExampleImagingSystem}.  Here, an incompletely monochromated and therefore partially coherent X-ray source (energy $E=\hbar\omega$) illuminates a thin object lying between the planes $\alpha$ and $\beta$, before passing through a linear shift-invariant imaging system (LSI) that is flanked by the planes $\gamma$ and $\delta$.  Free-space gaps exist between the exit-surface $\beta$ of the sample and the entrance-surface $\gamma$ of the LSI.  There is another gap between the exit-surface $\delta$ of the LSI and the surface $\varepsilon$ of the detector. This position-sensitive two-dimensional detector measures the intensity distribution that is output by the cascaded system.  

\subsubsection{Example for strictly monochromatic case}

If one can assume both a strictly monochromatic paraxial field and a thin object in Fig.~\ref{fig:ExampleImagingSystem}, the projection approximation can be used to map the complex field $\psi_{\omega}(x,y)$ over the entrance-surface $\alpha$ of the object, to the field 
\begin{equation}
\mathcal{T}_{\omega}(x,y)\psi_{\omega}(x,y)\textrm{ :  plane }\beta 
\end{equation}
over the exit surface $\beta$: see Eqs.~(\ref{eq:PhaseShiftProjectionApproximation}) and (\ref{eq:zz8a}). To subsequently propagate through vacuum by the distance $Z_1$ between the exit surface $\beta$ of the object and the entrance surface $\gamma$ of the linear shift-invariant imaging system (LSI system), one can apply the free-space diffraction operator $\mathcal{D}_{Z_1}$ to the field over the plane $\beta$: see Eqs.~(\ref{eq:FresnelDiffraction}) and (\ref{eq:DiffractionOperatorFresnelDiffraction}).  This gives:
\begin{equation}
\quad\quad\mathcal{D}_{Z_1}\mathcal{T}_{\omega}(x,y)\psi_{\omega}(x,y)\textrm{ :  plane }\gamma 
\end{equation}
over the entrance surface $\gamma$ of the LSI system.  Assuming the LSI system to be in a state characterised by the control parameters $(\tau_1,\tau_2,\cdots)$, one can then use Eq.~(\ref{eq:ArbitraryShiftInvariantLinearCoherentImagingSystem5}) to propagate to the plane $\delta$, over which the complex disturbance will be:
\begin{equation}
\quad\quad\mathcal{D}(\tau_1,\tau_2,\cdots)\mathcal{D}_{Z_1}\mathcal{T}_{\omega}(x,y)\psi_{\omega}(x,y)\textrm{ :  plane }\delta . 
\end{equation}
Free-space propagation through a distance $Z_2$ then gives the complex disturbance over the surface of the detector:
\begin{equation}
\quad\quad\mathcal{D}_{Z_2}\mathcal{D}(\tau_1,\tau_2,\cdots)\mathcal{D}_{Z_1}\mathcal{T}_{\omega}(x,y)\psi_{\omega}(x,y)\textrm{ :  plane }\varepsilon .
\end{equation}
The squared modulus of the above expression gives the intensity measured by the detector.  Alternatively, one could calculate the phase $\phi_{\omega}(x,y,z=\varepsilon)$ of the above expression, to determine the phase of the coherent X-ray wave-fronts impinging upon the detector.  Other quantities can also be derived from the above expression for the complex field $\psi_{\omega}(x,y,z=\varepsilon)$ over the plane $\varepsilon$ coinciding with the surface of the detector, such as the transverse energy-flow vector (Poynting vector)
\begin{equation}
    \quad\quad{\bf S}_{\perp}(x,y) \propto |\psi_{\omega}(x,y,z=\varepsilon)|^2 \, \nabla_{\perp}\phi_{\omega}(x,y,z=\varepsilon)
\end{equation}
or other related quantities such as the angular-momentum density and the vorticity of the transverse energy-flow vector.  See e.g.~Berry (2009) for how these last-mentioned quantities may be calculated. 

\subsubsection{Example for partially-coherent case} 

The above case showed how to propagate the strictly monochromatic field $\psi_{\omega}(x,y)$ over the entrance surface $\alpha$ of the object in Fig.~\ref{fig:ExampleImagingSystem}, through both the object and an LSI system, to the surface $\varepsilon$ of a two-dimensional detector.  If the field is partially-coherent, then for each angular frequency $\omega$ corresponding to each energy $E$ via $E=\hbar\omega$, one can instead have an ensemble of strictly monochromatic fields of the form specified by Eqs.~(\ref{eq:EnsembleStrictlyMonochromaticFields1}) and (\ref{eq:EnsembleStrictlyMonochromaticFields2}).  Each field $\psi_{\omega}^{(j)}(x,y)$ is propagated through the optical system in exactly the same way as for the coherent case, with the associated statistical weights $c_j$ being unchanged via propagation through the optical system.  Note, however, that while the weights $c_j$ do not change, the normalisation of the individual wave-functions may change, as they may be transmitted through the sample or LSI system with different efficiency (Detlefs, 2019). This immediately yields an expression for converting the ensemble of strictly monochromatic input fields (over plane $\alpha$) in Eq.~(\ref{eq:EnsembleStrictlyMonochromaticFields1}), to the following ensemble of output fields over the plane $\varepsilon$: 
\begin{equation}
\quad\quad\{\mathcal{D}_{Z_2}\mathcal{D}(\tau_1,\tau_2,\cdots)\mathcal{D}_{Z_1}\mathcal{T}_{\omega}(x,y)\psi_{\omega}^{(j)}(x,y),c_j\}.
\label{eq:EnsembleofFieldsOverDetector}
\end{equation}
The resulting statistical ensemble can be used to calculate a number of derived quantities of interest:

\begin{itemize}
\item The cross-spectral density $W_\omega(x_1,y_1,x_2,y_2)$ (see Eq.~(\ref{eq:Coherence1}));
\item The spectral density $S_\omega(x,y)$ (see Eq.~(\ref{eq:GettingSfromW})), which may be integrated over $\omega$ to give the total spectral density (see Eq.~(\ref{eq:Coherence7}));
\item The position-dependent ensemble-averaged angular momentum density and vorticity, via ensemble averaging the expressions for these quantities that appear in a coherent-optics context in Berry (2009);
\item The Wigner function, the ambiguity function, the generalised radiance function etc.~are two-point correlation functions that can all be derived from the ensemble in Eq.~(\ref{eq:EnsembleofFieldsOverDetector}). See e.g.~Alonso (2011) for a description of how the Wigner, ambiguity, generalised-radiance and related functions may be calculated from the statistical ensemble of fields in Eq.~(\ref{eq:EnsembleofFieldsOverDetector}).

\end{itemize}

For a recent example of the use of the space--frequency formalism to model coherence transport through a modern source of partially coherent X-rays, namely the undulator in a modern synchrotron storage ring, see Paganin \& S\'anchez del R\'{\i}o (2019), together with references therein.

It is worth pointing out that the two-point  space--frequency formalism may be generalised to the case of four-point correlation functions, six-point correlation functions and so on (Mandel \& Wolf, 1995).  Four-point correlation functions are of importance in the X-ray version of the Hanbury Brown--Twiss effect (Kunimune et al., 1997).  Four-point correlations at the field level, in the guise of two-point correlations at the intensity level, also play an important role in the field of X-ray ghost imaging (Yu et al., 2016; Pelliccia, Rack, Scheel, Cantelli, \& Paganin, 2016; Schori \& Shwartz, 2017a; Zhang, He, Wu, Chen, \& Wang, 2018; Pelliccia et al., 2018; Schori, Borodin, Tamasaku, \& Shwartz, 2018; Ceddia \& Paganin, 2018; Kingston et al., 2018; Kingston, Myers, Pelliccia, Svalbe, \& Paganin, 2019; Kim et al., 2020). While interesting on account of the fact that it is an imaging modality in which no photon that passes through a sample is ever detected by a position-sensitive detector, X-ray ghost imaging will not be discussed any further here.  We also note that correlation functions of various orders are of pivotal importance for X-ray quantum optics (Adams et al., 2013; Kuznetsova \& Kocharovskaya, 2017; Li, Medvedev, Chapman, \& Shih, 2017; Schori et al., 2017b; Schori, Borodin, Tamasaku, \& Shwartz, 2018; Sofer, Strizhevsky, Schori, Tamasaku, \& Shwartz, 2019; Volkovich \& Shwartz, 2020).  In X-ray quantum optics, as its name implies, X-rays are treated on a manifestly quantum-mechanical level.  This may be done, for example, via a quantum electrodynamics formalism that describes the X-ray photon field via a Heisenberg field operator expressed in terms of photon creation and destruction operators (Mandel \& Wolf, 1995; Mandl \& Shaw, 2010). The blossoming of quantum X-ray optics in coming years is an obvious emerging research opportunity for coming generations, as one can readily ascertain by studying the most recent literature.

\subsubsection{Remark regarding the breakdown of certain simplifying assumptions}

Here, we revisit our previously-stated assumptions that the object being imaged is ``static, non-magnetic, elastically scattering and sufficiently gently spatially varying for the projection approximation to be valid''.  In making such simplifying assumptions, many very interesting physical scenarios are excluded from consideration. 

Assuming the sample to be static ignores potentially important effects such as a moving sample, or the time-dependent accumulation of radiation damage.  If the exposure time can be made sufficiently short for the sample to be considered essentially static during the exposure, then the effects of sample motion or accumulated damage may be neglected.  One example of this fact is the ``diffraction before destruction'' aspect of biomolecular diffractive imaging using intense femtosecond X-ray pulses (Neutze, Wouts, van der Spoel, Weckert, \& Hajdu, 2000).  As another example, of the fact that one can image dynamic objects provided that they do not evolve appreciably during the time in which an image is acquired, Olbinado et al.~(2017) report megahertz frame-rate X-ray phase contrast imaging of transient processes such as the propagation of cracks in glass and the propagation of shock fronts in water.  As a last example, Garc\'{i}a-Moreno et al.~(2019) report dynamic X-ray phase-contrast tomography at a rate of over 200 tomographic reconstructions per second, for a sample of foaming molten aluminium. 

The assumption that the sample is non-magnetic allows polarisation effects such as magnetic circular dichroism (for, say, the case of circularly-polarised X-rays) to be ignored, but there will be scenarios in which such effects are important.  

The assumption of elastic scattering enables the assumption that the energy of the X-rays is not changed upon passing through the sample.  However, this assumption breaks down in the very important context of X-ray fluorescence. Inelastic scatter is also relevant when imaging very thick samples, for example in the context of medical X-ray imaging or non-destructive testing.

If the projection approximation is not applicable, then, as we have already seen, more general formalisms such as the multi-slice formulation may be employed.  Other formalisms, for the thick-object case where the projection approximation is not appropriate, include the Born-series expansion, the scattering-matrix approach, Darwin's theory of dynamical diffraction and Kato's theory of dynamical diffraction (Authier, 2001).  The dynamical theory of X-ray diffraction is replete with additional examples of important scattering scenarios for which the projection approximation is certainly not applicable.  We nuance this remark with the statement that, strictly speaking, the projection approximation is a dynamical theory, if ``dynamical diffraction theory'' is taken to be synonymous with ``any theory for X-ray scattering in which the X-rays are scattered multiple times''.  This is because X-rays continuously acquire phase and amplitude shifts at every point on their path through a sample, under the projection approximation, and are therefore scattered infinitely many times upon traversing a sample of non-zero thickness.  Hence the projection approximation is a dynamical theory of X-ray diffraction, strictly speaking, although it would almost never be described as such in the literature on dynamical X-ray diffraction.  

\subsubsection{Remark regarding partial coherence and unresolved speckle}

We have nowhere mentioned speckles (Goodman, 2007), in our discussions relating to partial coherence.  This is remiss, given that the concept of unresolved speckle lies at the heart of most phenomena relating to partial coherence (Vartanyants \& Robinson, 2003;  Nugent, Tran, \& Roberts, 2003; Paganin, 2006; Nesterets, 2008; Paganin \& S\'anchez del R\'{\i}o, 2019; Paganin \& Morgan, 2019).  Typical scalar partially-coherent X-ray fields are littered with a profusion of speckles that are in most cases too rapidly varying in both space and time to be resolved using existing detectors.  The idea that one averages over such speckles, with a spatial average over the detector element and a temporal average over the detection time, gives a means for visualising and indeed modelling partially-coherent X-ray fields.  With current computing power, and increased computing power projected for the future, such direct spatio-temporal modelling of partially-coherent fields---in terms of underpinning fluctuating speckle fields---forms a very promising and conceptually illuminating avenue for future research.  Note that, at the scalar-field level for realistic partially-coherent sources, such spatio-temporal speckle fields are typically turbulent (Alperin, Grotelueschen, \& Siemens, 2019).  They are permeated by complicated fractal random networks of tangled nodal-line-threaded phase singularities associated with vortices (Paganin, 2006; O'Holleren, Dennis, Flossmann, \& Padgett, 2008).  For more on X-ray phase vortices and their associated dynamically evolving nodal lines (``threads of darkness''), see e.g.~Chapter 5 of Paganin (2006), together with references therein.  If one ascends to a vectorial description of the classical X-ray electromagnetic field for partially-coherent X-ray imaging scenarios, the speckles in the instantaneous energy density will remain, but the phase vortices will be threaded by a random network of vectorial singularities.  See e.g.~Nye (1999), and references therein, for a discussion of the singularities of vector fields such as the electric and magnetic fields.  Phase vortices may also be present in X-ray correlation functions such as the cross-spectral density, which can themselves be speckled: see Pelliccia \& Paganin (2012) and Paganin \& S\'anchez del R\'{\i}o (2019), together with references therein.

\subsection{Fokker--Planck equation for paraxial X-ray optics}

Recall the notion of propagation-based phase contrast, that was illustrated conceptually in Fig.~\ref{fig:RRR1}, and illustrated mathematically via the finite-difference form of the transport-of-intensity equation that was written down in Eq.~(\ref{eq:TIE_for_defocus}). Supposing a thin phase--amplitude sample to have been placed in the plane corresponding to $\delta z = 0$, we can view the first term on the right-side of Eq.~(\ref{eq:TIE_for_defocus}) as conveying information regarding the attenuation properties of the sample, with the second term conveying information regarding the refractive properties of the sample.  Previously, we described these two sample attributes as ``attenuation'' and ``refraction''.  Both attributes can yield contrast in an associated image of the sample, namely ``attenuation contrast'' (or ``absorption contrast'') and ``phase contrast'', respectively.  Both attributes underpinned many of our preceding discussions.

The attributes of attenuation and refraction may be augmented with a third sample attribute.  This third attribute is related to the presence of unresolved micro-structure in the sample.  This spatially-unresolved sample micro-structure, which may be disordered or partially ordered, leads to small-angle X-ray scattering (SAXS) that is a function of position.  This third attribute is an important contrast mechanism, which can be accessed using a variety of X-ray imaging modalities (Morrison \& Browne, 1992; Harding \& Schreiber, 1999; Pagot et al., 2003; Levine \& Long, 2004; Rigon, Arfelli, \& Menk, 2007; Pfeiffer et al., 2008; Bech et al., 2010; Yashiro, Terui, Kawabata, \& Momose, 2010; Yashiro et al., 2011; Lynch et al., 2011; Modregger et al., 2012; Strobl, 2014).  Note that, for the purposes of the present chapter, we make no distinction between SAXS and USAXS, with the latter acronym referring to ultra-small-angle X-ray scattering.  In many respects the position-dependent SAXS, that features at several points in our narrative, would be better termed ``position-dependent USAXS'', for reasons that the reader can ascertain by studying the above-cited references.

\begin{figure}
\includegraphics[scale=1.5]{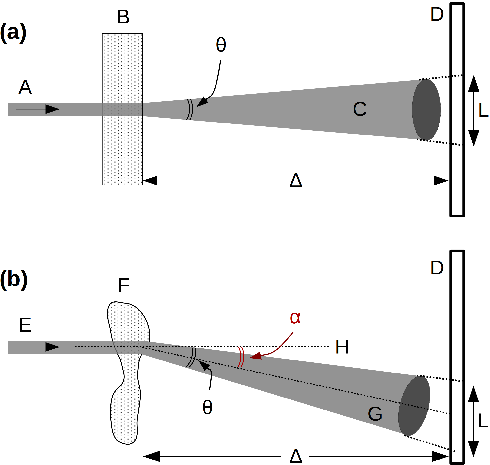}
\caption{Illustration of the mechanisms described by the Fokker--Planck formulation of paraxial X-ray imaging in the small-defocus regime.  (a) When a thin monochromatic X-ray beam $A$ illuminates a slab-like sample $B$ containing unresolved disordered (or partially ordered) micro-structure, the resulting transmitted beam $C$ is broadened into a fan of small-angle X-ray scattering (SAXS). (b) More generally, the SAXS fan will also be refracted and attenuated.  The Fokker--Planck equation models all three mechanisms (attenuation, refraction and SAXS).}
\label{fig:FokkerPlanck}
\end{figure}

The essence of SAXS (see e.g.~Kratky \& Glatter (1982)), for the purposes of the present discussion, is sketched in Fig.~\ref{fig:FokkerPlanck}(a).  Suppose a monochromatic X-ray plane wave is incident from the left in this figure, but restrict consideration to a thin beamlet (``pencil beam'') $A$ embedded within such a broad-field beam.  Suppose, further, that this pencil beam illuminates a statistically homogeneous slab $B$ that contains spatially-unresolved micro-structure, namely micro-structure that is smaller than the point spread function of a position-sensitive detector $D$.  Note that, by assumption, we assume the detector $D$ to be placed at a distance $\Delta$ downstream of the sample, which is sufficiently small for the corresponding image to be in the near field (Fresnel number or unity or more).  This near-field condition implies that sample structures that are smaller than the point spread function will not be resolved. As a result of passing through the slab $B$, the unresolved micro-structure will cause the pencil beam to be broadened into a fan (``SAXS fan'' $C$) with opening angle $\theta$.  There are many ways to understand this broadening, including but not limited to (i) multiple refraction/scattering of the traversing X-rays upon the random structures within the sample, and (ii) the creation of a ``gas'' of unresolved speckles that are smeared out upon spatial averaging (``coarse graining'') by the position-sensitive detector.  Irrespective of the viewpoint via which we conceptualise the formation of the SAXS fan $C$, its opening angle $\theta$ implies that when the fan traverses the distance $\Delta$ there will be an associated blur width $L$ over the entrance surface of the detector $D$.  Since the slab $B$ is statistically homogeneous, by assumption, the SAXS-fan opening angle $\theta$ will be independent of the transverse position $(x,y)$ of the X-ray pencil illuminating the slab. 

This last-mentioned statement invites an evident generalisation: if the slab $B$ in Fig.~\ref{fig:FokkerPlanck}(a) is not statistically homogeneous, but rather contains unresolved spatially-random micro-structure whose statistical properties change with location within the sample, then $\theta$ will no longer be independent of transverse position $(x,y)$.  One would then have a position-dependent SAXS fan with position-dependent opening angle $\theta(x,y)$, that will have an associated position-dependent blur width
\begin{equation}
    L(x,y)= \theta(x,y) \Delta
\end{equation}
over the surface of the detector $D$.  This generalisation invites further generalisation, since if the sample is now allowed to have spatially-unresolved random micro-structure whose statistical properties vary with location within the sample, then it is natural to also allow the sample to exhibit spatially-resolved variations in macro-structure that are modelled by an attenuation and refraction (phase shift) that both vary with transverse position $(x,y)$.  

So, under the model considered above, a thin sample may be considered to have three attributes, insofar as its ``imprint'' upon a traversing monochromatic X-ray beam is concerned: position-dependent attenuation, position-dependent phase shift and position-dependent SAXS-fan opening angle. This corresponds to conceptually splitting the X-ray energy flow, downstream of the sample, into two ``channels''.  The first of these channels corresponds to what might be termed ``coherent energy flow'', with its associated attenuation and phase contrast that may be modelled, in the small-$\Delta$ regime, by the finite difference form of the transport-of-intensity equation that was given in Eq.~(\ref{eq:TIE_for_defocus}).  The second channel corresponds to diffusive energy flow associated with SAXS (see e.g.~Paganin \& Morgan (2019), together with references therein).  See Fig.~\ref{fig:FokkerPlanck}(b), which shows the incident monochromatic X-ray pencil beam $E$ being attenuated (absorption contrast) by the sample $F$, as well as being refracted via a deflection angle $\alpha$ with respect to the local optical axis $H$ (propagation-based phase contrast) and diffused into a local-SAXS fan $G$.  While only the diffusive channel has been shown in the diagram, superposed over this there will be the coherent channel which propagates as if there were no micro-structure present.  How can all of these effects be modelled in a paraxial-imaging context?  

To incorporate the diffusive energy-flow channel into the transport-of-intensity equation, we need an additional term to augment Eq.~(\ref{eq:TIE_for_defocus}).  Denoting this additional term by $\mathscr{B}(x,y,\Delta)$, we write the modified transport-of-intensity equation as:
\begin{eqnarray}\label{eq:FokkerPlanck1}
I(x,y,z+\Delta) \approx I(x,y,z) -\frac{\Delta
}{k}\nabla_{\perp}\cdot\left[I(x,y,z)\nabla_{\perp}\phi(x,y,z)\right]+\mathscr{B}(x,y,\Delta). 
\end{eqnarray}
Many different levels of approximation, for $\mathscr{B}(x,y,\Delta)$, may be considered.  Four such approximations, ordered in terms of increasing degree of generality, are outlined below. 

 For a first level of approximation to the $\mathscr{B}(x,y,\Delta)$ term in Eq.~(\ref{eq:FokkerPlanck1}), we assume the position-dependent SAXS fans to be rotationally symmetric.  These rotationally-symmetric SAXS fans, emanating from the nominally-planar exit surface of the sample, are characterised by their transverse-position-dependent opening angles $\theta(x,y)$. We can then write: 
\begin{eqnarray}\label{eq:FokkerPlanck2}
\mathscr{B}(x,y,\Delta)=F(x,y)\Delta^2[\theta(x,y)]^2\nabla_{\perp}^2I(x,y,z), 
\end{eqnarray}
thereby transforming Eq.~(\ref{eq:FokkerPlanck1}) into the following Fokker--Planck equation (Risken, 1989) for paraxial X-ray imaging (Morgan \& Paganin, 2019; Paganin \& Morgan, 2019):
\begin{eqnarray}
\nonumber I(x,y,z+\Delta) \approx I(x,y,z) -\frac{\Delta
}{k}\nabla_{\perp}\cdot\left[I(x,y,z)\nabla_{\perp}\phi(x,y,z)\right] \\ +F(x,y)\Delta^2[\theta(x,y)]^2\nabla_{\perp}^2I(x,y,z). 
\label{eq:FokkerPlanck1-extended}
\end{eqnarray}
The above expression simultaneously incorporates both coherent and diffuse energy flow, downstream of the sample.  Here, $F(x,y)$ is a dimensionless function which quantifies the fraction of the incident X-rays that are converted to the diffuse local-SAXS channel upon traversing the sample.  Note that $F(x,y)$ can be related to the variances of the unresolved spatially-rapid phase fluctuations induced by the sample's micro-structure (Paganin \& Morgan, 2019), but this point will not be further pursued here.  Note also that we are implicitly assuming $F(x,y) \ll 1$ in the present analysis, although the formalism is easily extended to the case of stronger diffuse scattering (Paganin \& Morgan, 2019).  

Returning back to Eq.~(\ref{eq:FokkerPlanck1-extended}), there are many ways---separate from the first-principles derivation in Paganin \& Morgan (2019), which will not be detailed here---by which one can justify the form for the SAXS-induced blurring that is given on the far right-hand side of this expression.  We briefly outline three such conceptual pathways here, all of which begin by making the following trivial re-arrangement of Eq.~(\ref{eq:FokkerPlanck1-extended}):
\begin{eqnarray}\label{eq:FokkerPlanck3}
I(x,y,z+\Delta) \approx [1+\Delta^2 D(x,y)\nabla_{\perp}^2]I(x,y,z)  -\frac{\Delta
}{k}\nabla_{\perp}\cdot\left[I(x,y,z)\nabla_{\perp}\phi(x,y,z)\right].
\end{eqnarray}
Here, we have introduced the dimensionless diffusion coefficient
\begin{eqnarray}\label{eq:FokkerPlanck4}
D(x,y)=F(x,y)[\theta(x,y)]^2.
\end{eqnarray}
We are now ready to tally our three different justifications, for the manner in which position-dependent SAXS fans have been modelled in the Fokker--Planck equation for paraxial X-ray imaging:
\begin{itemize}
    \item The first term on the right-hand side of Eq.~(\ref{eq:FokkerPlanck3}) may be viewed as a forward-finite-difference form of a diffusion-equation evolution, which blurs (diffuses) each point in $I(x,y,z)$ over a transverse length scale which dimensional reasoning gives as $\Delta\sqrt{D(x,y)}$. This local blurring, or local diffusion, is due to the local transverse-position-dependent SAXS fans described earlier in this section. 
    \item If $D(x,y)$ were to be independent of position, then the Fourier derivative theorem would imply that the Fourier transform with respect to $x$ and $y$, of the differential operator $1+\Delta^2 D \nabla_{\perp}^2$ on the right-hand side of Eq.~(\ref{eq:FokkerPlanck3}), would be given by the multiplicative low-pass filter (inverted parabola in Fourier space) $1-\Delta^2 D (k_x^2+k_y^2)$.  Simple dimensional analysis shows this low-pass filter to correspond to real-space blurring over a transverse length scale of $\Delta\sqrt{D}$.  If, rather than considering the dimensionless diffusion coefficient to be strictly independent of transverse position,  we instead consider it to be a slowly-varying function $D(x,y)$ of transverse position, then we are again led to the form given in the first term on the right-hand side of Eq.~(\ref{eq:FokkerPlanck3}).  
    \item Those familiar with Laplacian-based unsharp-mask image sharpening will recognise $1-\Delta^2 D(x,y)\nabla_{\perp}^2$ as an operator that locally sharpens an image (Subbarao, Wei, \& Surya, 1995).  Using the binomial approximation to estimate the inverse operator as
\begin{equation}
[1-\Delta^2 D(x,y)\nabla_{\perp}^2]^{-1} \approx 1+\Delta^2 D(x,y)\nabla_{\perp}^2,
\end{equation}
we conclude that $1+\Delta^2 D(x,y)\nabla_{\perp}^2$ is an operator that effects local blurring, in the first term on the right-hand side of Eq.~(\ref{eq:FokkerPlanck3}).  
\end{itemize}

So far, we have considered the local SAXS fans to be rotationally symmetric, leading to the isotropic-diffusion Fokker--Planck equation given in Eq.~(\ref{eq:FokkerPlanck1-extended}).  This is very easily extended to the more general ``directional dark field'' case (Jensen et al., 2010a,b; Yashiro et al., 2011) where the position-dependent SAXS fans are not rotationally symmetric, but rather have an elliptical cross section. To achieve this, the scalar dimensionless diffusion coefficient $D(x,y)$ can be replaced by a $2\times 2$ matrix, namely a  rank-two diffusion tensor, as follows:
\begin{equation}
    D(x,y)\longrightarrow \begin{bmatrix}
D_{xx}(x,y) & D_{xy}(x,y) \\
D_{xy}(x,y) & D_{yy}(x,y) \\
\end{bmatrix}.
\label{eq:RankTwoDiffusionTensor}
\end{equation}
Equipped with this diffusion tensor, Eq.~(\ref{eq:FokkerPlanck1-extended}) generalises to the following anisotropic Fokker--Planck equation for paraxial X-ray imaging in the presence of both coherent and diffusive energy-flow channels (Morgan \& Paganin, 2019; Paganin \& Morgan, 2019):
\begin{equation}
\begin{split}
I(x,y,z + \Delta) \approx I(x,y,z) - \frac{\Delta
}{k}\nabla_{\perp}\cdot\left[I(x,y,z)\nabla_{\perp}\phi(x,y,z)\right]  \quad\quad\quad\quad\quad\quad\quad\quad\quad\quad\quad\quad \\ + \Delta^2 \left[ D_{xx}(x,y)\frac{\partial^2}{\partial x^2} I(x,y,z)
 + D_{yy}(x,y)\frac{\partial^2}{\partial y^2} I(x,y,z)   + 2 D_{xy}(x,y)\frac{\partial^2}{\partial x \partial y} I(x,y,z)\right].
\end{split}
\label{eq:FokkerPlanck5}
\end{equation}

The two levels of approximation given above---namely (i) the locally-isotropic Fokker--Planck equation (Eq.~(\ref{eq:FokkerPlanck1-extended})) for rotationally-symmetric position-dependent SAXS fans and (ii) the locally-anisotropic Fokker--Planck equation (Eq.~(\ref{eq:FokkerPlanck5})) for non-rotationally-symmetric locally-elliptical position-dependent SAXS fans---may be easily augmented by two additional formulations that are still more general in scope.  (iii) Along these lines, the  rank-two diffusion tensor in Eq.~(\ref{eq:RankTwoDiffusionTensor}) may be considered as merely the first member of an infinite hierarchy of progressively higher-rank diffusion tensors.  Each member of this tensor hierarchy is related to a particular moment of the local SAXS fans (Modregger et al., 2017; Modregger, Endrizzi, \& Olivo, 2018).  This leads to a Kramers--Moyal equation (Risken, 1999), which generalises the Fokker--Planck equation of paraxial X-ray optics.  The progressively higher orders of transverse spatial derivative, in this infinite-order Kramers--Moyal partial differential equation for paraxial X-ray optics (Morgan \& Paganin, 2019; Paganin \& Morgan, 2019), enable the structure of the position-dependent SAXS fans to be taken into account. (iv) Lastly, in the most general formulation of those that will be mentioned here, one can simply model $\mathscr{B}(x,y,\Delta)$ by the SAXS fan emerging from each point on the exit surface of the sample.  In this case, we can represent $\mathscr{B}(x,y,\Delta)$ via the linear integral transform:
\begin{eqnarray}\label{eq:FokkerPlanck6}
\mathscr{B}(x,y,\Delta) = F(x,y)\iint K(x,y,x',y',\Delta) I(x',y',z)\, dx'\,dy'.
\end{eqnarray}
Here, $K(x,y,x',y',\Delta)$ represents the SAXS fan illuminating the detector plane with coordinates $(x,y,z+\Delta)$, emanating from each point $(x',y',z)$ over the nominally-planar exit surface of the sample.  Alternatively, we may use collision-theory language (Bransden \& Joachain, 1989) and instead speak of $K(x,y,x',y',\Delta)$ as being proportional to the  local differential cross section for position-dependent diffuse scattering by the sample.

We close this section by noting that, while the above outline of the Fokker--Planck formalism for paraxial X-ray optics has been couched in terms of the propagation-based modality for X-ray phase-contrast imaging, the formalism can also be applied to other modalities such as grating-based methods, speckle-tracking methods and edge-illumination methods (Morgan \& Paganin, 2019; Paganin \& Morgan, 2019; Pavlov et al., 2020).  We also note that, by incorporating an additive constant into the Fokker--Planck local-SAXS diffusion coefficient, the effects of source-size blur and detector-induced blur can also be incorporated, for all of the above-mentioned modalities (Beltran, Paganin, \& Pelliccia, 2018). 

\section{The inverse problem: retrieving sample information from X-ray phase-contrast images}

In the preceding section, we considered the forward problem of X-ray phase contrast imaging.  As we saw, forward problems may be viewed in general terms as seeking to deduce effects that result from specified causes.  Thus, for example, we saw that a typical forward problem in X-ray imaging might be to take a known X-ray wave-field that illuminates a known sample in the context of a known fully-coherent or partially-coherent imaging system, and thereby deduce the image of that sample that would be formed over the surface of a specified position-sensitive detector.  

Modelling a particular forward problem allows us to subsequently pose the corresponding ``inverse problem''. Loosely speaking, inverse problems seek to determine causes from effects (Bertero and Boccacci, 1998; Sabatier, 2000).  Examples include: 

\begin{enumerate}

\item Schr\"{o}dinger's inferring of his famous equation, based on data available at the time, such as the measured energy levels of the hydrogen atom; 

\item Determining the position at which a guitar string of known length is plucked, given a measurement of the spectrum of different sound pitches created by the plucked string; 

\item Determining the molecular structure of an unknown protein from X-ray, electron, or neutron diffraction data;

\item Determining both the magnitude and the phase of the two-dimensional projected complex refractive index arising from coherent X-ray illumination of a sample, under the projection approximation, for known experimental parameters such as X-ray wavelength, source-to-detector distance etc., given one or more measured propagation-based phase-contrast intensity images;

\item Using the two-dimensional reconstructions from the preceding item in this list, for a number of different angular orientations of a sample, in order to determine the three-dimensional complex refractive index of that sample via computed tomography.

\end{enumerate}

If the underlying fundamental physics equations are known, and enough reasonable initial data are specified, the forward problems of classical physics are typically soluble. This broad statement is based on the fact that, in performing an experiment to model a given classical-physics scenario, nature always chooses a ``solution''---namely the actual physical state for a classical system at a given specified time in its future---given a specified starting state of the system.  This solution will not necessarily be in one-to-one correspondence with the starting point.  For instance, in dissipative systems with a point-like attractor---like a pendulum in the presence of friction, which will progressively come to a halt in the vertical position, regardless of its starting point---a family of different state-space trajectories may converge upon a single point in state space (Ruelle, 1989).  The solution may also exhibit sensitive dependence upon initial conditions, e.g.~in non-dissipative chaotic systems with strange attractors: see Ruelle (1989), together with popular accounts of the ``butterfly effect''.  Such subtleties however do not change the fact that, classically speaking, ``nature always chooses a solution'', provided that the initial conditions have been completely specified.  Moreover, if systems of equations, which model a given scenario in the physical world, evolve into states that are singular (e.g.~the infinite energy densities associated with ray caustics in geometric optics), then this is a signature that a more general theory is needed.  For examples of this idea of the breakdown of optical-physics theories giving a window to more general theories, see Berry \& Upstill (1980), Berry (1998), and Paganin (2006).  Again, the existence of singularities in a physical model does not contradict the earlier statement regarding nature always finding a solution.  Also, there may be the more subtle problem that, for a given system of equations, it may not be rigorously known whether solutions---to the equations as posed---even exist for certain specified classes of initial condition.  For example, such questions remain outstanding for the Navier--Stokes equations of classical fluid mechanics (Kreiss \& Lorenz, 1989).  Again, such interesting subtleties do not contradict our earlier statement that the forward problems of classical physics are typically soluble.        

Inverse problems are harder to solve, in general, than their associated forward problems.  Solutions to specified inverse problems do not necessarily exist.  Even if they do exist, they may not be unique.  Even if a unique solution exists, it may not be stable with respect to perturbations in the data due to realistic amounts of experimental noise, and other imperfections that will be present in any real experiment.  If an inverse problem is indeed such that there exists a unique solution that is stable with respect to perturbations in the input data, it is said to be ``well posed in the sense of Hadamard'' (Hadamard, 1923; Kress, 1989; Sabatier, 2000).  While the property of well-posedness is often desirable from both an analytic and aesthetic perspective, the class of inverse problems, of scientific and practical relevance, is rather broader than the class of inverse problems that are well posed in the sense of Hadamard.  In this latter context, various forms of iterative optimisation method are very powerful, although a treatment of such methods lies beyond the scope of our discussions.

By way of an outline for what follows, the present section is devoted to the inverse problem of retrieving sample information given one or more X-ray phase-contrast images of the sample.  We first consider a simpler problem, namely a field-level inverse problem in which one seeks to reconstruct a complex disturbance that is input into a linear imaging system, given one or more output complex disturbances that correspond to one or more states of the imaging system.  We then consider the much more realistic (and harder!) inverse problem of phase retrieval, in which one seeks to reconstruct the phase (or, more generally, both the phase and the intensity) of an input field, given one or more output intensity images as data.  Such a phase-retrieval problem is considered from several different perspectives in the present section.  These perspectives include an approach that is based on contrast transfer functions, an approach based on the transport-of-intensity equation, and an approach that is based on a very simple model in which the complex transfer function is considered to be a linear function of spatial frequency.  In the course of these discussions, several other important topics will emerge, including: (i) viewing optical imaging systems in terms of the flow of optical information; (ii) viewing the inverse problem of phase-contrast imaging as a form of virtual optics, in which a computational system forms an intrinsic component of an imaging system; (iii) the important role of prior knowledge in addressing the inverse problems of phase retrieval, (iv) a simple explicit calculation showing how the presence of unresolved sample micro-structure may be associated with position-dependent fans of diffuse X-ray scatter, which complements our earlier discussions on the same topic.

\subsection{Two inverse problems}

We open this sub-section by revising what we have learned so far regarding the forward problem of imaging using generalised shift-invariant linear (phase contrast) imaging systems.  We separately consider the forward and inverse problems for such imaging systems at the levels of (i) complex scalar fields, (ii) intensities (i.e., the squared moduli of the complex scalar fields).  Note that the former inverse problem is somewhat idealised, since complex X-ray wave-fields are not measured directly, because their period of oscillation in time is too rapid. Rather, it is time-averaged intensities that are directly measured by X-ray detectors, with the time average being taken over the acquisition time of the detector.

\subsubsection{A field-level inverse problem}

At the field level for an arbitrary linear shift-invariant imaging system, we learned that the input field may be related to the output via Eq.~(\ref{eq:ArbitraryShiftInvariantLinearCoherentImagingSystem3}), with the input-to-output operator $\mathcal{D}(\tau_1,\tau_2,\cdots)$ given by the Fourier-space filtration in  Eq.~(\ref{eq:ArbitraryShiftInvariantLinearCoherentImagingSystem5}).  The associated inverse problem, namely the determination of the input field given the output field, may be somewhat na\"{i}vely solved by:
\begin{eqnarray}\label{eq:TransferFunctionInverseProblem}
\psi_{\rm
IN}(x,y)= \mathcal{F}^{-1} \frac{1}{T(k_x,k_y,\tau_1,\tau_2,\cdots)}\mathcal{F} \psi_{\rm OUT}(x,y,\tau_1,\tau_2,\cdots). 
\end{eqnarray}

Unfortunately, very often there will be division-by-zero issues associated with spatial frequencies $(k_x,k_y)$ at which the transfer function $T(k_x,k_y,\tau_1,\tau_2,\cdots)$ vanishes.  This amounts to information loss in the forward problem, which leads to instability in the associated inverse problem.  Sometimes one can ``regularise'' the above expression by making the replacement
\begin{equation}
\frac{1}{T}\longrightarrow \frac{T^*}{|T|^2+\aleph},
\label{eq:Regulariser}
\end{equation}
where $\aleph$ is a small positive real number.  A more sophisticated solution is to consider several outputs associated with $N>1$ different states of the imaging system, leading to the following solution to the field-level inverse problem (Schiske, 2002):
\begin{eqnarray}\label{eq:TransferFunctionInverseProblem2}
\psi_{\rm
IN}(x,y)= \mathcal{F}^{-1}\sum_{j=1}^{j=N} \frac{T_j^*(k_x,k_y)}{\sum_{p=1}^{p=N} |T_p(k_x,k_y)|^2}\mathcal{F} \psi^{(j)}_{\rm OUT}(x,y). 
\end{eqnarray}
\noindent Here, $T_j(k_x,k_y)$ denotes the transfer function associated with the $jth$ state of the imaging system, and $\psi^{(j)}_{\rm OUT}(x,y)$ denotes the corresponding output.  The above expression will have no division-by-zero issues if $\sum_{p=1}^{p=N} |T_p(k_x,k_y)|^2$ is non-zero at every spatial frequency $(k_x,k_y)$.  The basic idea, which underpins Eq.~(\ref{eq:TransferFunctionInverseProblem2}), is to weight the different estimates that each imaging-system state gives for the reconstructed field (in Fourier space) by multiplicative factors that (i) sum to unity, and (ii) have a magnitude that is proportional to $|T_j|^2$.  If division-by-zero issues remain, one can always regularise the above expression, or increase the number of different states of the imaging system that are utilised. One can also leverage suitable prior knowledge regarding the sample.  

\subsubsection{A phase-retrieval inverse problem}

The inverse problem of phase retrieval, or more properly of phase--amplitude retrieval, seeks to reconstruct both the intensity and phase of an input field, given as data only the intensity of the output field corresponding to one or more states $(\tau_1,\tau_2,\cdots)$ of the imaging system.  This problem is more difficult, and often vastly more difficult, than the previously-considered field-level inverse problem.  Indeed, no closed-form solution exists in general, to the phase--amplitude retrieval problem.    

Note the evident parallels with the concept of inline holography as conceived by Gabor (1948), in which imaging is viewed as a two-step process, namely data recording, followed by reconstruction.  The ``holographic'' spirit of this latter point implicitly runs through many of our subsequent discussions regarding phase retrieval.  

Before proceeding, however, we make the following general remark, which again has parallels with inline holography. Since imperfect shift-invariant aberrated imaging systems typically yield measurable output images $I_{\textrm{OUT}}(x,y)$ that are influenced by the phase of the input complex field $\psi_{\textrm{IN}}(x,y)$, the output intensity may be viewed as containing encrypted or encoded information regarding the phase of the input field.  Under this view, the phase-retrieval problem corresponds to seeking a means to decrypt or decode one or more measured output-intensity maps, so as to infer the phase distribution (or, more generally, both the phase and the amplitude/intensity) of the input field. 

We devote the remainder of this sub-section to a very simple illustrative example regarding the core concepts of phase retrieval, treated in a fairly general albeit idealised setting.  To this end, return consideration yet again to the idea of a linear shift-invariant optical imaging system. As we have already seen, the action of such a system may be modelled via the Fourier-space filtration that was given in Eqs.~(\ref{eq:ArbitraryShiftInvariantLinearCoherentImagingSystem3}) and (\ref{eq:ArbitraryShiftInvariantLinearCoherentImagingSystem5}).  This Fourier-space filtration is equivalent to the convolution form given in Eq.~(\ref{eq:ArbitraryShiftInvariantLinearCoherentImagingSystem2}), and for the moment it is more convenient for us to work with this latter formulation.  

For simplicity, let us further suppose that the wave-field input into our system is uniform in intensity, corresponding for example to uniform illumination of a thin perfectly-transparent object (``pure phase object'') by $z$-directed monochromatic scalar plane waves. The action of our linear shift-invariant system, upon the input wave-field $\exp[i\phi(x,y)]$, will yield an output wave-field $\psi_{\textrm{OUT}}(x,y,\tau)$ that is given by Eq.~(\ref{eq:ArbitraryShiftInvariantLinearCoherentImagingSystem2}) as:
\begin{eqnarray}
  \psi_{\textrm{OUT}}(x,y,\tau)=\exp[i\phi(x,y)] \otimes G(x,y,\tau).
\label{eq:LSI-for-pure-phase-object}
\end{eqnarray}
We remind the reader that $G(x,y,\tau)$ denotes the Green function (also known as the complex point spread function, the complex impulse response, the generalised wave-field propagator, and the generalised Huygens wavelet) associated with the shift-invariant linear imaging system in a state that is characterised by the real control parameter $\tau$.  Also, we note that we here need only one control parameter $\tau$, rather than the previously-considered vector of control parameters $\{\tau_1,\tau_2,\cdots\}$.

Assuming, further, that the phase object is weak in the sense that 
\begin{equation}
    |\phi(x,y)| \ll 1,
\end{equation}
we may write the following approximation that is of first order in the phase shift induced by the object:
\begin{equation}
    \exp[i\phi(x,y)] \approx 1 + i \phi(x,y).
\end{equation}
With the above approximation, Eq.~(\ref{eq:LSI-for-pure-phase-object}) becomes:
\begin{eqnarray}
  \psi_{\textrm{OUT}}(x,y,\tau) \approx \Omega + i [\phi(x,y) \otimes G(x,y,\tau)].
\label{eq:LSI-for-weak-pure-phase-object}
\end{eqnarray}
In writing the above expression, we have introduced the notation:  
\begin{equation}
    1 \otimes G(x,y,\tau)=\Omega.
\end{equation}

Physically, $\Omega$ is a complex number that quantifies the phase--amplitude shift imparted by the system upon a normally-incident plane wave that traverses such a system.  With only a small loss of generality, we henceforth assume $\Omega$ to be real.  The magnitude of $\Omega$ will be taken to lie between zero and unity, since the system will typically filter but not amplify a wave field that passes through it.  If $|\Omega|=0$ then one has a ``dark field'' imaging system that completely blocks the unscattered incident light, so that the image of an object will appear as being immersed within a dark background, if that object is smaller than the field of view of the imaging system (Gage, 1920).  In such a dark-field imaging modality, all of the detected photons in the registered image correspond to photons that have been scattered by the sample, since all unscattered photons are blocked by the imaging system (Cowley, 1995). Conversely, if $|\Omega|\ne 0$ then one has a ``bright field'' imaging system, since the regions that do not contain the sample will not be completely dark.  For bright field imaging systems, both scattered and unscattered photons contribute to the registered image.  As an aside, we point out that the above usage of the term ``dark field'' is more general, and more historically consistent (Gage, 1920), than the use of the term ``dark-field'' given earlier in the present chapter, in the more restricted context of local fans of small-angle X-ray scattering.  

Returning to the main thread of the argument, we compute the intensity $I_{\textrm{OUT}}(x,y,\tau)$ that is output by the imaging system, by taking the squared modulus of Eq.~(\ref{eq:LSI-for-weak-pure-phase-object}) and then discarding the term that is quadratic in the sample phase.  This gives, for the bright-field case where $\Omega$ is non-zero,  
\begin{equation}
I_{\textrm{OUT}}(x,y,\tau) \approx |\Omega|^2-2\Omega [\phi(x,y)\otimes G_{\textrm{imag}}(x,y,\tau)].
\label{eq:LSI-phase-contrast-for-WPOA}
\end{equation}
Here, we have introduced the notation $G_{\textrm{imag}}(x,y,\tau)$ to denote the imaginary part of the Green function $G(x,y,\tau)$. Equation (\ref{eq:LSI-phase-contrast-for-WPOA}) demonstrates that our shift-invariant linear imaging system exhibits phase contrast, since the output intensity is a function of the input phase.  Moreover, the contrast of the output intensity is a linear function of the input phase (Zernike, 1942; Cowley, 1995). 

Recall the following two-dimensional form of the convolution theorem of Fourier analysis:
\begin{equation}
a(x,y)\otimes b(x,y) = 2 \pi \mathcal{F}^{-1}\{\mathcal{F}[a(x,y)]\mathcal{F}[b(x,y)]\},
\end{equation}
where $a(x,y)$ and $b(x,y)$ are otherwise-arbitrary functions that are sufficiently well behaved for their Fourier transforms to exist (see e.g.~p.~394 of Paganin (2006)).  Using this form of the convolution theorem, Eq.~(\ref{eq:LSI-phase-contrast-for-WPOA}) may be written as the Fourier-space filtration:
\begin{equation}
I_{\textrm{OUT}}(x,y,\tau) \approx |\Omega|^2 + \mathcal{F}^{-1} C (k_x,k_y,\tau) \mathcal{F} \phi(x,y).
\label{eq:phase-contrast-transfer-function-1}
\end{equation}
Here, we have defined the ``phase-contrast transfer function'' $C(k_x,k_y,\tau)$, associated with the linear shift-invariant imaging system in the state $\tau$, to be:
\begin{equation}
C(k_x,k_y,\tau) = - 4 \pi \Omega \mathcal{F} G_{\textrm{imag}}(x,y,\tau).
\label{eq:phase-contrast-transfer-function-2}
\end{equation}
For more information on the concept of a phase-contrast transfer function, see e.g.~Cowley (1995) and Pogany, Gao, \& Wilkins (1997), together with the historically important study of Zernike (1942). Note, also, that the effects of partial coherence may be taken into account by simply multiplying the phase-contrast transfer function by a ``coherence envelope'' that is unity at the Fourier-space origin, and decays at some characteristic cutoff spatial frequency, corresponding to the loss of spatial detail that is finer than the resolution limit of a partially-coherent phase-contrast imaging system (Pogany, Gao, \& Wilkins, 1997; Spence, 2003; Nesterets \& Gureyev, 2016).

The use of Eq.~(\ref{eq:phase-contrast-transfer-function-1}), to model the forward problem of forming a phase-contrast image of a weak pure phase object using an arbitrary shift-invariant linear imaging system, establishes the groundwork needed to tackle the subsequent inverse problem of phase retrieval. 

As a first approach to this inverse problem, suppose that one phase-contrast image $I_{\textrm{OUT}}(x,y,\tau)$ has been measured, corresponding to one particular state $\tau$ of the shift-invariant linear imaging system.  In an attempt to decode this image to extract the phase map input into the system, we can proceed in an analogous manner to  Eq.~(\ref{eq:TransferFunctionInverseProblem}), to write down the following na\"{i}ve solution to Eq.~(\ref{eq:phase-contrast-transfer-function-1}):
\begin{equation}
 \phi(x,y) = \mathcal{F}^{-1}\left\{ \frac{\mathcal{F}[I_{\textrm{OUT}}(x,y,\tau)-|\Omega|^2]}{C(k_x,k_y,\tau)}\right \}.   
\end{equation}
Once again, however, such an approach is typically (but not always) unsatisfactory on account of the division-by-zero blowup that will occur at all points $(k_x,k_y)$ in Fourier space, for which $C(k_x,k_y,\tau)$ vanishes.  This instability corresponds to a genuine loss of information, since, in the adopted model, such spatial frequencies are completely blocked by the imaging system.  The regularised version 
\begin{eqnarray}
 \phi(x,y) = \mathcal{F}^{-1}\left\{ \frac{C^*(k_x,k_y,\tau) \mathcal{F}[I_{\textrm{OUT}}(x,y,\tau)-|\Omega|^2]}{|C(k_x,k_y,\tau)|^2+\aleph}\right \}, \quad \aleph > 0, 
\end{eqnarray}
avoids such blowups (cf.~Eq.~(\ref{eq:Regulariser})), but such  palliating of division-by-zero divergences does not alter the fact that there has been genuine information loss in forming the phase-contrast image.

One means of overcoming such genuine information loss is to take several phase-contrast images, corresponding to different states of the imaging system.  Suppose, then, that we have a set of $N$ output images $\{I_{\textrm{OUT}}(x,y,\tau_j)\}$,  with $j=1,\cdots,N$ corresponding to $N$ different states of a linear shift-invariant imaging system.  Following the same lines that led from Eq.~(\ref{eq:TransferFunctionInverseProblem}) to Eq.~(\ref{eq:TransferFunctionInverseProblem2}), we may then write down the following estimate for the retrieved phase as (Saxton, 1994; Cloetens et al., 1999; Schiske, 2002; Paganin, Barty, McMahon, \& Nugent, 2004c; Gureyev, Pogany, Paganin, \& Wilkins, 2004a):
\begin{eqnarray}
\phi(x,y)= \mathcal{F}^{-1}\sum_{j=1}^{j=N} \frac{C^*(k_x,k_y,\tau_j)}{\sum_{p=1}^{p=N} |C(k_x,k_y,\tau_p)|^2}\mathcal{F} [I_{\textrm{OUT}}(x,y,\tau_j)-|\Omega_j|^2]. 
\end{eqnarray}
One very natural way, to generate a series of different states of a coherent imaging system in the context of phase retrieval, is to take intensity measurements over a sequence of parallel planes downstream of the exit-surface of a sample.  Such image sequences are known as through-focal series (Saxton, 1994; Cloetens et al., 1999), and are important in high-resolution X-ray phase retrieval (Cloetens et al,. 1999; Gureyev, Pogany, Paganin, \& Wilkins, 2004a; Gureyev, Davis, Pogany, Mayo, \& Wilkins, 2004b; Yu et al., 2017; Yu et al., 2018; Kuan et al., 2020).  Another means, to generate a series of different states of a coherent imaging system in the context of phase retrieval, is to tune the so-called ``coherent aberrations'' (e.g.~defocus, coma, astigmatism, spherical aberration) that may may be used to model such a system (Allen, Oxley, \& Paganin, 2001).  As previously mentioned, such ``coherent aberrations'' are associated with non-zero real values for the coefficients $\{\alpha_{m,n}\}$ that were introduced in Eq.~(\ref{eq:ArbitraryShiftInvariantLinearCoherentImagingSystem7}).

While the above simple indicative example has not made any use of prior knowledge that one might have about the sample, such prior knowledge can be incorporated into phase-retrieval strategies, particular when iterative refinement is involved. We shall have a little more to say, regarding the importance of prior knowledge in the context of phase retrieval, in the next sub-section.

We close the present sub-section by expanding a little, upon our earlier remark that we may consider the phase-retrieval process as decoding information that has been encoded into an X-ray beam's intensity, in the process of forming one or more phase-contrast images of that object.  Under this ``information optics'' view (Shannon, 1948a, 1948b; Gabor, 1961; Yaroslavsky \& Eden, 1996; Yu, 2017), we may think of a phase-contrast X-ray imaging beam as a pipeline that streams information.  Thus, as the X-ray beam traverses the sample, information regarding that sample is encoded into the beam.  Subsequent evolution of this information-laden X-ray beam, through the analogue information processing channel that may be associated with a phase-contrast imaging system, yields phase-contrast intensity maps that are measured via a position-sensitive detector.  At this point, the flowing analogue information, encoded in the photons of the propagating X-ray beam, is interfaced to the digital information recorded by the computer that stores the recorded intensity information.  Subsequent information processing, such as phase retrieval, occurs at the digital level.  This second step may be thought of as embodying ``virtual optics'' that augments the hardware optical elements used to form the recorded phase-contrast images.  The phase retrieval (and other image reconstruction steps), that proceed at a digital level, elevate the computer to a ``software lens'' component of an X-ray optical imaging system.  Of course, the idea of such ``computational imaging'' has long been a reality, with this being a special case of a decoder in Shannon's mathematical theory of communication (Shannon, 1948a, 1948b).  Computational imaging in an X-ray setting includes crystallography (Hammond, 2009), X-ray computed tomography (Natterer, 1986), X-ray interferometry (Bonse \& Hart, 1965), inline X-ray holography (Aoki \& Kikuta, 1974), X-ray coherent diffractive imaging (Miao, Charalambous, Kirz, \& Sayre, 1999), X-ray ptychography (Pfeiffer, 2018) etc.  In all such virtual-optics systems, the computer forms an intrinsic component of the imaging system, with (i) hardware optics streaming and manipulating X-ray optical information at the field level, supplemented by (ii) digital computers that stream and manipulate digital optical intensity information at the level of bits (Paganin et al., 2004a).  

\subsubsection{Use of prior knowledge in the inverse problem of phase-contrast imaging}

In the preceding sub-sections, we have mentioned the possibility that prior knowledge may be utilised in the context of solving inverse problems.  The importance of this observation must be emphasised, since one often knows a great deal about a sample before it is imaged.  Judicious use of such prior knowledge may allow one to improve the quality of one's image, or obtain an image of an acceptable level of quality using a smaller dose to the sample than would otherwise be needed. Examples of prior knowledge, that may be used in inverse problems of imaging (including but not limited to the inverse problem of phase retrieval), include:

\begin{enumerate}
    \item The sample is known to be thin, in the sense that the projection approximation may be safely assumed to hold;
    \item The sample is known to be a weak pure phase object;
    \item The sample is known to be composed of a single material of possibly varying density;
    \item The sample is known to be binary, i.e.~throughout its volume its complex refractive index can only take one of two different values;
    \item The sample is known to be composed of a small number of known elements.  For example, the elemental composition of a micro-electronics component or soft biological tissues may be known;
    \item The sample is known to be composed of a small number of homogeneous materials, each of which have a known complex refractive index;
    \item The sample is known to be a manufactured product, such as a mechanical component or a micro-electronics circuit, that is to be either accepted or rejected in the process of undergoing quality-control assessment;  
    \item The sample is known to be confined within a known volume of space (compact support);
    \item The sample is known to be elastically scattering, hence it can be safely assumed to not change the energy of the X-rays that pass through it during the imaging process.
\end{enumerate}

\subsection{Phase retrieval methods based on refraction}

Here we give a cursory description of a few phase-contrast methods that have received attention in recent years, and are currently used in the X-ray imaging community. This briefest of sketches does not cover all of the important existing techniques, nor does it do justice to the experimental details of the various techniques. For a review on the subject, providing insights on the relative strengths and limitations of many methods, see Wilkins et al.~(2014). Bravin, Coan, \& Suortti (2013) not so long ago published a review discussing pre-clinical and clinical applications of phase-contrast imaging.  For an extensive suite of recent reviews written by a number of leaders in the field, see the chapters on phase-contrast X-ray imaging in the volume edited by Russo (2018).  

Our rather modest approach, here, is to look at the different techniques following the transport-of-intensity equation (TIE), and specifically its version valid for small object-to-detector propagation distances $\delta z$, as written in Eq.~(\ref{eq:TIE_for_defocus}). To make our approach clearer, we expand the divergence operator on the right hand side of Eq.~(\ref{eq:TIE_for_defocus}):
\begin{eqnarray}\label{eq:TIE_for_defocus2}
I(x,y,z+\delta z) \approx I(x,y,z) -\frac{\delta
z}{k} \left[ \nabla_{\perp}I(x,y,z) \cdot \nabla_{\perp}\phi(x,y,z) +  I(x,y,z) \nabla_{\perp}^{2} \phi(x,y,z) \right]. \quad\quad
\end{eqnarray}

Under the approximations used to derive this finite-difference version of the TIE, the terms in the square brackets describe the phase-contrast contribution to the image. (i) The phase gradient corresponds to the direction of a local streamline (Fig.~\ref{fig:XRayWaveFronts}), whereas (ii) the Laplacian measures the curvature of the wave front (Fig.~\ref{fig:RRR1}). Stated differently: (i) The first term in square brackets contains the (transverse) phase gradient, and represents a prism-like effect that transversely displaces optical energy in a manner proportional to the local deflection angles $\theta_x(x,y)$ and $\theta_y(x,y)$ in the $x$ and $y$ directions, namely:
\begin{equation}\label{eq:DeflectionaAnglesPhaseGradient}
(\theta_x(x,y,z),\theta_y(x,y,z))=\frac{\nabla_{\perp}\phi(x,y,z)}{k}.    
\end{equation}
(ii) The second term in the square brackets, on the right-hand side of Eq.~(\ref{eq:TIE_for_defocus2}), is a lensing term.  This lensing term contains the transverse Laplacian, which describes the local concentration or rarefaction of optical energy density (and hence intensity) due to the sample locally focusing or defocusing the X-ray radiation streaming through it (Bremmer, 1952; Wilkins, Gureyev, Gao, Pogany, \& Stevenson, 1996).  This last-mentioned comment may be visualised via  features ${\bf 1}$ and ${\bf 2}$ in Fig.~\ref{fig:RRR1}. With the exception of direct methods to measure the phase---such as interferometry---many (but certainly not all!) commonly used phase-contrast methods measure phase derivatives, and many such methods can be described using Eq.~(\ref{eq:TIE_for_defocus2}).

Let us elaborate a little, on this last point. The following methods provide image contrast dependent upon the first derivative of the phase in the transverse plane (first term in the square bracket of Eq.~(\ref{eq:TIE_for_defocus2})):
\begin{itemize}
 
    \item X-ray grating-based imaging (Momose et al., 2003; Weitkamp et al., 2005; Pfeiffer et al., 2008);
    \item analyser-based X-ray imaging (F\"{o}rster, Goetz, \& Zaumseil, 1980; Somenkov, Tkalich, \& Shil'shtein, 1991; Davis, Gao, Gureyev, Stevenson, \& Wilkins, 1995; Ingal \& Beliaevskaya, 1995; Chapman et al., 1997; Wernick et al., 2003; Rigon, Arfelli, \& Menk, 2007);
    \item X-ray edge illumination (Olivo, Ignatyev, Munro, \& Speller, 2011; Munro et al., 2013; Pelliccia \& Paganin, 2013a; Diemoz {\em et al.}, 2017);
    \item X-ray speckle tracking (B\'{e}rujon, Ziegler, Cerbino, \& Peverini, 2012; Morgan, Paganin, \& Siu, 2012; Zdora, 2018).
    \end{itemize}
 Conversely, we may augment the above list of dot points with
\begin{itemize}
\item propagation-based methods (Snigirev, Snigireva, Kohn, Kuznetsov, \& Schelokov, 1995; Snigirev, Snigireva, Kohn, \& Kuznetsov, 1996a; Cloetens, Barrett, Baruchel, Guigay, \& Schlenker, 1996; Wilkins, Gureyev, Gao, Pogany, \& Stevenson, 1996; Nugent, Gureyev, Cookson, Paganin, \& Barnea, 1996)
\end{itemize}
which measure the second derivative of the phase, described by the second term in the square brackets of Eq.~(\ref{eq:TIE_for_defocus2}).

Before going into some more detailed analysis it is worth pointing out two general facts about phase-contrast X-ray imaging, which descend straight  from Eq.~(\ref{eq:TIE_for_defocus2}).
\begin{description}
\item[Fact 1] Most phase-contrast imaging techniques require propagation, or may be construed as requiring propagation. 
\item [Fact 2] Both the gradient and the Laplacian of the phase can be present, at the same time, in phase-contrast images.
\end{description}

Fact 1 follows from the observation that the square-bracketed term of Eq.~(\ref{eq:TIE_for_defocus2}) vanishes if $\delta z=0$. More physically, phase-contrast signal---for cases when it is not generated by interferometry---is generated by refraction. X-rays passing through a sample are refracted as well as absorbed. Normally refraction effects goes unnoticed as the refraction angle is extremely small. To become appreciable, measuring refraction requires the detector to be placed some distance away from the sample to analyse the wave front.

Fact 2 is strictly speaking correct only for methods that are sensitive to the phase gradient. Since all of these methods still require propagation, they will always measure a combination of the gradient and the Laplacian of the phase (Pavlov et al., 2004, 2005; Diemoz et al., 2017). 

\subsection{One method for X-ray phase-contrast imaging employing free-space propagation}

Here we outline a simple indicative example of the inverse problem of X-ray phase retrieval, in the specific setting of propagation-based phase contrast.  To gain an overview of the plethora of other approaches that have been developed to address the same question, see e.g.~the review article by Wilkins et al.~(2014), together with the suite of relevant chapters from the collection edited by Russo~(2018).

Consider propagation based X-ray phase-contrast imaging of a single-material object with projected thickness $T_{\textrm{object}}(x,y)$ that is normally illuminated by monochromated plane waves of uniform intensity $I_0$. The refractive index decrement $\delta$ and linear attenuation coefficient $\mu$ are both assumed to be known, for the X-ray energy $E=\hbar\omega$ that is being employed. We again drop explicit dependence on $\omega$, for clarity.  The projection-approximation expressions, in Eqs.~(\ref{eq:PhaseShiftProjectionApproximation}) and (\ref{eq:zz8a}), enable both the phase and the amplitude at the exit surface of the object to be obtained from $T_{\textrm{object}}(x,y)$. Hence, for a single-material object illuminated by normally incident plane waves of uniform intensity $I_0$, the phase shift in Eq.~(\ref{eq:PhaseShiftProjectionApproximation}) becomes
\begin{equation}
\Delta\phi(x,y,z=z_0)=-k \, \delta T_{\textrm{object}}(x,y),    
\end{equation}
and the absorption-contrast intensity map in Eq.~(\ref{eq:zz8a}) becomes
\begin{equation}
I(x,y,z=z_0)=I_0 \exp[-\mu T_{\textrm{object}}(x,y)].    
\end{equation}
In writing the above expressions, we have taken the nominally-planar exit surface of the sample to be the plane $z=z_0$ (see Fig.~\ref{fig:ProjectionApproximation}).

The above relations offer the logical possibility that the projected thickness of a single-material sample may be obtained from a single propagation-based phase-contrast image $I(x,y,z=z_0+\Delta)$, obtained at a distance $\Delta$ downstream of the object, with this distance being sufficiently small for the corresponding Fresnel number $N_F$ in Eq.~(\ref{eq:FresnelNumberFreeSpacePropagation}) to be much greater than unity.  This corresponds to the ``single edge fringe'' regime exemplified by the propagation-based phase-contrast images in the bottom right of Fig.~\ref{fig:RRR2}.  With the previously mentioned approximations, but no further approximations of any kind, the transport-of-intensity equation  (Eq.~(\ref{eq:TIE})) may be solved exactly, to give the projected thickness of the sample from a single propagation-based phase-contrast image (Paganin, Mayo, Gureyev, Miller, \& Wilkins, 2002):
\begin{equation}\label{eq:SPEX}
T_{\textrm{object}}(x,y)=-\frac{1}{\mu}\log_e\left(\mathcal{F}^{-1}
\left\{\frac{\mathcal{F}\left[I(x,y,z=z_0+\Delta)\right]/I_0}
{1+(\delta\Delta/\mu)(k_x^2+k_y^2)}\right\}
\right).
\end{equation}

The above algorithm has been widely utilised.  Its advantages, bought at the price of the previously stated strong assumptions, include simplicity, speed, very significant noise robustness and the ability to process time-dependent objects frame-by-frame.  While the method provides quantitative results when its key assumptions are sufficiently well met, qualitative reconstructions obtained under a broader set of conditions are often of utility where non-quantitative morphological information is sufficient.  It is also worth pointing out that, when the method is utilised in a tomographic context, its domain of utility broadens since many objects may be viewed as locally composed of a single material of interest, in three spatial dimensions, that cannot be described as composed of a single material in projection (Beltran, Paganin, Uesugi, \& Kitchen, 2010; Beltran et al., 2011).  
In such a tomographic setting, the algorithm is sufficiently robust with respect to noise that Beltran and colleagues noted, in the previously cited references, that it can exhibit signal-to-noise ratio (SNR) boosts of up to 200.  More recent studies have shown that this SNR boost has, as an approximate upper limit, 0.3 $\delta/\beta$ if Poisson statistics are assumed (Nesterets \& Gureyev, 2014; Gureyev et al., 2014).  The SNR boost may be traded off against significant reductions in image-acquisition time or radiation dose (Beltran et al., 2011; Gureyev et al., 2014; Nesterets \& Gureyev, 2014; Kitchen et al., 2017).   

A variant, of the method in Eq.~(\ref{eq:SPEX}), has been developed for analyser based phase-contrast imaging and other phase-contrast imaging systems that yield first-derivative phase contrast (Paganin, Gureyev, Pavlov, Lewis, \& Kitchen, 2004b).  Another variant has has been developed for phase-contrast imaging systems that simultaneously yield both first-derivative and second-derivative phase contrast (Pavlov et al., 2004, 2005).  For a multiple-image variant that improves both the signal-to-noise ratio and the resolution in comparison to what would be achieved by using Eq.~(\ref{eq:SPEX}), see e.g.~Yu et al.~(2018) and Kuan et al.~(2020), together with references therein.  We also refer the reader to Gureyev, Pogany, Paganin, \& Wilkins (2004a) and Paganin et al.~(2020), together with references therein.

\subsection{Phase-gradient methods}

In this category we find methods such as analyser-based X-ray imaging (ABI) (F\"{o}rster, Goetz, \& Zaumseil, 1980; Somenkov, Tkalich, \& Shil'shtein, 1991; Davis, Gao, Gureyev, Stevenson, \& Wilkins, 1995; Ingal \& Beliaevskaya, 1995; Chapman et al., 1997; Wernick et al., 2003; Rigon, Arfelli, \& Menk, 2007), grating interferometry (GI) and its variants (Momose et al., 2003; Weitkamp et al., 2005; Pfeiffer et al., 2008), edge illumination (EI) and its variants (Olivo, Ignatyev, Munro, \& Speller, 2011; Munro et al., 2013; Pelliccia \& Paganin, 2013a; Diemoz {\em et al.}, 2017), and X-ray speckle tracking (B\'{e}rujon, Ziegler, Cerbino, \& Peverini, 2012; Morgan, Paganin, \& Siu, 2012; Zdora, 2018). Scanning methods using a focused beam as a probe (Sayre \& Chapman, 1995; Schneider, 1998) also yield phase gradients and can be included in this description.

Looking once again at Eq.~(\ref{eq:TIE_for_defocus2}), we immediately understand what all of these methods have in common. To measure the transverse phase gradient $\nabla_{\perp} \phi(x,y,z)$ one must introduce a transverse intensity gradient $\nabla_{\perp} I(x,y,z)$ and allow for some propagation distance $\delta z$. Interestingly, such an intensity gradient can be introduced in a single frame or throughout multiple frames. Techniques such as speckle tracking, or single-image phase retrieval using a grating before the object, work by introducing spatial intensity variations in the field of view. Methods such as ABI or grating interferometry, for instance when using diffraction-enhanced imaging or fringe scanning respectively, rely on intensity gradients generated across several images. In this case the phase retrieval method will require more than one image to work. Typically single-image methods are quicker and enable lower X-ray dose, while multi-image methods can attain better spatial resolution.

Another case of multi-image phase retrieval is represented by scanning methods. In this case the intensity gradient is given by the beam itself, that can be shaped either before (i.e.~in STXM, Scanning Transmission X-ray Microscopy) or after the sample (EI) to yield the desired phase gradient.

Given the common basis we are here describing, it is not surprising that different methods may share similar approaches. We can go a step further and introduce a common formalism to deal with the plurality of methods that rely on phase gradients using, once again, the transfer function formalism.

\begin{figure}
\includegraphics[scale=1.3]{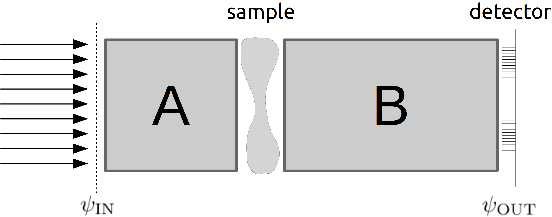}
\caption{Cascaded optical imaging system, in which an incident X-ray field $\psi_{\textrm{IN}}$ traverses an optical element $A$, before passing through a sample.  Upon exiting the sample, the X-rays pass through a second optical element $B$.  This leads to the field $\psi_{\textrm{OUT}}$ over the surface of a position-sensitive detector.}
\label{fig:TransferFunctionPhaseGardient}
\end{figure}

With the help of the drawing in Fig.~\ref{fig:TransferFunctionPhaseGardient}, we generalise the formalism introduced in a preceding section, to include any optical element or sequence of such elements that may be present in the beam path. Let us identify $\psi_{\rm IN}(x,y)$ with the complex wave field at a suitable plane before the sample. Note that, in order to account for optical elements placed \textit{before} the sample, the plane of $\psi_{\rm IN}(x,y)$ need not be defined immediately before the sample. Other optical elements can be placed between the sample and the detector, hence the output wave field $\psi_{\rm OUT}(x,y)$ is defined at the entrance plane of the detector.

Using the transfer function formalism, we can write a generic input--output equation of the form
\begin{equation} 
\psi_{\rm OUT}(x,y,\tau_1,\tau_2,\cdots)={\mathcal D}_{\Delta}
 \left[ \prod_{i=1}^{i=N} {\mathcal D}^{(i)} (\tau_1^{(i)}, \tau_2^{(i)}, \cdots) M_{i}(x,y) \right] \psi_{\rm IN}(x,y).
\end{equation}
Reading from right to left, each of the $M_{i}$ represents the transmission function of a ``mask'', be it a grating or grid, a speckle-forming mask, or the sample itself (assuming the projection approximation to be valid throughout). A given $M_{i}$ can also be the unit operator if no mask is present. Similarly, each of the generalised diffraction operators ${\mathcal D}^{(i)}$ can be a free-space propagation operator, an arbitrary aberrated shift-invariant imaging-system operator, or an operator for an arbitrary linear imaging system that is not shift invariant. The product symbol means that a number of these operators may be chained together, via successive multiplications from the left, according to how many individual optical elements are present in the beam path. Finally, the ${\mathcal D}_{\Delta}$ operator corresponds to free-space propagation, from the last optical or transmission element to the detector entrance plane.  Here, $\Delta$ denotes the distance from the exit-surface of the chain of cascaded optical elements, to the entrance surface of the position-sensitive detector.

Multiplicative masks introduce the intensity gradient that is required to visualise a phase gradient (see Eq.~(\ref{eq:TIE_for_defocus2}) once again). As already mentioned, phase contrast associated with the second derivative of the phase is also generally present in such systems.

Phase-retrieval procedures often involve comparing intensity maps measured with and without the sample, and may involve scanning one of the optical elements, for instance an analyzer crystal or a grating. The presence of the optical elements---and of the scanning process if required---makes phase-gradient methods more involved than free-space propagation. On the upside however, phase-gradient methods are generally well suited to decouple phase contrast, attenuation contrast, and also unresolved phase contrast which is usually referred to as scattering contrast (local fans of small-angle X-ray scattering).

To gain a better understanding of such a decoupling process, let us put forward a highly simplified albeit rather generic ``toy model'', for describing the action of phase-gradient methods. Let us assume that the plurality of optical elements in the beam path can be replaced by a single element of the type described in Eq.~(\ref{eq:ArbitraryShiftInvariantLinearCoherentImagingSystem5}).  This allows us to write the relation between input and output fields as:
\begin{equation}
\psi_{\rm OUT}(x) = \mathcal{F}^{-1}T(k_x)\mathcal{F}\psi_{\rm IN}(x).
\label{eq:SimplifiedShiftInvariantLinearCoherentImagingSystem}
\end{equation}
For simplicity we are here assuming dependence on a single transverse spatial frequency coordinate $k_{x}$. We further assume that the wave-function $\psi_{\rm IN}(x)$, upon which the imaging-system operator acts, describes the field immediately after the sample. Let us write $\psi_{\rm IN}(x)$ as 
\begin{equation} 
\psi_{\rm IN}(x) = \sqrt{I_{\rm IN}(x)} \mathrm{e}^{i \phi(x)} = \sqrt{I_{\rm IN}(x)} \mathrm{e}^{i \left[ \phi_{1}(x) + \phi_{2}(x)\right]}.
\label{eq:SplittingThePhase}
\end{equation}
We have here split the phase term into two components, namely $\phi_{1}(x)$ and $\phi_{2}(x)$.  These two components represent the slowly-varying and the rapidly-varying components of the phase, respectively, with the latter being small in amplitude.  For more detail on this decomposition, see e.g.~Nesterets (2008) and Nesterets \& Gureyev (2016), together with references therein. The stated difference in rapidity of spatial oscillations, between the two components $\phi_{1}(x)$ and $\phi_{2}(x)$, is a construction that depends upon the spatial resolution of the imaging system. Under this view, the slowly-varying component is that which can be spatially resolved by the imaging system. The rapidly-varying components cannot however be spatially resolved by the system, and produce, as we shall see shortly, the scattering contrast.  This is a crucially important point, so let us underline it here: the rapidly varying component $\phi_{2}(x)$ yields a detectable position-dependent signal, here loosely spoken of as ``scattering contrast'', even though the rapid fluctuations in $\phi_{2}(x)$ are unresolved. Such contrast is very closely related to the local-SAXS contrast that we described earlier, in the different but related context of the Fokker--Planck model for paraxial X-ray imaging.

To appreciate the essence of a large class of available phase-gradient methods, let us introduce a further simplification. Assume that the transfer function $T$ depends only linearly upon the spatial frequency (Paganin, Gureyev, Pavlov, Lewis, \& Kitchen, 2004b):
\begin{equation} 
T(k_{x}) = 1 + \tau k_{x}.
 \label{eq:qqqqqq}
\end{equation}
Such a linear dependence is not uncommon: linearising the flank of a rocking curve in ABI (Chapman et al., 1997), or the transmission function of a grating interferometer away from the extreme points (Pelliccia et al., 2013b) are examples of this approximation.  Note that the parameter $\tau$ is in general complex, as it is defined in regard to the complex-valued transfer function $T(k_{x})$, and will have units of length.    

Following the treatment in Paganin, Gureyev, Pavlov, Lewis, \& Kitchen (2004b) as well as Sec.~4.4.3 of Paganin (2006), Eqs.~(\ref{eq:SimplifiedShiftInvariantLinearCoherentImagingSystem}) and (\ref{eq:qqqqqq}) imply that the intensity at the output plane is
\begin{equation}
\frac{I_{\textrm{OUT}}(x)}{I_{\textrm{IN}}(x)} = 1 + 2 \mathrm{Re}[\tau] \frac{\partial \phi(x)}{\partial x}+ \mathrm{Im}[\tau] \frac{\partial}{\partial x} \mathrm{ln}I_{\textrm{IN}}(x) + 
|\tau|^{2} \left[ \frac{\partial \phi(x)}{\partial x} \right]^{2} 
+ \frac{|\tau|^{2}}{4} \left[ \frac{\partial}{\partial x} \mathrm{ln} I_{\textrm{IN}}(x) \right]^{2}.   
\label{Eq:differentialPhaseContrast}
\end{equation}
The first term, on the right side of this equation, describes conventional attenuation contrast.  The second term describes differential phase contrast, insofar as it is proportional to the transverse derivative of the phase.  The third term describes differential attenuation contrast, which is sensitive to the derivative of the attenuation.  The fourth term is a quadratic function of the differential phase contrast, with the fifth term being quadratic in the differential attenuation contrast.  

The previous equation describes the ideal case in which the spatial resolution of the imaging system is arbitrarily good. In practice, this resolution is finite. While such a statement may sound obvious, it causes a qualitative difference in the measurable signal. Let us explore this point in more detail.

By recalling Eq.~(\ref{eq:SplittingThePhase}), phase variations that fall below the resolution give a qualitatively different type of contrast, despite being generated by the same underlying physical process. To appreciate this point, we need to model a realistic acquisition process, whereby the incident intensity is averaged over the extent of the Point Spread Function (PSF) of the detector system (Nesterets, 2008). Such an average, which is here denoted by an overline, takes the form
\begin{equation}
\olsi{\frac{\partial \phi(x)}{\partial x}} = \olsi{\frac{\partial \left[\phi_{1}(x)+\phi_{2}(x)\right]}{\partial x}} \approx \frac{\partial \phi_{1}(x)}{\partial x}.
\end{equation}
The average over the PSF cancels the rapidly-changing component of the phase, leaving the slow-changing component only. This, once again, is conventional differential phase-contrast imaging, but nuanced to incorporate the realistic assumption of finite spatial resolution.

The quadratic term $[\partial\phi(x)/\partial x]^2$ in Eq.~(\ref{Eq:differentialPhaseContrast}) behaves differently, when we smear over the breadth of a detector PSF. In this case the term involving $\phi_2(x)$ does not disappear:
\begin{equation}\label{eq:qaz}
\olsi{\left[ \frac{\partial \phi(x)}{\partial x} \right]^{2} } = \olsi{\left[ \frac{\partial \phi_{1}(x)}{\partial x} \right]^{2} }+ \olsi{\left[ \frac{\partial \phi_{2}(x)}{\partial x} \right]^{2} } + 2\olsi{ \frac{\partial \phi_1(x)}{\partial x} \frac{\partial \phi_2(x)}{\partial x} }
\approx 
\olsi{\left[ \frac{\partial \phi_{2}(x)}{\partial x} \right]^{2} }.
\end{equation}
To justify the previous expression, let us recall that we can neglect the first term after the first equals sign, as the slow-varying component of the phase has (by definition) a small transverse derivative, hence its square is negligible compared to the other terms. We note parenthetically that, if the assumption in the previous sentence is not valid, we can very easily retain the term in question, in the following analysis.  Note, also, that the cross term in Eq.~(\ref{eq:qaz}) is a covariance that averages to zero as, in this case, it is a rapidly-changing component that becomes negligible once averaged over the PSF. Lastly, we both note and emphasise the term $\overline{[\partial\phi_2(x)/\partial x]^2}$ on the far right of the above equation, which may be identified with the local variance of the rapidly-fluctuating signal $\partial\phi_2(x)/\partial x$.  This position-dependent signal $\overline{[\partial\phi_2(x)/\partial x]^2}$ in Eq.~(\ref{eq:qaz}) is a form of SAXS-induced ``dark field'' signal that embodies diffuse scatter (Rigon, Arfelli, \& Menk, 2007; Modregger et al., 2017).  As pointed out earlier, this signal can contain important information regarding a sample, distinct from the other attributes of position-dependent attenuation and position-dependent phase shifts (Morrison \& Browne, 1992; Harding \& Schreiber, 1999; Pagot et al., 2003; Levine \& Long, 2004; Rigon, Arfelli, \& Menk, 2007; Pfeiffer et al., 2008; Bech et al., 2010; Yashiro, Terui, Kawabata, \& Momose, 2010; Yashiro et al., 2011; Lynch et al., 2011; Modregger et al., 2012; Strobl, 2014).   

To examine this last-mentioned point in a little more detail, recall from Eq.~(\ref{eq:DeflectionaAnglesPhaseGradient}) that $\partial\phi_2(x)/\partial x$ is proportional to the X-ray deflection angle $\theta_{x,\phi_2}(x)$ that is due to $\phi_2(x)$.  Hence $\overline{[\partial\phi_2(x)/\partial x]^2}$ is proportional to the variance ${\textrm{Var}}[\theta_{x,\phi_2}(x)]$ of the deflection angle that is introduced by the spatially-unresolved rapid phase fluctuations:
\begin{equation} \label{eq:VarianceOfDeflectionAngle}
\olsi{\left[ \frac{\partial \phi_{2}(x)}{\partial x} \right]^{2} } = k^2 \olsi{\left[ \theta_{x,\phi_2}(x) \right]^{2} } = k^2 {\textrm{Var}}[\theta_{x,\phi_2}(x)].    
\end{equation}
We can associate this variance, of the local X-ray angular deflections that are induced by the rapidly varying phase fluctuations, with the previously discussed SAXS-fan opening angle $\theta(x)$ (see Fig.~\ref{fig:FokkerPlanck}):
\begin{equation}
    {\textrm{Var}}[\theta_{x,\phi_2}(x)]\equiv[\theta(x)]^2.
\end{equation}

Bearing all of the above in mind, the general expression for the measurable intensity, namely Eq.~(\ref{Eq:differentialPhaseContrast}), becomes the following after we perform a smearing over the detector PSF:
\begin{equation}\label{eq:LinearTransferFunctionDF}
{\frac{~\olsi{I_{\textrm{OUT}}(x)}~}{I_{\textrm{IN}}(x)}} = 1 + 2 \mathrm{Re}[\tau] \frac{\partial \phi_{1}(x)}{\partial x} + \mathrm{Im}[\tau]{ \frac{\partial}{\partial x} \mathrm{ln}I_{\textrm{IN}}(x) }  + \frac{|\tau|^{2}}{4} {\left[ \frac{\partial}{\partial x} \mathrm{ln} I_{\textrm{IN}}(x) \right]^{2}}+ 
 k^2 |\tau|^{2} [\theta(x)]^2 .
\end{equation}
Note, that in obtaining the above expression, we have assumed $I_{\textrm{IN}}(x)$ to be sufficiently slowly varying with respect to $x$, that we may make the approximation $\olsi{I_{\textrm{IN}}(x)}\approx I_{\textrm{IN}}(x)$. The square of the differential phase (which now appears implicitly in the final term in the above expression) now contains only the fast-changing component, which is the one that falls below the resolution limit of the imaging system. While its individual highly-rapid fluctuations are not spatially resolved, this ``wave-field roughness'' term does produce contrast which, as anticipated, is qualitatively different from differential phase contrast. This term has the same quadratic dependence on SAXS-fan opening angle that was written down in an analyser-crystal phase-contrast setting by Rigon, Arfelli, \& Menk (2007).  The same quadratic dependence appears in a Fokker--Planck context in Eq.~(\ref{eq:FokkerPlanck4}), with a similar quadratic dependence given in Gureyev et al.~(2020).

We have described a simple model to explain the contrast mechanisms that are accessible by phase-gradient methods. What all these methods have in common is a transfer function that varies with the spatial frequency in some way. In the model described above, we have assumed a particularly simple linear relation (Eq.~(\ref{eq:qqqqqq})) which readily shows the dependence of the measured intensity upon the phase derivatives, without over-complicating the mathematical treatment. This simple approach naturally shows the phase-gradient contrast arising from the resolvable phase structure in the sample, as well as the scattering contrast generated by the unresolvable micro and nanostructures.  

To solve the inverse problem based on Eq.~(\ref{eq:LinearTransferFunctionDF}), consider the linear imaging system to be in two different states $A$ and $B$, characterised by the respective $\tau$ values $\tau_A,\tau_B$ in Eq.~(\ref{eq:qqqqqq}). If both input and output images $\overline{I_{\textrm{IN}}(x)}$ and $\overline{I_{\textrm{OUT}}(x)}$ are measured for each of the two states of the imaging system, then the corresponding pair of instances of Eq.~(\ref{eq:LinearTransferFunctionDF}) may be viewed as simultaneous linear equations for the two unknown quantities $\partial\phi_1(x)/\partial x$ and
\begin{equation}
[\theta(x)]^2=\frac{1}{k^2}\overline{\left[\frac{\partial\phi_2(x)}{\partial x}\right]^2}.
\end{equation}
These unknown quantities can then be recovered using  matrix inversion, provided that $\tau_A,\tau_B$ are chosen such that that the corresponding matrix of coefficients is non-singular, in the system of linear equations whose solution is sought.  If more than two images are taken, $\partial\phi_1(x)/\partial x$ and $\theta(x)$ may instead be recovered via least-squares refinement (Press, Teukolsky, Vetterling, \& Flannery, 2007) or other iterative approaches.  Recovery of the term $\theta(x)$ amounts to recovery of a position-dependent measure of the diffuse scatter (position-dependent small-angle scattering signal) that is induced by the X-rays traversing the spatially unresolved micro-structure permeating the sample volume.  If the output but not the input images are measured, and additional states of the imaging system are employed, then the inverse problem can be solved using a simple variant of the  method outlined in Paganin, Gureyev, Pavlov, Lewis, \& Kitchen (2004b) and Rigon, Arfelli, \& Menk (2007).  Also, if the transfer function is expanded to second order rather than first order in spatial frequencies, then the analysis of Pavlov et al.~(2004) may be used to broaden the domain of applicability of the calculation presented above.   

\section*{Acknowledgements}

This chapter is an extensively reworked version of a tutorial and companion notes (Paganin \& Pelliccia, 2019) presented by the authors as three two-hour seminars.  These seminars were delivered to the European Synchrotron (ESRF) community in Grenoble, France, on May 31 -- June 2, 2017. We thank the European Synchrotron for facilitating and videotaping these lectures, as well as making them available at \url{https://bit.ly/2GdoVg8}. Both authors offer sincere thanks to Alexander Rack and the ESRF Directors for supporting their visits to ESRF in 2017 and 2018. We thank Carsten Detlefs (ESRF), for his close reading of the first version of this text, and for his detailed comments that led to many significant improvements.  

\section*{Dedication}

We humbly dedicate this chapter to the memory of Claudio Ferrero.

\section*{References}

Adams, B. W., Buth, C., Cavaletto, S. M., Evers, J., Harman, Z., Keitel, C. H,. P\'{a}lffy, A., Pic\'{o}n, A., R\"{o}hlsberger, R., Rostovtsev, Y., \&  Tamasaku. (2013). K. X-ray quantum optics. {\em Journal of Modern Optics}, {\em 60}, 2--21. \url{https://doi.org/10.1080/09500340.2012.752113}.

\medskip

Allen, L. J., Oxley, M. P., \& Paganin, D. (2001). Computational aberration correction for an arbitrary linear imaging system. {\em Physical Review Letters}, {\em 87}, 123902. \url{https://doi.org/10.1103/PhysRevLett.87.123902}.  

\medskip

Alonso, M. A. (2011). Wigner functions in optics: describing beams as ray bundles and pulses as particle ensembles. {\em Advances in Optics and Photonics}, {\em 3}, 272--365. \url{https://doi.org/10.1364/AOP.3.000272}.

\medskip

Alperin, S. N., Grotelueschen, A. L., \& Siemens, M. E. (2019). Quantum turbulent structure in light. {\em Physical Review Letters}, {\em 122}, 044301. \url{https://doi.org/10.1103/PhysRevLett.122.044301}. 

\medskip

Aoki, S., \& Kikuta, S. (1974). X-Ray Holographic Microscopy. {\em Japanese Journal of Applied Physics}, {\em 13}, 1385--1392. \url{https://doi.org/10.1143/JJAP.13.1385}.

\medskip

Authier, A. (2001). {\em Dynamical theory of x-ray diffraction}. Oxford: Oxford University Press.

\medskip

Bech, M., Bunk, O., Donath, T., Feidenhans'l, R., David, C., \& Pfeiffer, F. (2010). Quantitative x-ray dark-field computed tomography. {\em Physics in Medicine and Biology}, {\em 55}, 5529--5539. \url{https://doi.org/10.1088/0031-9155/55/18/017}.

\medskip

Beltran, M. A., Paganin, D. M., Uesugi, K., \&  Kitchen, M. J. (2010). 2D and 3D X-ray phase retrieval of multi-material objects using a single defocus distance. {\em Optics Express}, {\em 18}, 6423--6436. \url{https://doi.org/10.1364/OE.18.006423}.

\medskip 

Beltran, M. A., Paganin, D. M., Siu, K. K. W., Fouras, A., Hooper, S. B., Reser, D. H., \& Kitchen, M. J. (2011). Interface-specific X-ray phase retrieval tomography of complex biological organs. {\em Physics in Medicine and Biology}, {\em 56}, 7353--7369. \url{https://doi.org/10.1088/0031-9155/56/23/002}.

\medskip

 Beltran, M. A., Kitchen, M. J., Petersen, T. C., \& Paganin, D. M. (2015). Aberrations in shift-invariant linear optical imaging systems using partially coherent fields. {\em Optics Communications}, {\em 355}, 398--405. \url{https://doi.org/10.1016/j.optcom.2015.06.075}.

\medskip

Beltran, M. A., Paganin, D. M., \& Pelliccia, D. (2018). Phase-and-amplitude recovery from a single phase-contrast image using partially spatially coherent x-ray radiation. {\em Journal of Optics}, {\em 20}, 055605. \url{https://doi.org/10.1088/2040-8986/aabbdd}.

\medskip

Berry, M. V., \& Upstill, C. (1980). Catastrophe optics: morphologies of caustics and their diffraction patterns. {\em Progress in Optics}, {\em 18}, 257--346. \url{https://doi.org/10.1016/S0079-6638(08)70215-4}.

\medskip

Berry, M. V. (1998). Much ado about nothing: optical dislocation lines (phase singularities, zeros, and vortices). {\em Proceedings of SPIE}, {\em 3487}, 1--5. \url{https://doi.org/10.1117/12.317693}.

\medskip

Berry, M. V. (2009). Optical currents. {\em Journal of Optics A: Pure and Applied Optics}, {\em 11}, 094001. \url{https://doi.org/10.1088/1464-4258/11/9/094001}.

\medskip

Bertero, M., \& Boccacci, P. (1998). {\em Introduction to inverse problems in imaging}.  Bristol: Institute of Physics Publishing. 

\medskip

B\'{e}rujon, S., Ziegler, E., Cerbino, R., \& Peverini, L. (2012). Two-dimensional X-ray beam phase sensing. {\em Physical Review Letters}, {\em 108}, 158102. \url{https://doi.org/10.1103/PhysRevLett.108.158102}. 

\medskip

Born, M., \& Wolf, E. (1999). {\em Principles of optics: electromagnetic theory of propagation, interference and diffraction of light}. (7th ed.). Cambridge: Cambridge University Press.

\medskip

Bransden, B. H., \& Joachain, C. J. (1989). Introduction to quantum mechanics. Harlow: Longman Scientific \& Technical.

\medskip

Bravin, A., Coan, P., \&  Suortti, P. (2013). X-ray phase-contrast imaging: from pre-clinical applications towards clinics. {\em Physics in Medicine and Biology}, {\em 58}, R1--R35. \url{https://doi.org/10.1088/0031-9155/58/1/R1}.

\medskip

Bonse, U., \& Hart, M. (1965). An X-ray interferometer. {\em Applied Physics Letters}, {\em 6}, 155--156. \url{https://doi.org/10.1063/1.1754212}.  
\medskip

Bracewell, R. N. (1986). {\em The Fourier transform and its applications}. (2nd ed.). New York: McGraw--Hill.

\medskip

Bremmer, H. (1952). On the asymptotic evaluation of diffraction integrals with a special view to the theory of defocusing and optical contrast. {\em Physica}, {\em 18}, 469--485. \url{https://doi.org/10.1016/S0031-8914(52)80079-5}. 

\medskip

Ceddia, D., \& Paganin, D. M. (2018). Random-matrix bases, ghost imaging and x-ray phase contrast computational ghost imaging. {\em Physical Review A}, {\em 97}, 062119. \url{https://doi.org/10.1103/PhysRevA.97.062119}. 

\medskip

Chapman, D., Thomlinson, W., Johnston, R. E., Washburn, D., Pisano, E., Gm\"ur, N., Zhong, Z., Menk, R., Arfelli, F., \& Sayers, D. (1997). Diffraction enhanced x-ray imaging. {\em Physics in Medicine and Biology}, {\em 42}, 2015--2025. \url{https://doi.org/10.1088/0031-9155/42/11/001}.

\medskip

Cloetens, P., Barrett, R,. Baruchel, J., Guigay, J.-P., \& Schlenker, M. (1996). Phase objects in synchrotron radiation hard x-ray imaging. {\em Journal of Physics D: Applied Physics}, {\em 29}, 133--146. \url{https://doi.org/10.1088/0022-3727/29/1/023}. 

\medskip

Cloetens, P., Ludwig, W., Baruchel, J., Van Dyck, D., Van Landuyt, J., Guigay, J.-P., \& Schlenker, M. (1999). Holotomography: Quantitative phase tomography with micrometer resolution using hard synchrotron radiation x rays. {\em Applied Physics Letters}, {\em 75}, 2912--2914. \url{https://doi.org/10.1063/1.125225}.

\medskip

Cowley, J. M., \& Moodie, A. F. (1957). The scattering of electrons by atoms and crystals. I. A new theoretical approach. {\em Acta Crystallographica}, {\em 10}, 609--619. \url{https://doi.org/10.1107/S0365110X57002194}. 

\medskip

Cowley, J. M., \& Moodie, A. F. (1959). The scattering of electrons by atoms and crystals. III. Single-crystal diffraction patterns. {\em Acta Crystallographica}, {\em 12}, 360--367. \url{https://doi.org/10.1107/S0365110X59001104}. 

\medskip

Cowley, J.M. (1995). {\em Diffraction physics}. (3rd ed.). Amsterdam: Elsevier.

\medskip

Davis, T. J., Gao, D., Gureyev, T. E., Stevenson, A. W., \& Wilkins, S. W. (1995). Phase-contrast imaging of weakly absorbing materials using hard X-rays. {\em Nature}, {\em 373}, 595--598. \url{https://doi.org/10.1038/373595a0}. 

\medskip

Detlefs, C. (2019). Private communication to the authors.

\medskip

Diemoz, P. C., Hagen, C. K., Endrizzi, M., Minuti, M., Bellazzini, R., Urbani, L., De Coppi, P., \& Olivo, A. (2017). Single-shot X-ray phase-contrast computed tomography with nonmicrofocal laboratory sources.  {\em Physical Review Applied}, {\em 7}, 044029. \url{https://doi.org/10.1103/PhysRevApplied.7.044029}. 

\medskip

Dirac, P. A. M. (1935). {\em The principles of quantum mechanics}. (2nd ed.). Oxford: Oxford University Press.

\medskip

Dirac, P. A. M. (1945). On the analogy between classical and quantum mechanics. {\em Reviews of Modern Physics}, {\em 17}, 195--199. \url{https://doi.org/10.1103/RevModPhys.17.195}.

\medskip

D\"{o}ring, F., Robisch, A. L., Eberl, C., Osterhoff, M., Ruhlandt, A., Liese, T., Schlenkrich, F., Hoffmann, S., Bartels, M., Salditt, T., \& Krebs, H. U. (2013). Sub-5 nm hard x-ray point focusing by a combined Kirkpatrick--Baez mirror and multilayer zone plate. {\em Optics Express}, {\em 21}, 19311--19323. \url{https://doi.org/10.1364/OE.21.019311}.

\medskip

Du, M., Nashed, Y. S. G., Kandel, S., G\"{u}rsoy, D., \& Jacobsen, C. (2020). Three dimensions, two microscopes, one code: automatic differentiation for x-ray nanotomography beyond the depth of focus limit. {\em Science Advances}, {\em 6}, eaay3700. \url{https://doi.org/10.1126/sciadv.aay3700}.

\medskip

Duke, P. J. (2000). {\em Synchrotron radiation: production and properties}. Oxford: Oxford University Press.

\medskip

Ehrenberg, W. (1947). X-ray optics. {\em Nature}, {\em 160}, 330--331. \url{https://doi.org/10.1038/160330b0}. 

\medskip

Eisebitt, S., L\"{u}ning, L., Schlotter, W. F., L\"{o}rgen, M., Hellwig, O., Eberhardt, W., \& Stohr, J. (2004). Lensless imaging of magnetic nanostructures by X-ray spectro-holography.  {\em Nature}, {\em 432}, 885--888. \url{https://doi.org/10.1038/nature03139}.

\medskip

Erko, A. I., Aristov, V. V., \& Vidal, B. (1996). {\em Diffraction x-ray optics}. Bristol: Institute of Physics Publishing. 

\medskip

Feynman, R. P. (1948). Space--time approach to non-relativistic quantum mechanics. {\em Reviews of Modern Physics}, {\em 20}, 367--387. \url{https://doi.org/10.1103/RevModPhys.20.367}.

\medskip

Feynman, R. P., \& Hibbs, A. R. (1965). {\em Quantum mechanics and path integrals}. New York: McGraw--Hill.

\medskip

F\"{o}rster, E., Goetz, K., \& Zaumseil, P. (1980). Double crystal diffractometry for the characterization of targets for laser fusion experiments. {\em Kristall und Technik}, {\em 15}, 937--945. \url{https://doi.org/10.1002/crat.19800150812}. 

\medskip

Freund, P. G. O. (1986). {\em Introduction to supersymmetry}. Cambridge: Cambridge University Press.

\medskip

Gabor, D. (1948). A new microscopic principle. {\em Nature}, {\em 161}, 777--778. \url{https://doi.org/10.1038/161777a0}. 

\medskip

Gabor, D. (1961). Light and information. {\em Progress in Optics}, {\em 1}, 109--153. \url{https://doi.org/10.1016/S0079-6638(08)70609-7}. 

\medskip

Gage, S. H. (1920). Modern dark-field microscopy and the history of its development. {\em Transactions of the American Microscopical Society}, {\em 39}, 95--141. \url{https://doi.org/10.2307/3221838}. 

\medskip

 Garc\'{i}a-Moreno, F., Kamm, P. H., Neu, T. R., B\"{u}lk, F., Mokso, R., Schlep\"{u}tz, C. M., Stampanoni, M., \& Banhart, J. (2019). Using X-ray tomoscopy to explore the dynamics of foaming metal. {\em Nature Communications}, {\em 10}, 3762. \url{https://doi.org/10.1038/s41467-019-11521-1}.

\medskip

Giovannini, D., Romero, J., Poto\v{c}ek, V., Ferenczi, G., Speirits, F., Barnett, S. M., Faccio, D., \& Padgett, M. J. (2015). Spatially structured photons that travel in free space slower than the speed of light. {\em Science}, {\em 347}, 857--860. \url{https://doi.org/10.1126/science.aaa3035}.

\medskip

Goodman, J. W. (1985) {\em Statistical optics}. New York: Wiley.

\medskip

Goodman, J. W. (2005). {\em Introduction to Fourier optics}. (3rd ed.). Englewood Colorado: Roberts \& Company.

\medskip

Goodman, J. W. (2007). {\em Speckle phenomena in optics: theory and applications}. Englewood Colorado: Roberts \& Company.

\medskip

Gureyev, T. E., Pogany, A., Paganin, D. M., \& Wilkins, S. W. (2004a). Linear algorithms for phase retrieval in the Fresnel region. {\em Optics Communications}, {\em 231}, 53--70. \url{https://doi.org/10.1016/j.optcom.2003.12.020}.

\medskip

Gureyev, T. E., Davis, T. J., Pogany, A., Mayo, S. C., \& Wilkins, S. W. (2004b). Optical phase retrieval by use of first Born- and Rytov-type approximations. {\em Applied Optics}, {\em 43}, 2418--2430. \url{https://doi.org/10.1364/AO.43.002418}.

\medskip

Gureyev, T. E., Mayo, S. C., Myers, D. E., Nesterets, Ya., Paganin, D. M., Pogany, A., Stevenson, A. W., \& Wilkins, S. W. (2009). Refracting R\"{o}ntgen's rays: propagation-based x-ray phase contrast for biomedical imaging. {\em Journal of Applied Physics}, {\em 105}, 102005. \url{https://doi.org/10.1063/1.3115402}.

\medskip

Gureyev, T. E., Mayo, S. C., Nesterets, Ya. I., Mohammadi, S., Lockie, D., Menk, R. H., Arfelli, F., Pavlov, K. M., Kitchen, M. J., Zanconati, F., Dullin, C., \& Tromba, G. (2014). Investigation of the imaging quality of synchrotron-based phase-contrast mammographic tomography. {\em Journal of Physics D: Applied Physics}, {\em 47}, 365401. \url{https://doi.org/10.1088/0022-3727/47/36/365401}.

\medskip

Gureyev, T. E., Paganin, D. M., Arhatari, B. D., Taba, S. T., Lewis, S., Brennan, P. C., Quiney, H. M. (2020). Dark-field signal extraction in propagation-based phase-contrast imaging. {\em Physics in Medicine and Biology}, in press. \url{https://doi.org/10.1088/1361-6560/abac9d}.

\medskip

Hadamard, J. (1923). {\em Lectures on Cauchy's problem in linear partial differential equations}. New Haven: Yale University Press. 

\medskip

Hammond, C. (2009). {\em The basics of crystallography and diffraction}. (3rd ed.). Oxford: Oxford University Press.

\medskip

Harding, G., \& Schreiber, B. (1999). Coherent 
X-ray scatter imaging and its applications in
biomedical science and industry. {\em Radiation Physics and Chemistry}, {\em 56}, 229--245. \url{https://doi.org/10.1016/S0969-806X(99)00283-2}.

\medskip

Hecht, E. (1987). {\em Optics}. (2nd ed.). Reading: Addison--Wesley.

\medskip

Huang, K. (1987) {\em Statistical mechanics}. (2nd ed.). New York: Wiley.

\medskip

Ingal, V. N., \& Beliaevskaya, E. A. (1995). X-ray plane-wave topography observation of the phase contrast from a non-crystalline object. {\em Journal of Physics D: Applied Physics}, {\em 28}, 2314--2317. \url{https://doi.org/10.1088/0022-3727/28/11/012}.

\medskip

Jacobsen, C. (2019). {\em X-ray microscopy}. Cambridge: Cambridge University Press.

\medskip

Jap, B. K., \& Glaeser, R. M. (1978). The scattering of high-energy electrons. I. Feynman path-integral formulation. {\em Acta Crystallographica A}, {\em 34}, 94--102. \url{https://doi.org/10.1107/S0567739478000170}.

\medskip

Jensen, T. H., Bech, M., Bunk, O., Donath, T., David, C.,  Feidenhans'l, R., \& Pfeiffer, F. (2010a). Directional x-ray dark-field imaging. {\em Physics in Medicine and Biology}, {\em 55}, 3317--3323. \url{https://doi.org/10.1088/0031-9155/55/12/004}.

\medskip

Jensen, T. H., Bech, M., Zanette, I., Weitkamp, T., David, C., Deyhle, H., Rutishauser, S., Reznikova, E., Mohr, J., Feidenhans'l, R., \& Pfeiffer, F. (2010b). Directional x-ray dark-field imaging of strongly ordered systems. {\em Physical Review B}, {\em 82}, 214103. \url{https://doi.org/10.1103/PhysRevB.82.214103}.

\medskip

Keller, J. B. (1962). Geometrical theory of diffraction. {\em Journal of the Optical Society of America}, {\em 52}, 116--130. \url{https://doi.org/10.1364/JOSA.52.000116}. 

\medskip

Kim, Y. Y., Gelisio, L., Mercurio, G., Dziarzhytski, S., Beye, M., Bocklage, L., Classen, A., David, C., Gorobtsov, O. Yu., Khubbutdinov, R., Lazarev, S., Mukharamova, N., Obukhov, Yu. N., R\"{o}sner, B., Schlage, K., Zaluzhnyy, I. A., Brenner, G., R\"{o}hlsberger, R., von Zanthier, J., Wurth, W., \& Vartanyants, I. A. (2020). Ghost imaging at an XUV free-electron laser. {\em Physical Review A}, {\em 101}, 013820. \url{https://doi.org/10.1103/PhysRevA.101.013820}.

\medskip

Kingston, A. M., Pelliccia, D., Rack, A., Olbinado, M. P., Cheng, Y., Myers, G. R., \& Paganin, D. M. (2018). Ghost tomography. {\em Optica}, {\em 5}, 1516--1520. \url{https://doi.org/10.1364/OPTICA.5.001516}.  

\medskip

Kingston, A. M., Myers, G. R., Pelliccia, D., Svalbe, I. D., \&  Paganin, D. M. (2019). X-ray ghost-tomography: artefacts, dose distribution, and mask considerations. {\em IEEE Transactions on Computational Imaging}, {\em 5}, 136--149. \url{https://doi.org/10.1109/TCI.2018.2880337}. 

\medskip

Kirkland, E. J. (2010). {\em Advanced Computing in Electron Microscopy}. (2nd ed.). Berlin/Heidelberg: Springer.

\medskip

Kirkpatrick, P., \& Baez, A. V. (1948). Formation of optical images by x-rays. {\em Journal of the Optical Society of America}, {\em 38}, 766--774. \url{https://doi.org/10.1364/JOSA.38.000766}. 

\medskip

Kitchen, M. J., Buckley, G. A., Gureyev, T. E., Wallace, M. J., Andres-Thio, N., Uesugi, K., Yagi, N., \& Hooper, S. B. (2017). CT dose reduction factors in the thousands using X-ray phase contrast. {\em Scientific Reports}, {\em 7}, 15953. \url{https://doi.org/10.1038/s41598-017-16264-x}.

\medskip

Klein, A. G., \& Opat, G. I. (1976). Observation of $2\pi$ rotations by Fresnel diffraction of neutrons. {\em Physical Review Letters}, {\em 37}, 238--240. \url{https://doi.org/10.1103/PhysRevLett.37.238}. 

\medskip

Kratky, O., \& Glatter, O. (Eds). (1982). {\em Small angle x-ray scattering}. London: Academic Press.

\medskip

Kreiss, H. O., \& Lorenz, J. (1989). {\em Initial-boundary problems and the Navier--Stokes equations}.  San Diego: Academic Press.

\medskip

Kress, R. (1989). {\em Linear integral equations}. Berlin: Springer--Verlag.

\medskip

Kuan, A. T., Phelps, J. S., Thomas, L. A., Nguyen, T. M., Han, J., Chen, C.-L., Azevedo, A. W., Tuthill, J. C., Funke, J., Cloetens, P., Pacureanu, A., \& Lee, W.-C. A. (2020). Dense neuronal reconstruction through X-ray holographic nano-tomography. {\em Nature Neuroscience}, online Technical Report available at \url{https://doi.org/10.1038/s41593-020-0704-9}. 

\medskip

Kunimune, Y., Yoda, Y., Izumi, K., Yabashi, M., Zhang, X.-W., Harami, T., Ando, M., \& Kikuta, S. (1997). Two-photon correlations in x-rays from a synchrotron radiation source. {\em Journal of Synchrotron Radiation}, {\em 4}, 199--203. \url{https://doi.org/10.1107/S0909049597006912}.

\medskip

Kuznetsova, E., \& Kocharovskaya, O. (2017). Quantum optics with X-rays. {\em Nature Photonics}, {\em 11}, 685--686. \url{https://doi.org/10.1038/s41566-017-0034-y}.

\medskip

Levine, L. E., \& Long, G. G. (2004). X-ray imaging with ultra-small-angle X-ray
scattering as a contrast mechanism. {\em Journal of Applied Crystallography}, {\em 37}, 757--765. \url{https://doi.org/10.1107/S0021889804016073}.

\medskip

Li, K., Wojcik, M., \& Jacobsen, C. (2017). Multislice does it all---calculating the performance of nanofocusing X-ray optics. {\em Optics Express}, {\em 25}, 1831--1846. \url{https://doi.org/10.1364/OE.25.001831}.

\medskip

Li, Z., Medvedev, N., Chapman, H. N., \& Shih, Y. (2017). Radiation damage free ghost diffraction with atomic resolution. {\em Journal of Physics B: Atomic, Molecular and Optical Physics}, {\em 51}, 025503. \url{https://doi.org/10.1088/1361-6455/aa9737}.

\medskip

Lipson, S. G., \&  Lipson, H. (1981). {\em Optical physics}.  (2nd ed.). Cambridge: Cambridge University Press.  

\medskip

Lovesey, S. W., \& Collins, S. P. (1996). {\em X-ray scattering and absorption by magnetic materials}. Oxford: Oxford University Press. 

\medskip

Lynch, S. K., Pai, V., Auxier, J., Stein, A. F., Bennett, E. E., Kemble, C. K., Xiao, X., Lee, W.-K., Morgan, N. Y., \& Wen, H. H. (2011). Interpretation of dark-field contrast and particle-size selectivity in grating interferometers. {\em Applied Optics}, {\em 50}, 4310--4319. \url{https://doi.org/10.1364/AO.50.004310}.

\medskip

Madelung, E. (1927). Quantentheorie in hydrodynamischer Form. {\em Zeitschrift f\"{u}r Physik}, {\em 40}, 322--326. \url{https://doi.org/10.1007/BF01400372}. 

\medskip

Maggiore, M. (2005). {\em A modern introduction to quantum field theory}. Oxford: Oxford University Press.

\medskip

Mandel, L., \&  Wolf, E. (1995). {\em Optical coherence and quantum optics}.  Cambridge: Cambridge University Press.  

\medskip

Mandl, F., \& Shaw, G. (2010).  {\em Quantum field theory}. (2nd ed.). Chichester: Wiley.

\medskip

Martz, H. E., Kozioziemski, B. J., Lehman, S. K., Hau-Riege, S.,  Schneberk, D. J., \& Barty, A. (2007). Validation of radiographic simulation codes including x-ray phase effects for millimeter-size objects with micrometer structures. {\em Journal of the Optical Society of America A}, {\em 24}, 169--178. \url{https://doi.org/10.1364/JOSAA.24.000169}.

\medskip

Miao, J., Charalambous, P., Kirz, J., \& Sayre, D. (1999). Extending the methodology of X-ray crystallography to allow imaging of micrometre-sized non-crystalline specimens. {\em Nature}, {\em 400}, 342--344. \url{https://doi.org/10.1038/22498}.

\medskip

Modregger, P., Scattarella, F., Pinzer, B. R., David, C., Bellotti, R., \& Stampanoni, M. (2012). Imaging the ultrasmall-angle x-ray scattering distribution with grating interferometry. {\em Physical Review Letters}, {\em 108}, 048101. \url{https://doi.org/10.1103/PhysRevLett.108.048101}.

\medskip

Modregger, P., Kagias, M., Irvine, S. C., Br\"{o}nnimann, R., Jefimovs, K., Endrizzi, M., \& Olivo, A. (2017). Interpretation and utility of the moments of small-angle x-ray scattering distributions. {\em Physical Review Letters}, {\em 118}, 265501. \url{https://doi.org/10.1103/PhysRevLett.118.265501}. 

\medskip

Modregger, P., Endrizzi, M., \& Olivo, A. (2018). Direct access to the moments of scattering distributions in x-ray imaging. {\em Applied Physics Letters}, {\em 113}, 254101. \url{https://doi.org/10.1063/1.5054849}. 

\medskip

Momose, A., Kawamoto, S., Koyama, I., Hamaishi, Y., Takai, K., \& Suzuki, Y. (2003). Demonstration of X-ray Talbot interferometry. {\em Japanese Journal of Applied Physics}, {\em 42}, L866--L868. \url{https://doi.org/10.1143/JJAP.42.L866}.

\medskip

Morgan, K. S., Paganin, D. M., \& Siu, K. K. W. (2012). X-ray phase imaging with a paper analyzer. {\em Applied Physics Letters}, {\em 100}, 124102. \url{https://doi.org/10.1063/1.3694918}. 

\medskip

Morgan, K. S., \& Paganin, D. M. (2019). Applying the Fokker--Planck equation to grating-based x-ray phase and dark-field imaging. {\em Scientific Reports}, {\em 9}, 17465. \url{https://doi.org/10.1038/s41598-019-52283-6}.

\medskip

Morrison, G. R., \& Browne, M. T. (1992). Dark‐field imaging with the scanning transmission x‐ray microscope.  {\em Review of Scientific Instruments}, {\em 63}, 611--614.  \url{https://doi.org/10.1063/1.1143820}.

\medskip

M\"uller, P., Sch\"urmann, M., \& Guck, J. (2016). The theory of diffraction tomography.  Preprint available at \url{https://arxiv.org/abs/1507.00466}.

\medskip

Munro, P. R. T., Rigon, L., Ignatyev, K., Lopez, F. C. M., Dreossi, D., Speller, R. D., \&  Olivo, A. (2013). A quantitative, non-interferometric X-ray phase contrast imaging technique.  {\em Optics Express}, {\em 21}, 647--661. \url{https://doi.org/10.1364/OE.21.000647}.

\medskip

Munro, P. R. T. (2019). Rigorous multi-slice wave optical simulation of x-ray propagation in inhomogeneous space. {\em Journal of the Optical Society of America A}, {\em 36}, 1197--1208. \url{https://doi.org/10.1364/JOSAA.36.001197}. 

\medskip

Natterer, F. (1986).  {\em The mathematics of computerized tomography}. Chichester: Wiley.

\medskip

Nesterets, Ya. I. (2008). On the origins of decoherence and extinction contrast in phase-contrast imaging. {\em Optics Communications}, {\em 281}, 533--542. \url{https://doi.org/10.1016/j.optcom.2007.10.025}.

\medskip

Nesterets, Ya. I., \& Gureyev, T. E. (2014). Noise propagation in x-ray phase-contrast imaging and computed tomography. {\em Journal of Physics D: Applied Physics}, {\em 47}, 105402. \url{https://doi.org/10.1088/0022-3727/47/10/105402}.

\medskip

Nesterets, Ya. I., \& Gureyev, T. E. (2016). Partially coherent contrast-transfer-function
approximation. {\em Journal of the Optical Society of America A}, {\em 33}, 464--474. \url{https://doi.org/10.1364/JOSAA.33.000464}.

\medskip

Neuh\"{a}usler, U., Schneider, G., Ludwig, W., Meyer, M. A., Zschech, E., \& Hambach, D. (2003). X-ray microscopy in Zernike phase contrast mode at 4 keV photon energy with 60 nm resolution.  {\em Journal of Physics D: Applied Physics}, {\em 36}, A79--A82. \url{https://doi.org/10.1088/0022-3727/36/10A/316}.

\medskip

Neutze, R., Wouts, R., van der Spoel, D., Weckert, E., \& Hajdu, J. (2000). Potential for biomolecular imaging with femtosecond X-ray pulses. {\em Nature}, {\em 406}, 752--757. \url{https://doi.org/10.1038/35021099}. 

\medskip

Nugent, K. A., Gureyev, T. E., Cookson,  D., Paganin,  D., \& Barnea, Z. (1996). Quantitative phase imaging using hard x rays. {\em Physical Review Letters}, {\em 77}, 2961--2964. \url{https://doi.org/10.1103/PhysRevLett.77.2961}. 

\medskip

Nugent, K. A., Tran, C. Q., \& Roberts, A. (2003). Coherence transport through imperfect x-ray optical systems. {\em Optics Express}, {\em 11}, 2323--2328. \url{https://doi.org/10.1364/OE.11.002323}.

\medskip

Nye, J. F. (1999). {\em Natural focusing and fine structure of light}.  Bristol: Institute of Physics Publishing. 

\medskip

O'Holleran, K., Dennis, M. R., Flossmann, F., \& Padgett, M. J. (2008). Fractality of light's darkness. {\em Physical Review Letters}, {\em 100}, 053902. \url{https://doi.org/10.1103/PhysRevLett.100.053902}.

\medskip

Olbinado, M. P., Just, X., Gelet, J.-L., Lhuissier, P., Scheel, M., Vagovic, P., Sato, T., Graceffa, R., Schulz, J., Mancuso, A., Morse, J., \& Rack, A. MHz frame rate hard X-ray phase-contrast imaging using synchrotron radiation. {\em Optics Express}, {\em 25}, 13857--13871. \url{https://doi.org/10.1364/OE.25.013857}.

\medskip

Olivo, A., Ignatyev, K., Munro, P. R. T., \& Speller, R. D. (2011). Noninterferometric phase-contrast images obtained with incoherent x-ray sources. {\em Applied Optics}, {\em 50}, 1765--1769. \url{https://doi.org/10.1364/AO.50.001765}.

\medskip

Paganin, D., Mayo, S. C., Gureyev, T. E.,  Miller, P. R., \& Wilkins, S. W. (2002). Simultaneous phase and amplitude extraction from a single defocused image of a homogeneous object. {\em Journal of Microscopy}, {\em 206}, 33--40. \url{https://doi.org/10.1046/j.1365-2818.2002.01010.x}.  

\medskip

Paganin, D., Gureyev, T. E., Mayo, S. C., Stevenson, A. W., Nesterets, Ya. I., \& Wilkins, S. W. (2004a). X-ray omni microscopy. {\em Journal of Microscopy}, {\em 214}, 315--327. \url{ https://doi.org/10.1111/j.0022-2720.2004.01315.x}.

\medskip

Paganin, D., Gureyev, T. E., Pavlov, K. M., Lewis, R. A., \& Kitchen, M. (2004b). Phase retrieval using coherent imaging systems with linear transfer functions. {\em Optics Communications}, {\em 234}, 87--105. \url{https://doi.org/10.1016/j.optcom.2004.02.015}.

\medskip

Paganin, D., Barty, A., McMahon, P. J., \& Nugent, K. A. (2004c). Quantitative phase--amplitude microscopy. III. The effects of noise. {\em Journal of Microscopy}, {\em 214}, 51--61. \url{https://doi.org/10.1111/j.0022-2720.2004.01295.x}.  

\medskip

Paganin, D. M. (2006). {\em Coherent x-ray optics}. Oxford: Oxford University Press.

\medskip

Paganin, D. M., \& Gureyev, T. E. (2008). Phase contrast, phase retrieval and aberration balancing in shift-invariant linear imaging systems. {\em Optics Communications}, {\em 281}, 965--981. \url{https://doi.org/10.1016/j.optcom.2007.10.097}.

\medskip

Paganin, D. M., Petersen, T. C., \& Beltran, M. A. (2018). Propagation of fully coherent and partially coherent complex scalar fields in aberration space. {\em Physical Review A}, {\em 97}, 023835. \url{https://doi.org/10.1103/PhysRevA.97.023835}.

\medskip

Paganin, D. M., \& S\'anchez del R\'{\i}o, M. (2019). Speckled cross-spectral densities and their associated correlation singularities for a modern source of partially coherent x rays. {\em Physical Review A}, {\em 100}, 043813. \url{https://doi.org/10.1103/PhysRevA.100.043813}.

\medskip

Paganin, D. M., \& Morgan, K. S. (2019). X-ray Fokker--Planck equation for paraxial imaging. {\em Scientific Reports}, {\em 9}, 17537. \url{https://doi.org/10.1038/s41598-019-52284-5}.

\medskip

Paganin, D. M., \& Pelliccia, D. (2019). Tutorials on x-ray phase contrast imaging: some fundamentals and some conjectures on future developments.  Preprint available at \url{https://arxiv.org/abs/1902.00364}.

\medskip

Paganin, D. M., Favre-Nicolin, V., Mirone, A., Rack, A., Villanova, J., Olbinado, M. P., Fernandez, V., da Silva, J. C., \& Pelliccia, D. (2020). Boosting spatial resolution by incorporating periodic boundary conditions into single-distance hard-x-ray phase retrieval. {Journal of Optics}, {\em 22}, 115607. \url{https://doi.org/10.1088/2040-8986/abbab9}.

\medskip

Pagot, E., Cloetens, P., Fiedler, S., Bravin, A., Coan, P., Baruchel, J., H\"{a}rtwig, J., \& Thomlinson, W. (2003). A method to extract quantitative information in analyzer-based x-ray phase contrast imaging. {\em Applied Physics Letters}, {\em 82}, 3421--3423. \url{https://doi.org/10.1063/1.1575508}.  

\medskip

Pavlov, K. M., Gureyev, T. E., Paganin, D., Nesterets, Ya. I., Morgan, M. J., \& Lewis, R. A. (2004). Linear systems with slowly varying transfer functions and their application to x-ray phase-contrast imaging. {\em Journal of Physics D: Applied Physics}, {\em 37}, 2746--2750. \url{https://doi.org/10.1088/0022-3727/37/19/021}.

\medskip

Pavlov, K. M., Gureyev, T. E., Paganin, D., Nesterets, Ya. I.,  Kitchen, M. J., Siu, K. K. W., Gillam, J. E., Uesugi, K., Yagi, N.,  Morgan, M. J., \& Lewis, R. A. (2005). Unification of analyser-based and propagation-based X-ray phase-contrast imaging. {\em Nuclear Instruments and Methods in Physics Research A}, {\em 548}, 163--168. \url{https://doi.org/10.1016/j.nima.2005.03.084}.

\medskip

Pavlov, K. M., Paganin, D. M., Li, H., Berujon, S., Roug\'{e}-Labriet, H., \& Brun, E. (2020). X-ray multi-modal intrinsic-speckle-tracking. {\em Journal of Optics}, in press. \url{https://doi.org/10.1088/2040-8986/abc313}.

\medskip

Pelliccia, D., \& Paganin, D. M. (2012). Coherence vortices in vortex-free partially coherent x-ray fields. {\em Physical Review A}, {\em 86}, 015802. \url{https://doi.org/10.1103/PhysRevA.86.015802}. 

\medskip

Pelliccia, D., \& Paganin, D. M. (2013a). X-ray phase imaging with a laboratory source using selective reflection from a mirror.  {\em Optics Express}, {\em 21}, 9308--9314. \url{https://doi.org/10.1364/OE.21.009308}.

\medskip

Pelliccia, D., Rigon, L., Arfelli, F., Menk, R.-H., Bukreeva, I., \& Cedola, A. (2013b). A three-image algorithm for hard x-ray grating interferometry.  {\em Optics Express}, {\em 21}, 19401--19411. \url{https://doi.org/10.1364/OE.21.019401}.

\medskip

Pelliccia, D., Rack, A., Scheel, M., Cantelli, V., \& Paganin, D. M. (2016). Experimental x-ray ghost imaging. {\em Physical Review Letters}, {\em 117}, 113902. \url{https://doi.org/10.1103/PhysRevLett.117.113902}. 

\medskip

Pelliccia, D., Olbinado, M. P., Rack, A., Kingston, A. M., Myers, G. R., \& Paganin, D. M. (2018). Towards a practical implementation of X-ray ghost imaging with synchrotron light. {\em IUCrJ}, {\em 5}, 428--438. \url{https://doi.org/10.1107/S205225251800711X}.

\medskip

Pfeiffer, F., Bech, M., Bunk, O., Kraft, P., Eikenberry, E. F.,  Br\"onnimann, Ch., Gr\"unzweig, C., \& David, C. (2008). Hard-X-ray dark-field imaging using a grating interferometer. {\em Nature Materials}, {\em 7}, 134--137. \url{https://doi.org/10.1038/nmat2096}. 

\medskip

Pfeiffer, F. (2018). X-ray ptychography. {\em Nature Photonics}, {\em 12}, 9--17. \url{https://doi.org/10.1038/s41566-017-0072-5}.

\medskip

Pogany, A., Gao, D., \& Wilkins, S. W. (1997). Contrast and resolution in imaging with a microfocus x-ray source. {\em Review of Scientific Instruments}, {\em 68}, 2774--2782. \url{https://doi.org/10.1063/1.1148194}.

\medskip

Press, W. H., Teukolsky, S. A., Vetterling, W. T., \& Flannery, B. P. (2007). {\em Numerical recipes: the art of scientific computing}. (3rd ed.). Cambridge: Cambridge University Press.

\medskip

Rigon, L., Arfelli, F., \& Menk, R.-H. (2007). Three-image diffraction enhanced imaging algorithm to extract absorption, refraction and ultrasmall-angle scattering. {\em Applied Physics Letters}, {\em 90}, 114102. \url{https://doi.org/10.1063/1.2713147}.

\medskip

Risken, H. (1989). {\em The Fokker--Planck equation: methods of solution and applications}. (2nd ed.). Berlin: Springer Verlag.

\medskip

Ruelle, D. (1989). {\em Chaotic evolution and strange attractors}.  Cambridge: Cambridge University Press.

\medskip

Russo, P. (Ed.) (2018). {\em Handbook of x-ray imaging: physics and technology}.  Boca Raton: Taylor \& Francis, CRC Press.

\medskip

Sabatier, P. C. (2000). Past and future of inverse problems. {\em Journal of Mathematical Physics}, {\em 41}, 4082--4124.  \url{https://doi.org/10.1063/1.533336}.

\medskip

Saleh, B. E. A., \& Teich, M. C. (2007).  {\em Fundamentals of photonics}. (2nd ed.). Hoboken New Jersey: Wiley--Interscience.

\medskip

Saxton, W. O. (1994). What is the focus variation method? Is it new? Is it direct? {\em Ultramicroscopy}, {\em 55}, 171--181. \url{https://doi.org/10.1016/0304-3991(94)90168-6}.

\medskip

Sayre, D., \& Chapman, H. N. (1995). X-ray microscopy. {\em Acta Crystallographica A}, {\em 51}, 237--252. \url{https://doi.org/10.1107/S0108767394011803}.

\medskip

Schiske, P. (2002). Image reconstruction by means of focus series. {\em Journal of Microscopy}, {\em 207}, 154--154.  Note that this is a translation from the German, of a paper that first appeared in 1968, in the {\em Proceedings of the Fourth Regional Congress on Electron Microscopy}, Rome 1968, vol. 1, pp. 145--146.  \url{https://doi.org/10.1046/j.1365-2818.2002.01042.x}.

\medskip

Schneider, G. (1998). Cryo X-ray microscopy with high spatial resolution in amplitude and phase contrast. {\em Ultramicroscopy}, {\em 75}, 85--104. \url{https://doi.org/10.1016/S0304-3991(98)00054-0}.

\medskip

Schori, A., \& Shwartz, S. (2017a). X-ray ghost imaging with a laboratory source. {\em Optics Express}, {\em 25}, 14822--14828. \url{https://doi.org/10.1364/OE.25.014822}. 

\medskip

Schori, A., B\"{o}mer, C., Borodin, D., Collins, S. P., Detlefs, B., Moretti Sala, M., Yudovich, S., \& Shwartz, S. (2017b). Parametric down-conversion of x rays into the optical regime. {\em Physical Review Letters}, {\em 119}, 253902. \url{https://doi.org/10.1103/PhysRevLett.119.253902}.

\medskip

Schori, A., Borodin, D., Tamasaku, K., \& Shwartz, S. (2018). Ghost imaging with paired x-ray photons. {\em Physical Review A}, {\em 97}, 063804. \url{https://doi.org/10.1103/PhysRevA.97.063804}.

\medskip

Sears, F. W., \& Salinger, G. L. (1975). {\em Thermodynamics, kinetic theory and statistical thermodynamics}. (3rd ed.). Reading: Addison--Wesley.

\medskip

Shannon, C. E. (1948a).  A mathematical theory of communication. {\em Bell System Technical Journal}, {\em 27}, 379--423. \url{https://doi.org/10.1002/j.1538-7305.1948.tb01338.x}.

\medskip

Shannon, C. E. (1948b).  A mathematical theory of communication (cont.). {\em Bell System Technical Journal}, {\em 27}, 623--656. \url{https://doi.org/10.1002/j.1538-7305.1948.tb00917.x}.

\medskip

Snigirev, A., Snigireva, I., Kohn, V., Kuznetsov, S., \& Schelokov, I. (1995). On the possibilities of x-ray phase contrast microimaging by coherent high-energy synchrotron radiation. {\em Review of Scientific Instruments}, {\em 66}, 5486--5492. \url{https://doi.org/10.1063/1.1146073}. 

\medskip

Snigirev, A., Snigireva, I., Kohn, V. G., \& Kuznetsov, S. M. (1996a). On the requirements to the instrumentation for the new generation of the synchrotron radiation sources. Beryllium windows. {\em Nuclear Instruments and Methods in Physics Research A}, {\em 370}, 634--640. \url{https://doi.org/10.1016/0168-9002(95)00849-7}. 

\medskip

Snigirev, A., Kohn, V., Snigireva, I., \& Lengeler, B. (1996b). A compound refractive lens for focusing high-energy X-rays. {\em Nature}, {\em 384}, 49--51. \url{https://doi.org/10.1038/384049a0}. 

\medskip

Sofer, S., Strizhevsky, E., Schori, A., Tamasaku, K., \& Shwartz, S. (2019). Quantum enhanced x-ray detection. {\em Physical Review X}, {\em 9}, 031033. \url{https://doi.org/10.1103/PhysRevX.9.031033}.

\medskip

Somenkov, V. A., Tkalich, A. K., \& Shil'shtein, S. Sh. (1991). Refraction contrast in x-ray introscopy.  {\em Soviet Physics: Technical Physics}, {\em 36}, 1309--1311.

\medskip

Spence, J. C. H. (2003). {\em High-resolution electron microscopy}. (3rd ed.). Oxford: Oxford University Press.

\medskip

Strobl, M. (2014). General solution for quantitative dark-field contrast imaging with grating interferometers. {\em Scientific Reports}, {\em 4}, 7243. \url{https://doi.org/10.1038/srep07243}.

\medskip 

Subbarao, M., Wei, T.-C, \& Surya, G. (1995). Focused image recovery from two defocused images recorded with different camera settings. {\em IEEE Transactions on Image Processing}, {\em 4}, 1613--1628. \url{https://doi.org/10.1109/TIP.1995.8875998}.

\medskip

Teague, M. R. (1983). Deterministic phase retrieval: a Green's function solution. {\em Journal of the Optical Society of America}, {\em 73}, 1434--1441. \url{https://doi.org/10.1364/JOSA.73.001434}. 

\medskip

Thomson, M. (2013). {\em Modern particle physics}.  Cambridge: Cambridge University Press.  

\medskip

Tomie, T. (1994). X-ray lens. {\em Japanese Patent}, No. 6-045288.

\medskip

Tomie, T. (2010). The birth of the X-ray refractive lens. {\em Spectrochimica Acta Part B}, {\em 65}, 192--198. \url{https://dx.doi.org/10.1016/j.sab.2010.02.009}.

\medskip

Van Dyck, D. (1975). The path integral formalism as a new description for the diffraction of high-energy electrons in crystals. {\em physica status solidi (b)}, {\em 72}, 321--336. \url{https://doi.org/10.1002/pssb.2220720135}.

\medskip

Van Dyck, D. (1985). Image calculations in high-resolution electron microscopy: problems, progress, and prospects. {\em Advances in Electronics and Electron Physics}, {\em 65}, 295--355. \url{https://doi.org/10.1016/S0065-2539(08)60880-X}.

\medskip

Vartanyants, I. A., \& Robinson, I. K. (2003). Origins of decoherence in coherent X-ray diffraction experiments. {\em Optics Communications}, {\em 222}, 29--50. \url{https://doi.org/10.1016/S0030-4018(03)01558-X}.

\medskip

Volkovich, S., \& Shwartz, S. (2020). Subattosecond x-ray Hong--Ou--Mandel metrology. {\em Optics Letters}, {\em 45}, 2728--2731. \url{https://doi.org/10.1364/OL.382044}.

\medskip

Weitkamp, T., Diaz, A., David, C., Pfeiffer, F., Stampanoni, M., Cloetens, P., \& Ziegler, E. (2005). X-ray phase imaging with a grating interferometer. {\em Optics Express}, {\em 13}, 6296--6304. \url{https://doi.org/10.1364/OPEX.13.006296}.

\medskip

Wernick, M. N., Wirjadi, O., Chapman, D., Zhong, Z., Galatsanos, N. P., Yang, Y., Brankov, J. G., Oltulu, O., Anastasio, M. A., \&  Muehleman, C. (2003). Multiple-image radiography. {\em Physics in Medicine and Biology}, {\em 48}, 3875--3895. \url{https://doi.org/10.1088/0031-9155/48/23/006}.

\medskip 

White, V., \& Cerrina, F. (1992). Metal-less X-ray phase-shift masks for nanolithography. {\em Journal of Vacuum Science and Technology B}, {\em 10}, 3141--3144. \url{https://doi.org/10.1116/1.585944}. 

\medskip

Wilkins, S. W., Gureyev, T. E., Gao, D., Pogany, A., \& Stevenson, A. W. (1996). Phase-contrast imaging using polychromatic hard X-rays. {\em Nature}, {\em 384}, 335--338. \url{https://doi.org/10.1038/384335a0}. 

\medskip

Wilkins, S. W., Nesterets, Ya. I., Gureyev, T. E., Mayo, S. C., Pogany, A., \& Stevenson, A. W. (2014). On the evolution and relative merits of hard X-ray phase-contrast imaging methods. {\em Philosophical Transactions of the Royal Society A}, {\em 372}, 20130021. \url{https://doi.org/10.1098/rsta.2013.0021}. 

\medskip

Wolf, E. (1969). Three-dimensional structure determination of semi-transparent objects by holographic data. {\em Optics Communications}, {\em 1}, 153--156. \url{https://doi.org/10.1016/0030-4018(69)90052-2}. 

\medskip

Wolf, E. (1982). New theory of partial coherence in the space–frequency domain. Part I: spectra and cross spectra of steady-state sources. {\em Journal of the Optical Society of America}, {\em 72}, 343--351. \url{https://doi.org/10.1364/JOSA.72.000343}.

\medskip

Wolf, E. (2007). {\em Introduction to the theory of coherence and polarization of light}.  Cambridge: Cambridge University Press.  

\medskip

Yaroslavsky, L., \& Eden, M. (1996). {\em Fundamentals of digital optics}.  Boston: Birkh\"{a}user.  

\medskip

Yashiro, W., Terui, Y., Kawabata, K., \& Momose, A. (2010). On the origin of visibility contrast in
x-ray Talbot interferometry. {\em Optics Express}, {\em 18}, 16890--16901. \url{https://doi.org/10.1364/OE.18.016890}.

\medskip

Yashiro, W., Harasse, S., Kawabata, K., Kuwabara, H., Yamazaki, T., \& Momose, A. (2011). Distribution of unresolvable anisotropic microstructures revealed in visibility-contrast images using x-ray Talbot interferometry. {\em Physical Review B}, {\em 84}, 094106. \url{https://doi.org/10.1103/PhysRevB.84.094106}.

\medskip

Yu, H., Lu, R., Han, S., Xie, H., Du, G., Xiao, T., \& Zhu, D. (2016). Fourier-transform ghost imaging with hard x rays. {\em Physical Review Letters}, {\em 117}, 113901. \url{https://doi.org/10.1103/PhysRevLett.117.113901}.  

\medskip

Yu, F. T. S. (2017). {\em Entropy and information optics: connecting information and time}.  Boca Raton: Taylor \& Francis, CRC Press.

\medskip

Yu, B., Weber, L., Pacureanu, A., Langer, M., Olivier, C., Cloetens, P., \& Peyrin, F. (2017). Phase retrieval in 3D X-ray magnified phase nano CT: imaging bone tissue at the nanoscale. {\em 2017 IEEE 14th International Symposium on Biomedical Imaging (ISBI 2017)}, 56--59. \url{https://doi.org/10.1109/ISBI.2017.7950467}.

\medskip

Yu, B., Weber, L., Pacureanu, A., Langer, M., Olivier, C., Cloetens, P., \& Peyrin, F. (2018). Evaluation of phase retrieval approaches in magnified X-ray phase nano computerized tomography applied to bone tissue. {\em Optics Express}, {\em 26}, 11110--11124. \url{https://doi.org/10.1364/OE.26.011110}.

\medskip

Zdora, M.-C. (2018). State of the art of X-ray speckle-based phase-contrast and dark-field imaging. {\em Journal of Imaging}, {\em 4}, 60. \url{https://doi.org/10.3390/jimaging4050060}.

\medskip

Zernike, F. (1938). The concept of degree of coherence and its application to optical problems. {\em Physica}, {\em 5}, 785--795. \url{https://doi.org/10.1016/S0031-8914(38)80203-2}. 

\medskip

Zernike, F. (1942). Phase contrast, a new method for the microscopic observation of transparent objects. {\em Physica}, {\em 9}, 686--698. \url{https://doi.org/10.1016/S0031-8914(42)80035-X}. 

\medskip

Zhang, A.-X., He, Y.-H., Wu, L.-A., Chen, L.-M., \& Wang, B.-B. Tabletop x-ray ghost imaging with ultra-low radiation. (2018). {\em Optica}, {\em 5}, 374--377. \url{https://doi.org/10.1364/OPTICA.5.000374}.   

\end{document}